# The multifractal nature of turbulent energy dissipation

By CHARLES MENEVEAU† AND K. R. SREENIVASAN
Mason Laboratory, Yale University, New Haven, CT 06520, USA

The intermittency of the rate of turbulent energy dissipation $\epsilon$ is investigated experimentally, with special emphasis on its scale-similar facets. This is done using a general formulation in terms of multifractals, and by interpreting measurements in that light. The concept of multiplicative processes in turbulence is (heuristically) shown to lead to multifractal distributions, whose formalism is described in some detail. To prepare proper ground for the interpretation of experimental results, a variety of cascade models is reviewed and their physical contents are analysed qualitatively. Point-probe measurements of $\epsilon$ are made in several laboratory flows and in the atmospheric surface layer, using Taylor's frozen-flow hypothesis. The multifractal spectrum $f(\alpha)$ of $\epsilon$ is measured using different averaging techniques, and the results are shown to be in essential agreement among themselves and with our earlier ones. Also, long data sets obtained in two laboratory flows are used to obtain the *latent* part of the $f(\alpha)$ curve, confirming Mandelbrot's idea that it can in principle be obtained from linear cuts through a three-dimensional distribution. The tails of distributions of box-averaged dissipation are found to be of the square-root exponential type, and the implications of this finding for the $f(\alpha)$ distribution are discussed. A comparison of the results to a variety of cascade models shows that binomial models give the simplest possible mechanism that reproduces most of the observations. Generalizations to multinomial models are discussed.

## 1. Introduction

It has long been known (Batchelor & Townsend 1949) that small scales of turbulence are intermittent. The small-scale quantity that has received most attention is the rate of dissipation of kinetic energy, $\epsilon$. Figures 1(*a*) and 1(*b*) show experimental signals of a representative component of $\epsilon$ obtained respectively in a laboratory boundary layer and in the atmospheric surface layer. They illustrate the intermittent nature of $\epsilon$ and emphasize that it becomes increasingly conspicuous with increasing flow Reynolds number.

A conceptually appealing view, dating back to Obukhov (1962) and Kolmogorov (1962), visualizes the transfer of kinetic energy to the small scales as a self-similar cascade with an associated multiplicative process. This view is still at the heart of many phenomenological intermittency models. Based on the central-limit theorem, Kolmogorov (1962) and Obukhov (1962) proposed a lognormal distribution of the rate of dissipation (see also Yaglom 1966 and Gurvich & Yaglom 1967), while Novikov (1971) and Mandelbrot (1972) clarified inherent problems of the lognomal model. Another type of multiplicative intermittency model was proposed by

† Present address: Department of Mechanical Engineering, Johns Hopkins University, Baltimore, MD 21218, USA.

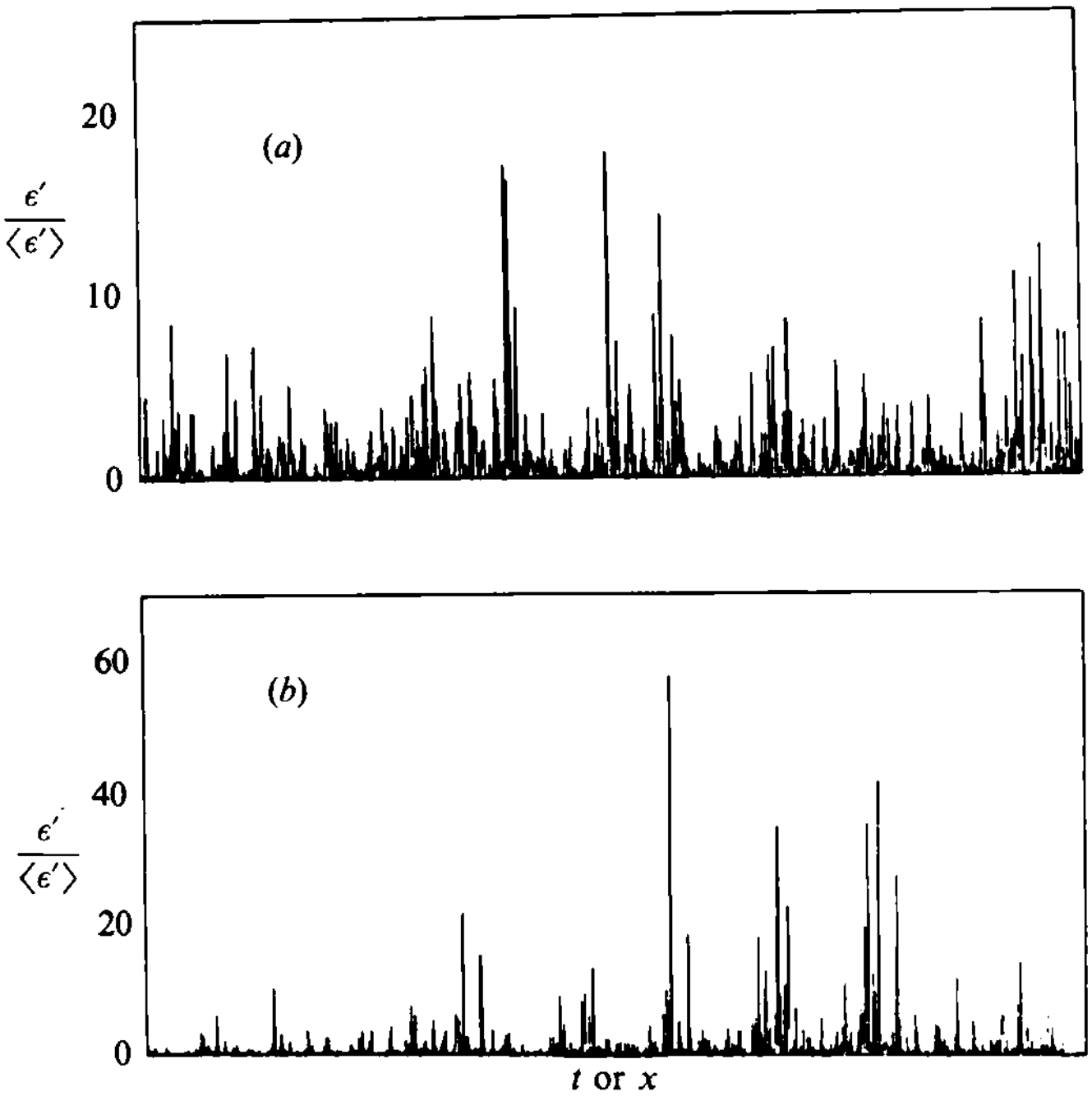

FIGURE 1. Typical signals of a representative component of $\epsilon$, namely $\epsilon' \sim (\mathrm{d}u_1/\mathrm{d}t)^2$ normalized by the mean: (*a*) was obtained in a laboratory boundary layer at a moderate Reynold number, and (*b*) in the atmospheric surface layer at a high Reynolds number. For a description of the experimental conditions, see §3.1 and table 1.

Novikov & Stewart (1964) and further generalized by Novikov (1969, 1971, 1990). Mandelbrot (1974) introduced the general cascade model of random curdling and interpreted the Novikov–Stewart model geometrically using notions of fractal geometry. More detailed physical and geometrical implications of this type of model were analysed by Frisch, Sulem & Nelkin (1978), who coined for their specific version the name $\beta$-model. Kraichnan (1974) used a cascade model similar to random curdling (but with a spatially less explicit structure) expressed in terms of band-limited velocity fluctuations. For a general analysis of the physical content of the ideas behind self-similar cascades, see Kraichnan (1974) and Nelkin (1989).

Of particular interest in Mandelbrot's (1974) analysis is the prediction that certain high-order moments computed from point-probe measurements will diverge at high Reynolds numbers. Schertzer & Lovejoy (1985) analysed data from atmospheric turbulence, with special emphasis on this prediction, and concluded that it is correct. They also proposed a model exhibiting such a behaviour (see also the review article by Levich 1987). Laboratory measurements, however, have not confirmed the prediction on the divergence of high-order moments (Anselmet *et al.* 1984; Gagne 1987).

In a parallel development that was silent on multistage cascades, different models for the geometry of dissipative structures were proposed in terms of sheets (Corrsin 1962) and tubes (Tennekes 1968). An ambitious but incomplete experimental investigation by Kuo & Corrsin (1972) suggested that the structure was somewhat more filament-like rather than blob-like or slab-like. A summary and discussion of these models, as well as a treatment of the fine structure via the application of the Hilbert transform, was given by Sreenivasan (1985). Numerical simulations using vortex methods (Chorin 1982) have confirmed the intermittent nature of turbulence activity and demonstrated the usefulness of fractal geometry in describing it. Interesting analogies with critical phenomena and polymer dynamics have been explored (Nelkin 1973; Mori 1980; Hentschel & Procaccia 1982; Chorin 1988*a*, *b*). Furthermore, Mori (1980) highlighted interesting connections between fractals and local expansion rates in the context of turbulence.

From an experimental point of view, little attention was given to Mandelbrot's general model of random curdling until recently. Numerous measurements of intermittency were made (e.g. Gibson, Stegen & Connell 1970; Tennekes & Wyngaard 1972; Frenkiel & Klebanoff 1975; McConnell 1976; Park 1976; Van Atta & Antonia 1980, to name but a few), and compared to either the lognormal or the $\beta$-model hypotheses. The inadequacy of lognormal models for high-order moments was demonstrated by Sreenivasan, Antonia & Danh (1977), and the high-order velocity structure function measurements of Anselmet *et al.* (1984) made it clear that both lognormal and $\beta$-models were inadequate.

To account for the observations, Frisch & Parisi (1985) introduced the idea of distributions of singularities, all lying on interwoven sets of varying fractal dimensions, and coined the name multifractal. They related such a description to the hierarchy of moment exponents originally proposed by Mandelbrot (1974) to characterize his random curdling model. This was advanced further by Benzi *et al.* (1984) who introduced the so-called random $\beta$-model and proposed its application to measures created on strange attractors in phase space. A similar path was taken by Hentschel & Procaccia (1983) who introduced the hierarchy of the so-called 'generalized dimensions' $D_q$, and Halsey *et al.* (1986), who coined the name $f(\alpha)$ for the set of fractal dimensions characterizing multifractals. Mandelbrot (1989) has further clarified some properties of $f(\alpha)$ in terms of his earlier work of 1974. In fact, much of Novikov's early work can, with hindsight, be cast in terms of multifractals.

We feel that the theory of multifractals has acquired a certain maturity at this point, permitting an intuitive understanding of multiplicative processes and of the intermittent distributions in turbulence. This feeling is due, in part, to the number of applications in physical sciences where multifractals and multiplicative processes have been found useful (see e.g. Paladin & Vulpiani 1987). In part, it is based on measurements on the multifractal nature of dissipation fields in turbulent flows (Meneveau & Sreenivasan 1987*a*, 1989; Sreenivasan & Meneveau 1986, 1988; Prasad, Meneveau & Sreenivasan 1988; Ramshankar 1988; Meneveau 1989). We therefore think that it is worthwhile consolidating results relating to the multifractal nature of $\epsilon$. With this in view, this paper expands some of the earlier work, and provides a careful account of the measurements. It reviews previous cascade models in a unified fashion and examines them in the context of multifractality. Finally, it presents a detailed analysis of the behaviour of high-order moments of $\epsilon$, and its implications for the observed intermittency.

The present measurements were made by stationary single-wire probes. Several flows studied here were created in the laboratory at moderate Reynolds numbers; the scaling range was limited but high-order moments could be measured accurately because of guaranteed statistical convergence. We also made measurements in the atmospheric surface layer for which the scaling range is large but high-order moments cannot be obtained accurately (because the data records required would be so long as to preclude stationary conditions). The focus will be on the scaling behaviour of the dissipation integrated over 'volumes' of sizes pertaining to the inertial range. As noted by Kraichnan (1974), such a variable is in itself not an inertial range quantity and need not follow the self-similar behaviour expected in the inertial range. It is therefore of considerable interest to explore whether such a variable does indeed exhibit self-similar behaviour and, if so, its relationship with other scaling exponents of the inertial range.

The paper is structured as follows. Section 2 provides an introduction to multiplicative processes. It includes basic definitions of multiplicative processes (§2.1), early cascade models (§2.2), the characterization of multifractals by singularity spectra and generalized dimensions (§2.3), Mandelbrot's formalism of random curdling (§2.4), some special cases of random curdling (§2.5), non-cascade models (§2.6) and, finally, practical considerations concerning the measurement of multifractal characteristics (§2.7). Section 3 presents experimental results concerning the multifractal nature of the dissipation field $\epsilon$ (approximated by the square of the single derivative of the streamwise velocity, obtained using Taylor's hypothesis). A detailed discussion of the power-law behaviour and of the convergence of moments as a function of averaging domain is presented. Also, experimental results concerning the scaling behaviour of high-order moments are analysed by studying the tails of the distribution functions of the dissipation, and their relationship with the multifractal spectrum. Section 4 presents an analysis of the measured multifractal spectrum of the field of dissipation, as well as a detailed comparison of the results to a variety of cascade models. A summary of conclusions is presented in §5.

## 2. Multiplicative processes and cascade theories: a review

### 2.1. *Multiplicative processes: general concept and definitions*

The basic ingredient of multiplicative processes is that large 'eddies' or fluid pieces transform or break down into smaller ones; the fragmented pieces themselves yield even smaller ones, and so on. This then defines pieces of different generations; the generation step will be denoted by $n$. To each piece is associated a characteristic linear dimension $r$ (for example, the diameter in the case of spherical eddies). We assume that the characteristic scale of a piece of the $n$th generation, $r(n)$, will be given by the product of $n$ numbers (to be called length multipliers $l_j$, $1 < j < n$), each of which is the ratio of consecutive lengthscales. In other words,

$$r(n) = r(0) \prod_{j=1}^{n} r(j)/r(j-1) = r(0) \prod_{j=1}^{n} l_j. \tag{2.1}$$

Another vital ingredient is the concept of a measure density which, in the present context, is the rate of dissipation per unit volume $\epsilon(\boldsymbol{x})$, where $\boldsymbol{x}$ belongs to the union of all the pieces. Of particular interest is the total dissipation $E_r$ in a certain piece $\Omega$ of size $r$. This will be given by the integral of $\epsilon(\boldsymbol{x})$ over the piece $\Omega$ as

$$E_r = \int_{x \in \Omega} \epsilon(\boldsymbol{x})\, \mathrm{d}^3 x. \tag{2.2}$$

When a piece $\Omega$ decays into smaller ones, each smaller piece can be thought of as receiving a fraction of $E_r$. Analogous to lengthscales, the total dissipation on a certain piece of size $r(n)$ of the $n$th generation will be given by the product of $n$ numbers (to be called measure multipliers $M_j$), each of which is the ratio of consecutive measures. That is,

$$E_{r(n)} = E_{r(0)} \prod_{j=1}^{n} E_{r(j)}/E_{r(j-1)} = E_{r(0)} \prod_{j=1}^{n} M_j. \tag{2.3}$$

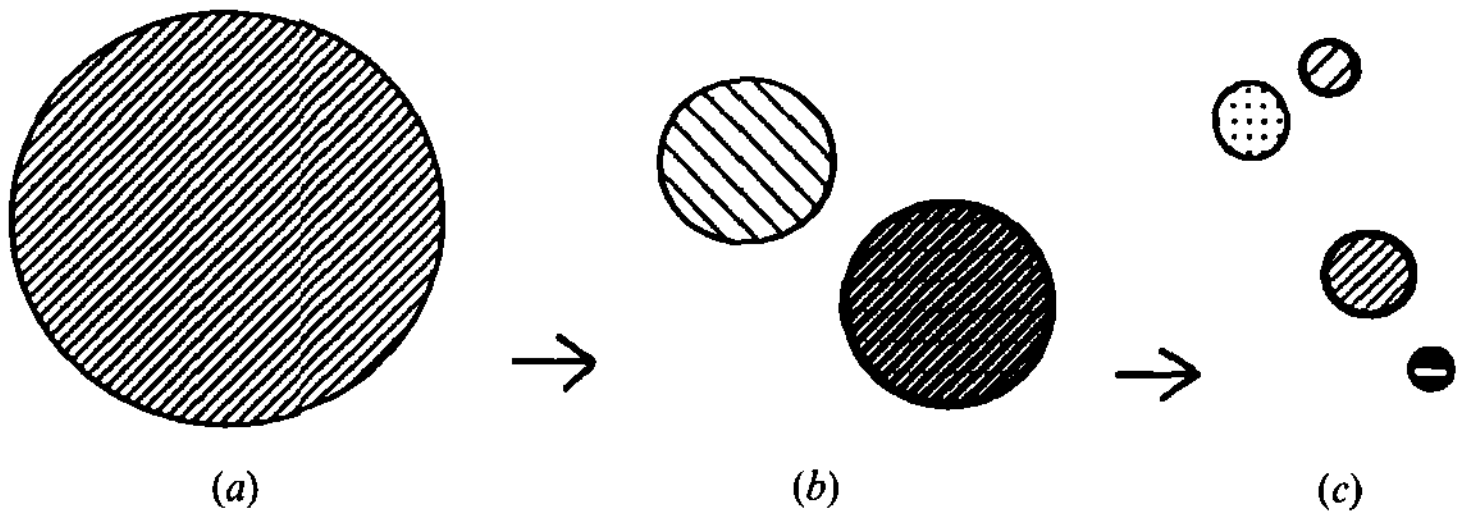

FIGURE 2. Schematic representation of an isotropic multiplicative process. A large piece, (*a*), is divided into two smaller pieces, (*b*). Both pieces or 'blobs' may have a different density of measure, as indicated by the different shading. After the next iteration of the multiplicative process, each piece of (*b*) is divided into even smaller pieces, (*c*), etc.

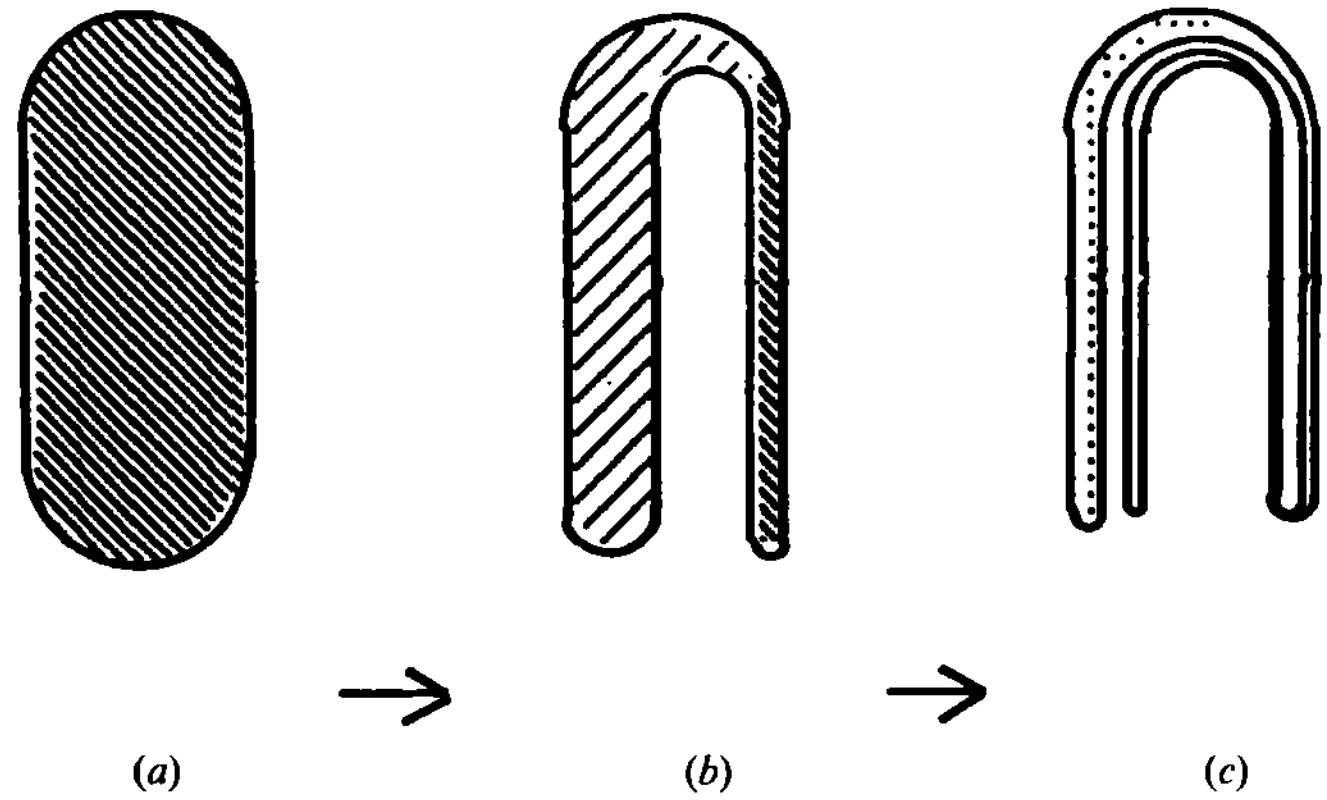

FIGURE 3. Schematic representation of three stages of a stretching and folding process. A piece (*a*), is stretched in the vertical direction, contracted unequally (thus accounting for the unequal thickness and measure) and then folded back to form the piece, (*b*). After another similar step, (*c*) is obtained.

It is then clear that Richardson's (1922) picture of turbulence cascade, in which 'blobs' or whorls of turbulent fluid break down into smaller pieces – each 'feeding on their velocity', i.e. receiving a certain fraction of the flux of kinetic energy from larger scales – is a possible multiplicative process. This is depicted schematically in figure 2.

Another generic process that occurs in nonlinear dynamical systems can be described qualitatively as the process of stretching and folding, typical of the evolution of a 'blob' of points corresponding to different initial conditions in phase space. This is also referred to as a horseshoe process. Figure 3 is a schematic of three stages of such a process, where the stretching in the vertical direction is followed by a folding. We adopt the view that the stretching and folding of turbulent fluid elements in physical space can be regarded qualitatively in the same spirit. The overall isotropy and simplicity of the breakdown of blobs is not present here, but it is a multiplicative process in the sense that the thickness and density of each of the pieces are products of successive multipliers.

Summarizing, a multiplicative process is one of fragmentation of a large piece into smaller ones, with each new piece receiving a fraction of the 'measure' of the larger unit, in such a way that the size and measure of a small piece are products of multipliers $l_j$ and $M_j$ ($M_j \geqslant 0, l_j \leqslant 1$) associated with its predecessors at all previous stages or generations. In the absence of definitive deductions from Navier–Stokes equations, the quantities $l_j$ and $M_j$ have to be considered random variables with a certain probability distribution. When such distribution functions do not depend on the level $j$ (or the characteristic size $r$), self-similarity will appear and, as will be seen below, power-law behaviour occurs in the moments and distribution function of $E_r$. Physically, this implies that as long as the eddy size is larger than the Kolmogorov scale $\eta$ and smaller than the integral scale $L$ of the flow, the precise dynamics resulting from the Navier–Stokes equations – which determines the multipliers – should be independent of viscosity and, far enough from physical boundaries, also independent of boundary conditions.

To clarify notation we stress that the index $j$ refers to different generations. The variables $M_j$ and $l_j$ assume different values at a particular generation at different locations. When such a distinction is necessary, it will be denoted by a second index $i$; e.g. $M_{j,i}$ is the measure multiplier corresponding to a piece at position $i$ of generation $j$.

## 2.2. *Some early cascade models*

### 2.2.1. *The 1941 theory of Kolmogorov*

This theory of universal, isotropic distribution of small scales of motion envisages a cascade where the only relevant quantity is the mean flux $\langle \epsilon \rangle$ or $\langle E_r \rangle$ of energy from large to small scales. This is a trivial multiplicative process in which the measure multipliers at a given stage are equal. There is no apparent dynamical reason for dismissing this possibility, but experience (see figure 1) precludes it. As pointed out by Kraichnan (1974), non-intermittent distributions of $\epsilon_r$ and $E_r$ can be produced only by strong spatial mixing of energy at all scales of motion, such that energy equilibration occurs as soon as it is transferred from any one scale to its offsprings. Kolmogorov's (1941) theory implies that the mixing is so large that all fluctuations in the inertial range are smoothed. By the definition of the inertial range, viscosity cannot be responsible for this equilibration. It could in principle occur by the action of pressure fluctuations, which are known (Batchelor 1953) to transfer energy from one velocity component to another at roughly the same scales. Dimensional arguments show that the typical timescale of this process is of order $\tau(r) \sim r/\Delta u_r$, where $\Delta u_r$ is a typical velocity increment over the distance $r$. This is also the timescale characterizing the decay of an eddy into its offspring, and one could therefore argue that there is barely enough time for equilibrating energy at a given scale. A certain degree of equilibration is likely to occur, but inhomogeneities at all scales remain because turbulence structures decay before the process is completed.

### 2.2.2 *The hypothesis of lognormality*

In order to account for the observed intermittency, it is natural to assume that the $M_j$ in (2.3) fluctuate according to some distribution. Taking the logarithm of (2.3), one can write

$$\ln [E_r/E_L] = \sum_{k=1}^{n} \ln (M_j), \tag{2.4}$$

where $E_L$ is the 'total' dissipation contained in pieces of fluid of size $L$ ($\sim r(0)$). Therefore, $\ln [E_r/E_L]$ is the sum of identically distributed random variables $\ln (M_j)$. For the sake of simplicity, let us assume that these random variables are finite (i.e. $M_j \neq 0$). Kolmogorov (1962) applied central-limit theorem to argue that $\ln (E_r)$ and $\ln (\epsilon_r)$ should have Gaussian distributions.

However, central-limit theorem cannot be applied to rare events, which are the ones that contribute most to high-order moments. This was noticed by Novikov (1971), who concluded that high-order moments cannot follow lognormal distribution. This was clarified further by Mandelbrot (1972). Furthermore, Orszag (1970) showed that if the moments followed lognormality, they could not uniquely determine the distribution. A further analysis of the rather unphysical conditions needed for asymptotic lognormal distributions can be found in Kraichnan (1974).

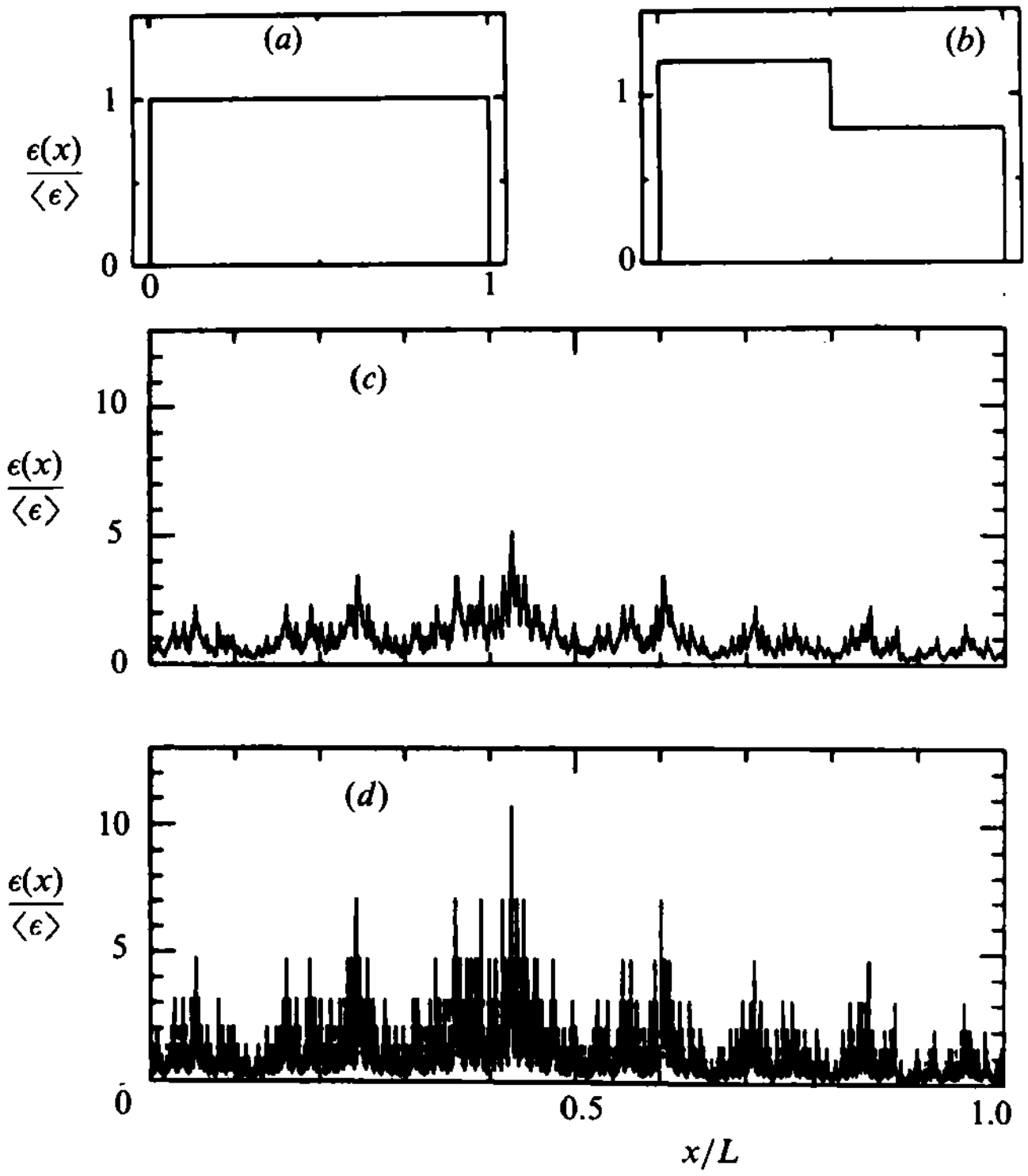

FIGURE 4. Binomial measure $\epsilon(x)/\langle\epsilon\rangle$ on the unit interval, using $M = 0.6$ or 0.4. (*a*) The original uniform distribution of density, and (*b*) after one fragmentation. The total dissipation on the two sides are 0.6 and 0.4, and the corresponding densities of $\epsilon(x)$ are 1.2 and 0.8. (*c*) $\epsilon(x)$ after 9 steps and (*d*) after 13 steps.

### 2.2.3. *β-model of fractally homogeneous turbulence*

In this model the multipliers $M_j$ are non-zero and equal on a fraction $\beta$ of the new offspring, but zero on the other fraction $(1-\beta)$ of the offspring (Novikov & Stewart 1964; Mandelbrot 1974; Frisch *et al.* 1978). Scaling properties appear again if $\beta$ is assumed to be independent of $r$. There is no mixing between the empty and non-empty regions. Therefore, this model corresponds to the assumption that the timescale of spatial mixing is much larger than that associated with the decay of eddies into smaller ones.

As will be seen in §3, the measured high-order moments of $E_r$ depart markedly from predictions of both lognormal and $\beta$-models. It is therefore necessary to study general multiplicative processes and their properties, to which the next two subsections are devoted.

### 2.3. *Multifractals and their characterization by singularity spectra and generalized dimensions*

The question addressed here is the following: Given a function $\epsilon(x)$ such as in figure 1, how best can one characterize it, and what can be said about the multiplicative process that generated it? It is apparent that the mean and variance of $\epsilon(x)$ or the variable $E_r$ contain little information about $\epsilon(x)$; furthermore, they are different for each cascade step. It has already been seen that lognormal and $\beta$-models are not general enough. The required quantifiers will be introduced via the simple example of a self-similar binomial process, but the formalism to be discussed is valid for general multiplicative processes.

The binomial process to be discussed here occurs in one dimension, where an initial segment of size $L$ is divided into two segments of equal length ($l_j = \frac{1}{2}$), and the $M_j$ have a bimodal distribution with only two possible values, say $M_1 = p_1 = 0.6$ or $M_2 = p_2 = 0.4$. That is, its distribution is given in terms of two $\delta$-functions as

$$p(M) = 0.5\{\delta(M-0.4)+\delta(M-0.6)\}, \tag{2.5}$$

independent of the cascade step $j$. For the present discussion, we additionally impose conservation of the measure at each step, which means that each piece gives rise to two pieces with the $M_j$ of both pieces always adding to unity. Whether the multiplier 0.4 (or 0.6) corresponds to the right or left offspring is selected at random. Figure 4 shows the density obtained by such a process after $n = 0$, 1, 9 and 13 iterations or generations. To make contact with dissipation later, we use the symbol $\epsilon$ to denote the measure. After $n$ steps, the size of each piece is $r/L = 2^{-n}$, and it is easy to see that $E_r$ can assume values given by

$$E_r/E_L = [p_1^{m/n} p_2^{1-m/n}]^n \quad (\text{where } m = 0, 1, \ldots, n).$$

Each such value of $E_r/E_L$ occurs $n!/[m!(n-m)!]$ times. Since $n = -\log_2(r/L)$, one can define a new (random) variable $\alpha$ according to

$$\alpha = \ln(E_r/E_L)/\ln(r/L) = -(m/n)\log_2 p_1 - (1-m/n)\log_2 p_2, \tag{2.7}$$

which now only depends on the ratio $m/n$ ($0 \leqslant m/n \leqslant 1$), rather than on $n$ itself. For illustrative purposes, $\alpha$ obtained from the binomial measure of figure 4 is shown in figure 5 after 9 and 13 iterations. We see that the random variable $\alpha$ fluctuates between limits that are independent of $r$ or $n$, which suggests that the process can now be characterized in terms of the distribution of the rescaled variable $\alpha$. For practical applications to follow, it is more convenient to define $\alpha$ as a local scaling or Hölder exponent (Mandelbrot 1989) according to

$$E_r/E_L \sim (r/L)^{\alpha} \quad \text{or} \quad \epsilon_r/\epsilon_L \sim (r/L)^{\alpha-d} \tag{2.8}$$

instead of as the ratio of logarithms (for the present example in one dimension, $d = 1$). The convenience one gains is that this eliminates worries about non-unity prefactors in (2.8), which in general make the convergence of $\alpha$ to a scale-independent variable rather slow (Meneveau & Sreenivasan 1989). Further, writing in this form emphasizes the fact that different values of $\alpha$ reflect different strengths of singularity as the box size tends to zero.

We now turn to the distribution of $\alpha$ itself. Figure 6 shows that $\Pi_r(\alpha)$, the (normalized) probability density function of $\alpha$ at the two chosen steps of the cascade, becomes narrower and more peaked with increasing $n$. Applying Stirling's formula to the binomial coefficient in the limit of very large $n$, it is apparent that the rescaled logarithmic distribution function $h(\alpha)$, defined as

$$h(\alpha) = \ln[\Pi_r(\alpha)]/\ln(L/r), \tag{2.9}$$

will tend asymptotically to

$$h(\alpha) = 1-(1-m/n)\log_2(n/m-1)+\log_2(n/m). \tag{2.10}$$

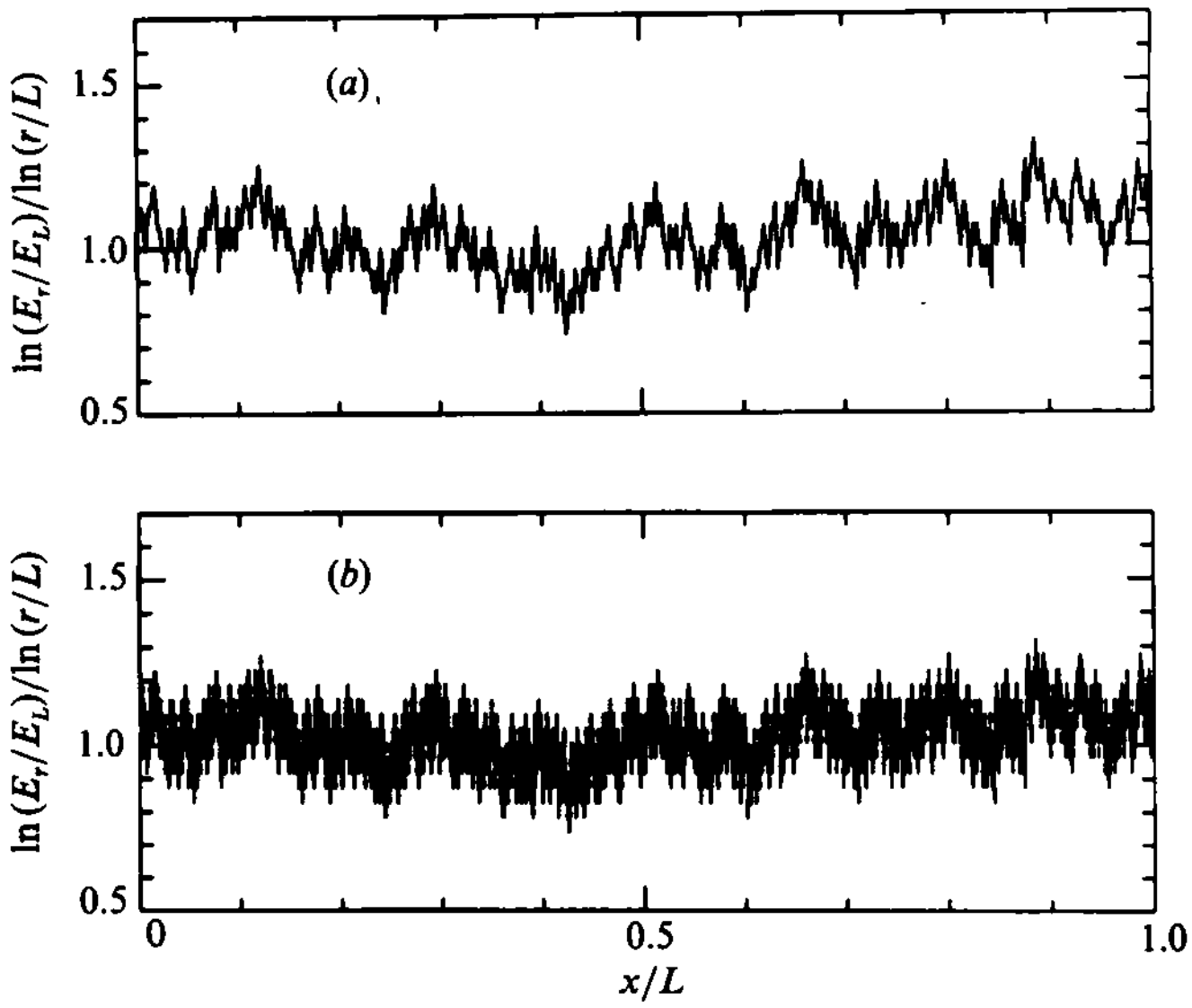

FIGURE 5. Local values of $\alpha = \ln(E_r/E_L)/\ln(r/L)$ for (*a*) $r/L = 2^{-9}$ after 9 steps, and (*b*) for $r/L = 2^{-13}$ after 13 steps.

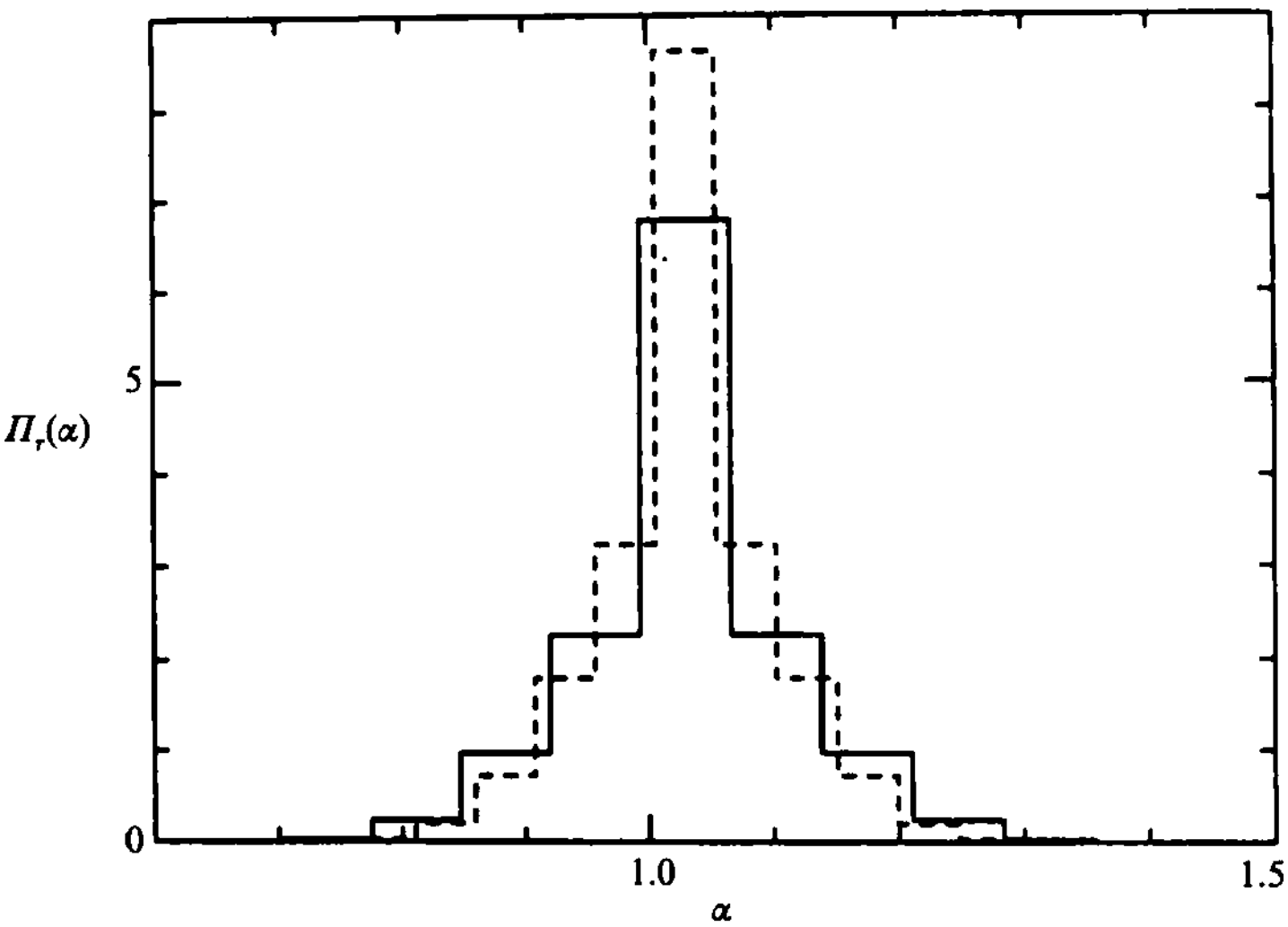

FIGURE 6. Normalized probability density $\Pi_r(\alpha)$ of the variable $\alpha$ of figure 5 for $r/L = 2^{-9}$ (solid line), and for $r/L = 2^{-13}$ (dashed line).

Again this depends only on the ratio $m/n$ and not on $n$ (or $r$). Thus by dividing the logarithm of the real distribution function by $n \sim \log_2(L/r)$, one obtains a conveniently rescaled, scale-invariant distribution function. Figure 7 shows $h(\alpha) = \ln[\Pi_r(\alpha)]/\ln(L/r)$ for the present binomial example, from which it is apparent that the distribution becomes scale-invariant (independent of $n$ or $r$) asymptotically. As pointed out by Mandelbrot (1989, where $h(\alpha)$ is called $\rho(\alpha)$), the convergence of such a function can be proved rigorously for any multiplicative process following a theorem due to Cramér.

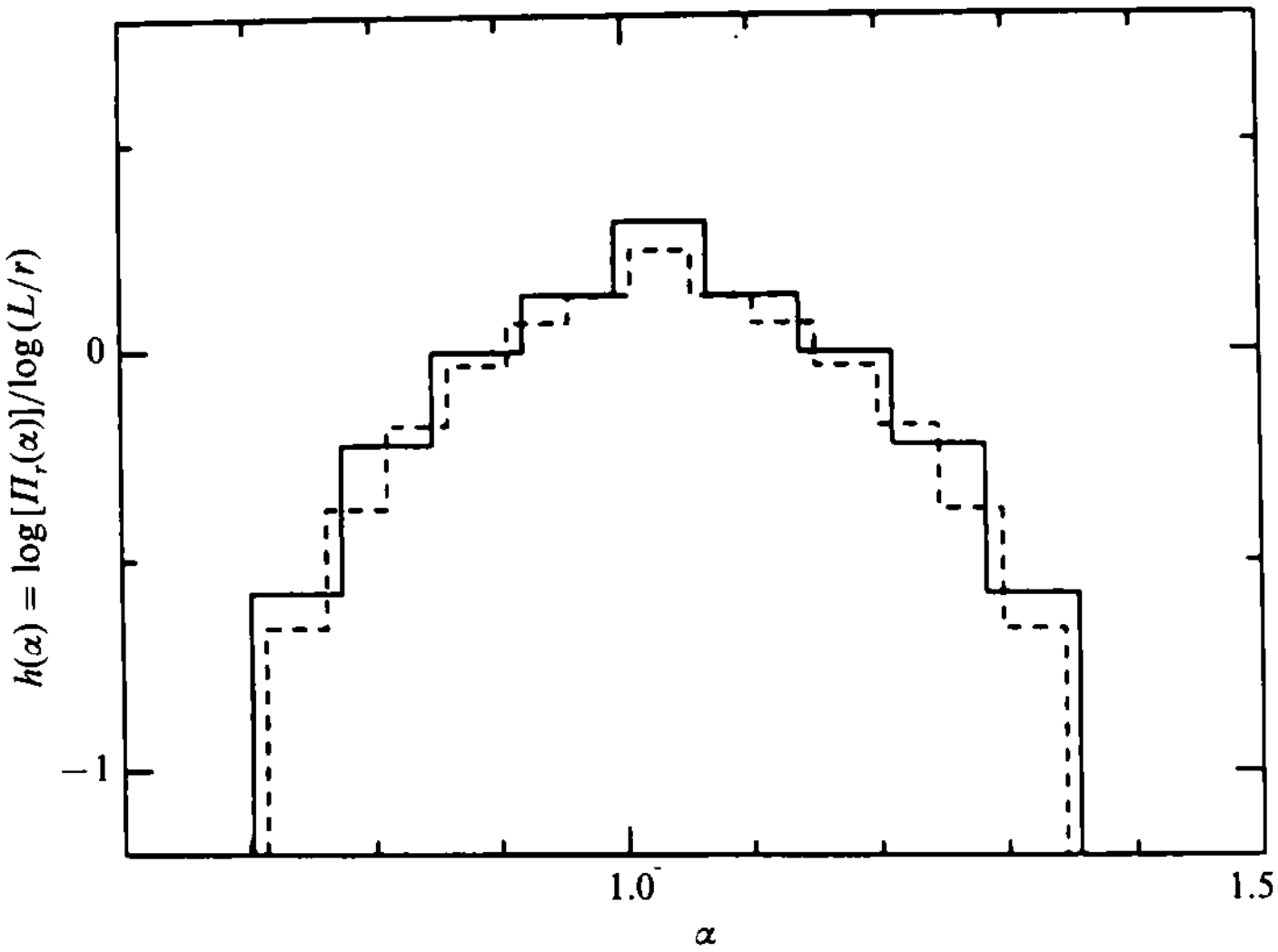

FIGURE 7. Logarithmic probability density $h(\alpha)$, normalized by $\log(L/r)$, for $r/L = 2^{-9}$ (solid line), and for $r/L = 2^{-13}$ (dashed line).

Now it is useful to ask the following question: Within how many boxes or pieces of size $r$ does the variable $\alpha$ assume values within a band of width $d\alpha$? For this purpose one has to multiply the probability $\Pi_r(\alpha)\,d\alpha$ by the total number of boxes present at a specified level of the process. The total number of pieces of size $r$ is equal to $r^{-1}$ for measures on a line as in figures 1 and 4, and in general equal to $(r/L)^{-d}$ in a $d$-dimensional space. The result is therefore

$$N_r(\alpha) = (r/L)^{-d}\Pi_r(\alpha). \tag{2.11}$$

(If the measure itself exists only on a fractal set of dimension $D < d$, the $d$ in (2.11) must be replaced by $D$.) It is now natural to define $f(\alpha)$ as the logarithm of $N_r(\alpha)$ normalized by $\ln(L/r)$. This implies that $f(\alpha) = h(\alpha) + d$ and that the scaling relation

$$N_r(\alpha)\,\mathrm{d}\alpha \sim \rho(\alpha)\,(r/L)^{-f(\alpha)}\,\mathrm{d}\alpha \tag{2.12}$$

holds. Here $\rho(\alpha)$ is some $\alpha$-dependent prefactor, not to be confused with the $\rho(\alpha)$ of Mandelbrot (1989). Instead of focusing on the scale-invariant distribution $h(\alpha)$, one can study the scale-invariant distribution $f(\alpha)$, the advantage being that a natural connection to fractal geometry can be made. This was recognized by Frisch & Parisi (1985) and further developed by Halsey *et al.* (1986), whose notation we use. We recall that a fractal set can be characterized by a dimension $D$ given by

$$N_r \sim (r/L)^{-D}, \tag{2.13}$$

where $N_r$ is the number of boxes of size $r$ needed to cover the set. Comparing (2.12) with (2.13), it is natural to interpret $f(\alpha)$ as the fractal dimension of the set with $\alpha$ values in a band $d\alpha$. Since in general $f(\alpha)$ can take on different values for different $\alpha$, measures $\epsilon(x)$ such as in figure 4 are called 'multifractal measures'.

Several comments are useful. Note that, in cascades, $N_r$ is also the total number of pieces resulting from the multiplicative process when they have reached a scale $r$. The dimension $D$ defined according to (2.13) corresponds to the Kolmogorov capacity, which may differ from the Hausdorff dimension. (For a discussion of various dimensions, see Farmer, Ott & Yorke 1983.) In general, one cannot exclude the possibility that $h(\alpha) < -d$. This means that there can be multiplicative processes for which a certain value of $\alpha$ will occur less and less often as the size $r$ is decreased. In such cases, $f(\alpha) < 0$ and cannot be interpreted as a dimension. This was noted by Frisch & Parisi (1985). Mandelbrot (1984, 1989) argued that this is no handicap in the statistical interpretation of multifractals. We shall expand on this in §2.4. Another comment relates to the rapidity with which the rescaled function $\ln[\Pi_r(\alpha)]/\ln(L/r)$ tends to the asymptotic distribution with decreasing $r$. This was treated in Meneveau & Sreenivasan (1989), where it was shown that logarithmic prefactors must in general be included in expressions like (2.12).

Summarizing up to this point, a measure resulting from a multiplicative process has a limiting scale-invariant distribution, and the relevant variable is a local exponent $\alpha$ whose distribution or relative frequency of occurrence is given in terms of $f(\alpha)$; $f(\alpha)$ can be interpreted geometrically in most cases as a fractal dimension. Since $\alpha$ characterizes the strength of the singularities, the curve $f(\alpha)$ may also be called the singularity spectrum.

Another way of characterizing a multiplicative measure is by means of moments. Returning to figure 4, it is apparent that the quantity $\langle \epsilon_r^2 \rangle$ increases as the cascade proceeds to smaller scales. However, it is easy to show that its logarithm divided by $\ln(r/L)$ is a constant, independent of the cascade step $n$. Following the thought that non-pathological distributions can be described by moments of all orders, it is useful to define the exponent $\tau(q)$ through the relation

$$\langle E_r^q \rangle \sim E_L^q (r/L)^{\tau(q)+D}. \tag{2.14}$$

For similar definitions of moment exponents (using different notations), see Novikov (1969) and Mandelbrot (1974). Alternatively, one can also consider the sum of $E_r^q$ over all (disjoint) boxes of size $r$ according to

$$\sum E_r^q \sim E_L^q (r/L)^{\tau(q)}. \tag{2.15}$$

Additionally, one can define (Hentschel & Procaccia 1983) the exponents $D_q$ as

$$D_q = \tau(q)/(q-1). \tag{2.16}$$

Hentschel & Procaccia (1983) showed that $D_0$ is the fractal dimension of the support of the measure, $D_1$ the information dimension and $D_2$ the so-called correlation dimension. Here, high positive values of $q$ emphasize regions of intense dissipation, while negative values of $q$ accentuate low-dissipation regions. The exponents $D_q$ are called 'generalized dimensions'. We relegate to Appendix A a discussion of the precise sense in which $D_q$ is to be interpreted as a dimension. For future reference, we write (2.14) also in terms of $\epsilon_r = E_r/r^d$, the mean dissipation in boxes of size $r$, according to

$$\langle \epsilon_r^q \rangle \sim \langle \epsilon \rangle^q (r/L)^{(q-1)(D_q-d)}. \tag{2.17}$$

Following Frisch & Parisi (1985) and Halsey *et al.* (1986), one can relate the exponents $D_q$, $\alpha$ and $f(\alpha)$ by evaluating the sum in (2.15) as an integral over all values of $\alpha$ as

$$\sum E_r^q \sim E_L^q \int \rho(\alpha)\,(r/L)^{q\alpha - f(\alpha)}\,\mathrm{d}\alpha \sim E_L^q (r/L)^{(q-1)D_q}. \tag{2.18}$$

We have used (2.12) in writing the first step. Using the method of steepest descent, one approximates the integrand in the limit of small $(r/L)$ by a Gaussian centred around the $\alpha$-value that minimizes $q\alpha - f(\alpha)$. The result is proportional to $r^{q\alpha - f(\alpha)}$ evaluated at an $\alpha$ such that

$$\partial f(\alpha)/\partial\alpha = q, \tag{2.19}$$

with the condition that $f''(\alpha) < 0$. Therefore, at this value of $\alpha$, one obtains

$$f[\alpha(q)] = q\alpha(q) - (q-1)\,D_q, \tag{2.20}$$

which, upon using (2.19), yields

$$\alpha(q) = \mathrm{d}/\mathrm{d}q[(q-1)\,D_q]. \tag{2.21}$$

These Legendre transformations (one replaces the local value of the function $\tau = (q-1)\,D_q$ by its slope $\alpha(q)$ and its intercept $f[\alpha(q)]$) relate the exponents $\alpha$, $f(\alpha)$ and $D_q$. The parameter $q$ selects a specific value of the variable $\alpha$ according to (2.21).

Here, a digression concerning a finer point is worthwhile. As is usual for systems with a small-scale cutoff, scaling relations such as (2.8) and (2.12) are not expected to be valid for $r$ smaller than $\eta$. In general, one may conjecture that (2.8) should be multiplied by a 'universal scaling function' $g_\alpha[r/\eta, \alpha]$, which has the property that for $r/\eta = x \gg 1$, $g_\alpha(x, \alpha) \rightarrow 1$, and for $x \rightarrow 0$, $g_\alpha(x, \alpha) \rightarrow x^{d-\alpha}$. Similarly, (2.12) should be multiplied by another scaling function $g_f[r/\eta, \alpha]$ with the condition that for $x \gg 1$, $g_f(x, \alpha) \rightarrow 1$, and for $x \rightarrow 0$, $g_f(x, \alpha) \rightarrow x^{f(\alpha)-d}$. In a similar fashion, (2.15) should also include a scaling function $g_q\{r/\eta, \alpha(q)\}$ with the property that for $x \gg 1$, $g_q[x, \alpha(q)] \rightarrow 1$, and for $x \rightarrow 0$, $g_q[x, \alpha(q)] \rightarrow x^{(q-1)(d-D_q)}$. The precise relation between $g_\alpha$, $g_f$ and $g_q$ probably depends on the prefactor $\rho(\alpha)$. The present work will not deal with such scaling functions. Such a study, which would be of interest in the context of the dissipative range of turbulent scales $(r < \eta)$, is left as a future task.

A useful characterization of intermittency is given in terms of the so-called intermittency exponent $\mu$. Several definitions exist which are not equivalent in general. Kolmogorov (1962) introduced $\mu$ as the rate of increase of the variance of $\log(\epsilon_r/\langle\epsilon\rangle)$ as a function of $\log(L/r)$ according to

$$\sigma^2_{\ln\epsilon} = \mu \ln(L/r). \tag{2.22}$$

In Appendix B we show that this intermittency exponent is related to the $D_q$-curve in the multifractal formalism according to

$$\mu = -\mathrm{d}^2[(q-1)\,D_q]/\mathrm{d}q^2|_{q=0}. \tag{2.23}$$

Another common definition of the intermittency exponent refers to the scaling exponent of the autocorrelation function of $\epsilon$ according to

$$\langle\epsilon(x)\,\epsilon(x+r)\rangle \sim \langle\epsilon\rangle^2 (r/L)^{-\mu'}. \tag{2.24}$$

If one uses $\langle\epsilon(x)\,\epsilon(x+r)\rangle \sim \langle\epsilon_r^2\rangle$ (Yaglom 1966; Cates & Deutsch 1987; Meneveau & Chhabra 1990), it is clear that

$$\mu' = d - D_2. \tag{2.25}$$

In general, $\mu \neq \mu'$.

Finally, we point out that early cascade models discussed before correspond to special cases of the multifractal description. More details are given in Appendix C.

### 2.4. *Random curdling*

In the last section, the multifractal formalism was motivated by considering a specific binomial distribution $p(M)$ of the multipliers $M$ in the multiplicative process, although (for the most part) the subsequent discussion was not constrained by the specifics of the model. For general distribution functions of the multipliers $M$, one obtains Mandelbrot's (1974) random curdling model. The model introduces important concepts concerning the experimental results of §3. In discussing it, we closely follow Mandelbrot (1974, 1984, 1989), and refer the reader to Kahane (1974), Peyriere (1974) and Kahane & Peyriere (1976) for rigorous proofs of several of the results.

Random curdling is a general multiplicative process, where a $\Delta$-dimensional 'piece' of size $r$ decays into $b^{\Delta}$ smaller pieces of equal (linear) size $rb^{-1}$; $b$ is the base of the process that can take any integer value. Although one is specifically thinking of three-dimensional space ($\Delta = 3$), we discuss cascades in some general $\Delta$-dimensional space (§2.4.1). Intersections of such cascades with lower dimensional subspaces of dimension $d < \Delta$ are of practical relevance, and the main results of such operations are discussed in §2.4.2. Details are relegated to Appendix A.

#### 2.4.1. *Conservative cascades in $\Delta$ dimensions*

A cascade is called conservative if the measure is conserved at each single step of the cascade, namely

$$\sum_i M_{j,i} = 1, \tag{2.26}$$

for all $j$, where the sum over $i$ extends to all $b^{\Delta}$ pieces created at a single cascade step. Since all $M_{j,i}$ are assumed to be positive, none can exceed unity.

When calculating the moment exponents $D_q$ of such measures, the dimensionality of the embedding domain will be indicated as a subscript on the exponents. For example, $D_{\Delta,q}$ stands for the $D_q$ exponents pertaining to the $\Delta$-dimensional domain. Now we focus on the statistics of the total dissipation or energy flux $E_{\Delta,r}$ in a box of size $r$ after the cascade has proceeded $k$ steps. We will assume that the size of the initial eddy is $L$ (comparable with the integral scale of the flow). Therefore

$$r/L = b^{-k}. \tag{2.27}$$

Following (2.3) the flux $E_{\Delta,r}$ in a given piece or box of size $r$ is the product of $k$ multipliers along the path on the hierarchical tree leading to the particular box. That is,

$$\frac{E_{\Delta,r}}{E_{\Delta,L}} = \prod_{j=1}^{k} M_j. \tag{2.28}$$

In order to calculate the moment exponents $D_{\Delta,q}$ one has to evaluate the sum of $E^q_{\Delta,r}$ over all $(L/r)^{\Delta}$ boxes. For this random model it is useful to define the exponents $D_{\Delta,q}$ according to

$$\langle \sum E^q_{\Delta,r} \rangle = E^q_{\Delta}(r/L)^{(q-1)D_{\Delta,q}}, \tag{2.29}$$

where $E_{\Delta} = E_{\Delta,L}$ and the averaging $\langle\ \rangle$ is performed over the distribution of the multipliers $M$. The average of the sum can be replaced by the averge of $E^q_{\Delta,r}$ multiplier by the total number of boxes. One then calculates the following average:

$$\left\langle \prod_{j=1}^{k} M^q_j \right\rangle = \left[ \int M^q p(M)\, \mathrm{d}M \right]^k = [\langle M^q \rangle]^k, \tag{2.30}$$

where one uses the assumptions that the distribution does not depend on $j$ and that the multipliers at different steps are uncorrelated. The index $j$ will generally be omitted from here onwards, unless explicitly required when denoting products at different cascade steps.

Combining (2.27)–(2.30), one arrives at the result that

$$D_{q,\Delta} = \log_b[b^{\Delta}\langle M^q\rangle]/(1-q). \tag{2.31}$$

Legendre transforms yield

$$\alpha_{\Delta}(q) = \frac{\langle M^q \log_b M\rangle}{\langle M^q\rangle}, \tag{2.32}$$

and

$$f_{\Delta}[\alpha(q)] = q\alpha_{\Delta}(q) + \log_b[b^{\Delta}\langle M^q\rangle]. \tag{2.33}$$

This illustrates the fact that $f_{\Delta}[\alpha(q)]$ depends on all the moments $\langle M^q\rangle$ of the distribution of $M$, and not just on the second-order moment of $\ln(M)$, as visualized in the lognormal case.

#### 2.4.2. *Lower-dimensional intersections*

The field of dissipation is three-dimensional, but most experiments examine only lower-dimensional intersections of it. It is therefore necessary to determine the relation between the properties of a $\Delta$-dimensional field and those of its $d$-dimensional intersection. Relegating details to Appendix A, we state the most important results: the $D_q$ exponents, as well as $\alpha$ and $f(\alpha)$, in $d$-dimensions are simply related to those in $\Delta$-dimensions according to

$$D_{q,d} = D_{q,\Delta} - (\Delta - d), \quad \alpha_d = \alpha_{\Delta} - (\Delta - d), \quad f_d(\alpha) = f_{\Delta}(\alpha) - (\Delta - d). \tag{2.34}$$

This means that by knowing the exponents in $\Delta$ dimensions, one can obtain the corresponding ones in $d$-dimensional cuts. A more basic question is the inverse problem of obtaining the exponents in the $\Delta$-space from those in the $d$-dimensional cut. From (2.34) it is apparent that the exponents for the $d$-dimensional cut can become negative, or $f_d(\alpha) < 0$. This does not present any problem in the statistical interpretation of multifractals, but the geometrical interpretation of $f_d(\alpha)$ as a dimension cannot be invoked. The cases when $D_{q,d} < 0$ and $\alpha_d < 0$ present more difficulties because it turns out that such values cannot be measured directly. This is related to possible divergence of certain moments along the $d$-dimensional cut (see Appendix A).

There are thus three distinct regions of the $f(\alpha)$ curve signifying different properties. It is useful to indicate the current nomenclature for each of them (Mandelbrot 1989). The region $f(\alpha) > 0$ is called the manifest part, while the region $f(\alpha) < 0$, $\alpha > 0$ is called the latent part; that with $f(\alpha)$ and $\alpha$ both negative is called the virtual part.

### 2.5. *Some special cases of random curdling*

Special cases can be obtained by assuming specific distributions for the multipliers $M$.

#### 2.5.1. *Hyperbolic or $\alpha$-model*

Motivated by the possibility that moments of $\epsilon$ might diverge on linear cuts, Schertzer & Lovejoy (1985) introduced the so-called $\alpha$-model, which is a simple example of a non-conservative cascade in one dimension. Here the multipliers $M$ can adopt two distinct values $M_0$ and $M_1$, with probabilities $P$ and $1-P$. Therefore,

$$p(M) = P\delta(M-M_0) + (1-P)\,\delta(M-M_1). \tag{2.35}$$

Since the condition $\langle M \rangle = b^{-d}$ must hold, there are three free parameters in this model: $M_0$, $P$ and $b$. By conveniently selecting them, one can produce divergence of moments for one-dimensional cuts (see Appendix A). We will return to this point in §4.

### 2.5.2. *Binomial model (p-model)*

One can in principle reduce the number of free parameters even further by fixing the numbers $b$ and $P$. By assuming that each offspring can have two distinct multipliers ($b = 2$) with the same probability, Meneveau & Sreenivasan (1987*b*) proposed a binomial, or two-scale Cantor measure, model.

The choice $b = 2$ was made essentially in accordance with the conventional wisdom that the energy transfer seems to be local in wavenumber space, and involves wavenumbers whose sizes are not disparate. Novikov (1971) gave a somewhat obscure justification of this choice on the basis of the quadratic nonlinearity of the Navier–Stokes equations. In three-dimensional space the cascade is assumed to be conservative and an eddy of size $r$ decays into $b^3 = 8$ new eddies of size $\frac{1}{2}r$. The only free parameter is $M_0$. In accordance with the literature on generalized Cantor measures (e.g. Halsey *et al.* 1986), multipliers corresponding to one-dimensional sections of this model were called $p_1$ and $1-p_1$. This implies that each piece receives either a fraction $M_0 = \frac{1}{4}p_1$ or $M_1 = \frac{1}{4}(1-p_1)$ of the flux of kinetic energy. Therefore, the '$p$-model' pertains to

$$p(M) = \tfrac{1}{2}\{\delta(M-M_0)+\delta(M-M_1)\}, \tag{2.36}$$

for which we have

$$\tau(q) = -\log_2[p_1^q+(1-p_1)^q]+(d-1)(q-1). \tag{2.37}$$

This model is intermediate between Kolmogorov's (1941) model and the $\beta$-model, in the sense that it allows for inhomogeneities to be partially mixed during the cascade.

Recently, a simple probabilistic model for the multiplier distribution has been proposed by Chhabra & Sreenivasan (1990).

## 2.6. *Non-fractal models of intermittency*

Several other recent models do not fall within the class of spatially self-similar cascades. These will be briefly reviewed here.

Inspired by the numerical results of Siggia (1978), Nakano & Nelkin (1985) proposed an intermittency model in which the energy transfer to smaller scales occurs in temporal bursts that are spatially extended as opposed to the nested spatial inhomogeneities envisioned in the fractal models. By assuming a certain scaling form of such bursts, characterized by a single exponent related to their speed of propagation, the scaling exponents $\tau(q)$ can be computed (Nakano 1988*b*) if one replaces the spatial averaging in (2.14) by a temporal one. Nakano has shown that the $f(\alpha)$ curve of such a model consists of two single points, and the model predicts no intermittency corrections to the $-\frac{5}{3}$ spectrum. It must be stressed that $\alpha$ and $f(\alpha)$ in this model do not correspond to geometric quantities as they do in the usual multifractal formalism, but arise rather as scaling exponents of time averages.

Another model, proposed by Yakhot, She & Orszag (1989) on the basis of renormalization group treatment of the randomly stirred Navier–Stokes equations, relaxes the conservation of flux of kinetic energy to smaller scales, and assumes that a fixed fraction of the flux at each cascade step proceeds directly to the smallest scale $\eta$. In other words, the flux $\Delta u_r^3/r$ (where $\Delta u_r = |u(x)-u(x-r)|$) differs from $\langle \epsilon \rangle$ and equals

$$\Delta u_r^3/r \sim \langle \epsilon \rangle (r/\eta)^{-\mu/2}. \tag{2.38}$$

This model does not obey the condition $\langle \Delta u_r^3/r \rangle = -\frac{4}{5}\langle \epsilon \rangle$ which arises from the Kármán–Howarth equation (see e.g. Monin & Yaglom 1971). However, proceeding further by replacing $\epsilon_r$ in (2.17) by $\Delta u_r^3/r$, we get

$$\tau(q) = -\tfrac{1}{2}\mu q + (q-1)\,d. \tag{2.39}$$

The situation that $\tau(q=1) \neq 0$ shows that the cascade is not conservative even on the average (as opposed to the non-conservative cascades of Appendix A which are conservative on the average).

Recently, Hosokawa (1989) proposed that the dissipation is distributed with a square-root exponential distribution. This was motivated by the numerical result that turbulent vorticity, $\omega$, is distributed exponentially; this feature is also born out by experiments (Sreenivasan & Fan 1989). The distribution has only one free parameter which can be fixed by the global mean $\langle \epsilon \rangle$. If the exponential behaviour occurs for all box sizes $r$, moments $\langle \epsilon_r^q \rangle$ for all $q$ cannot obey the scaling of the form (2.14). Thus, square-root exponential distributions at all $r$ are incompatible with multifractality. As will be seen in §3.3, the observation that the *tails* of $\epsilon_r$ might have a square-root exponential distribution is quite significant when analysing divergence of high-order moments. The contradiction with multifractality disappears if only the tail is square-root exponential.

A similar situation arises if the distribution of $\epsilon_r$ obeys gamma statistics (Andrews *et al.* 1989). This distribution has one more free parameter than that considered by Hosokawa (1989), and can be selected to produce the right power-law behaviour of the second moment. As observed by Andrews *et al.* (1989), it follows that higher-order moments do not obey exact power laws. Again, this occurs because the entire distribution is prescribed, which decays too quickly to produce any scale-invariant power-law behaviour of moments.

Another non-fractal model of intermittency has been proposed recently by Kraichnan (1990).

### 2.7. *Measuring $D_q$, $\alpha$ and $f(\alpha)$ in practice*

A practical question concerns the measurement of the exponents introduced in §2.3. Usually one does not know exactly the prefactors in (2.7), (2.11) and (2.14) because the precise value of $L$ is ambiguous, but they can be eliminated by taking ratios at two different scales $r$. The generalization of this procedure is to use many different scales and generate log–log plots whose slope (if there is a linear region) will be the exponent sought.

In many applications one does not know the measure at different levels of the cascade, but only at scales corresponding to the last cascade step. Under certain circumstances (spelled out in Appendix A), $E_r$ can be obtained by adding the measure in all the smaller boxes contained in the size $r$, a procedure that can be repeated for arbitrary $r$. This then allows the construction of the appropriate log–log plots.

In general, one also does not know the size and exact position of the pieces that resulted from the original multiplicative process. In the binomial example used here, we know that the process occurs on pieces of size $2^{-n}$ starting at the origin, but if we are given an $\epsilon(x)$ at a certain level of an unknown multiplicative process, we do not have this knowledge. It turns out that it is possible to use boxes of sizes (and positions) different from the 'natural partition' $2^{-n}$ and, for most cases of interest, the results will be unaffected (except for the appearance of oscillations as described below).

Thus, what renders the whole multifractal formalism applicable to real measurements is that we can obtain the multifractal exponents given the measure $\epsilon(x)$ at a single cascade step (usually at scales corresponding to an 'inner cutoff') by analysing $\epsilon(x)$ with varying degrees of resolution. This is very similar to the situation for simple fractal sets whose fractal dimension can be measured by looking at the set with varying resolutions (using e.g. arbitrarily placed boxes). The difference, however, is that we have to examine in addition the *intensity* or *density* $\epsilon(x)$ with varying degrees of resolution.

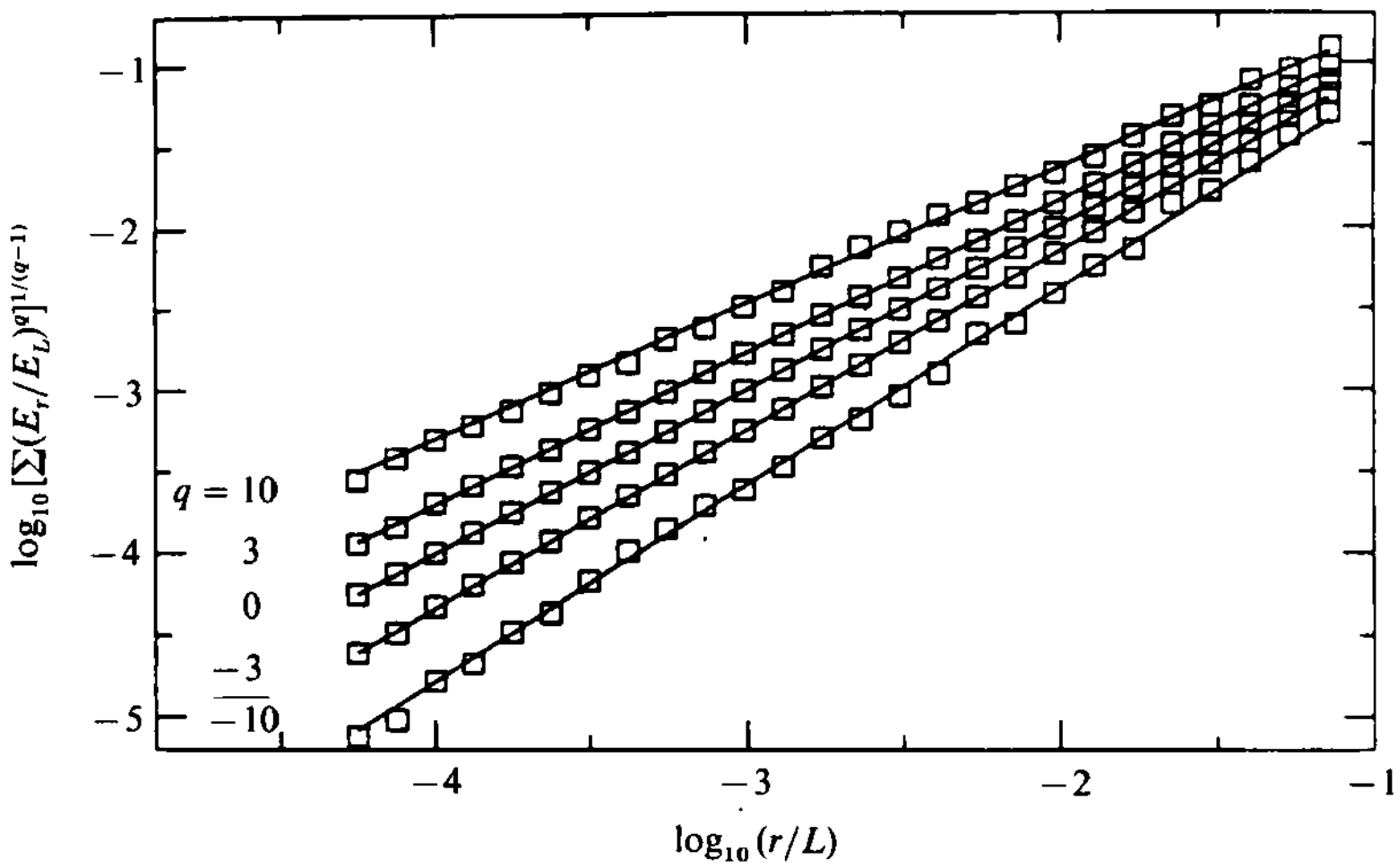

FIGURE 8. Logarithmic plots of $[\sum(E_r/E_L)^q]^{1/(q-1)}$ *vs.* the box size $(r/L)$ for five different values of $q$. The solid lines are least-square fits through the points. The slope of these lines is the measured value of $D_q$. The slight oscillation of the points around the power law is expected, and is due to the phenomenon of lacunarity.

To illustrate these points, consider our binomial process iterated 17 times, so that the smallest pieces are of size $2^{-17}$. Although we do know here the details of the cascade, we shall pretend – in analogy with the experimental situation – that we have access to the measure only at this particular level. Now, we compute $E_r$ as the integral of the measure over segments of different sizes $r$, where $r$ is larger than $2^{-17}$. The values of $r$ are logarithmically spaced. Again, to simulate the ignorance inherent in experiments, we deliberately choose box sizes different from the 'natural partition' on a binary base, arbitrarily the base 1.1.

Figure 8 shows the double logarithmic plots of $[\sum(E_r/E_L)^q]^{1/(q-1)}$ *vs.* $r/L$ for different values of $q$. The solid lines are least-square fits to the points, whose slopes (according to (2.14)) are $D_q$. The use of box sizes different from $2^{-m}$ induces slight oscillations around the basic power-law structure. This is related to the notion of *lacunarity* (Mandelbrot 1982; Smith, Fournier & Spiegel 1986; Novikov 1969 is an early reference), and introduces a small error in the determination of scaling exponents from log–log plots (Badii & Politi 1984; Arneodo, Grasseau & Kostelich 1987).

Figure 9(*a*) shows the resulting curve of $D_q$ *vs.* $q$ along with the analytical result

$$D_q = \log_2[p_1^q + p_2^q]/(1-q). \tag{2.40}$$

The two are in good agreement in spite of our ignorance about the binary process. Applying transformations (2.19) and (2.20) to the measured $D_q$ one obtains the $f(\alpha)$ curve shown in figure 9(*b*), where the continuous curve is the analytical result. The good agreement again emphasizes that measures such as in figure 4 can be characterized adequately by their scaling properties obtained from a particular step of the cascade process.

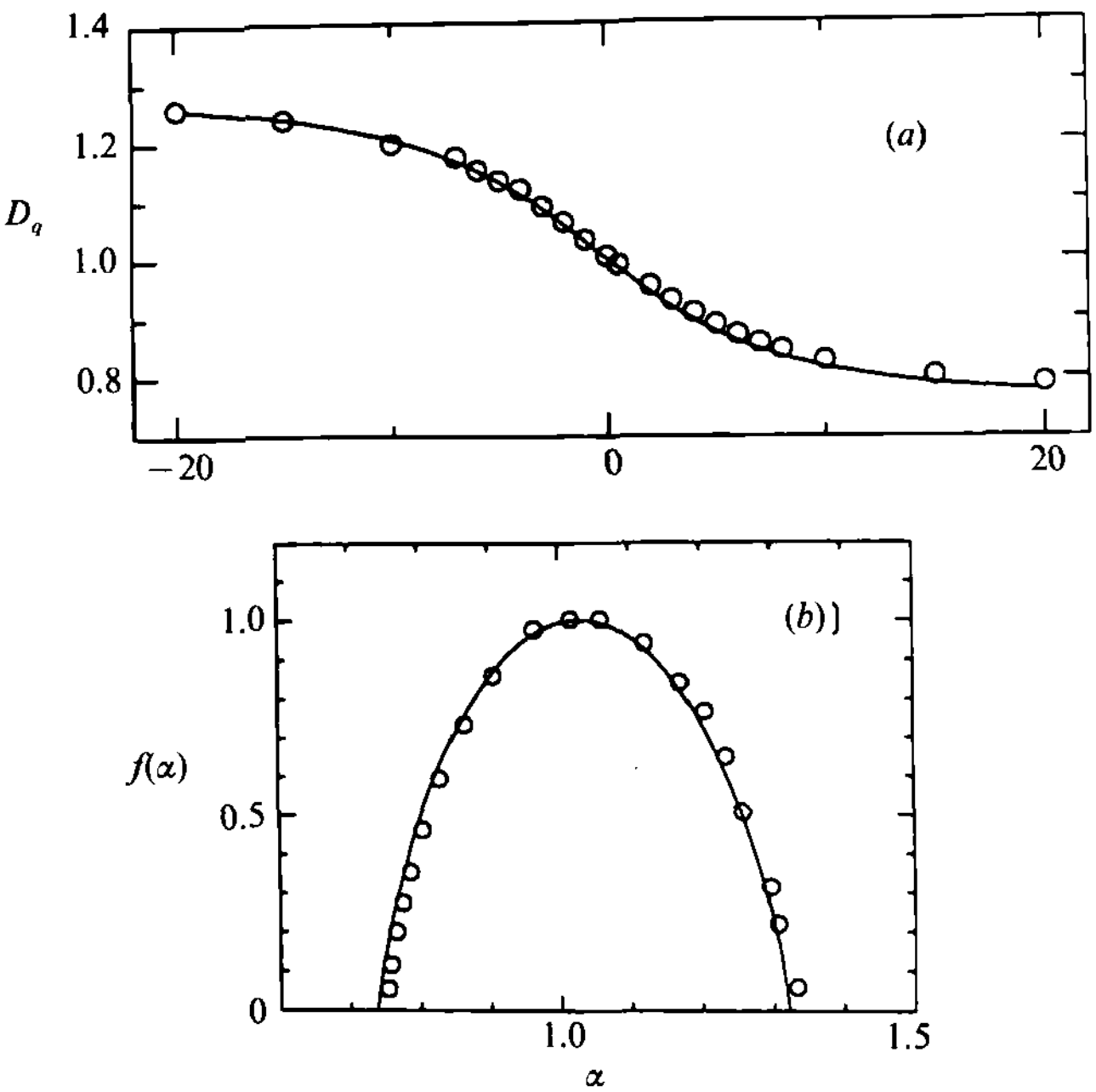

FIGURE 9. (a) $D_q$ curve of the binomial measure with $M = 0.6$ or $0.4$; (b) $f(\alpha)$ obtained from the $D_q$ curve using the Legendre transform. Circles are obtained from slopes of plots like figure 8, and the solid line is the theoretical prediction. It is seen that the 'blind' procedure used in boxing the data introduces small errors in $D_q$ and $f(\alpha)$.

For methods of obtaining $f(\alpha)$ directly without involving the moment exponents $D_q$, see Meneveau & Sreenivasan (1989), Chhabra & Jensen (1989) and Chhabra, Jensen & Sreenivasan (1989). For issues related to computing the $f(\alpha)$ curve directly from the multiplier distribution, see Chhabra & Sreenivasan (1990).

## 3. Experiments on the multifractal distribution of $\epsilon$

This section deals with the experimental exploration of the multifractal distribution of $\epsilon$, the dissipation rate of turbulent kinetic energy. Owing to experimental restrictions, we use one-dimensional cuts of a single term of $\epsilon$. Further, as is usually the practice, we resort to Taylor's frozen-flow hypothesis and analyse flows that have a convective velocity that is large compared with turbulent fluctuations. There is a vast literature on the validity of Taylor's hypothesis (e.g. Lumley 1965; Antonia, Chambers & Phan-Thien 1980), primarily directed towards possible corrections required when interpreting the frequency spectra as wavenumber spectra. In order to minimize data manipulations prior to the analysis, we do not attempt such corrections here, which are, in any case, not without problems, especially for low Reynolds number; see, for instance, Siggia (1981). Prasad *et al.* (1988) and Prasad & Sreenivasan (1990*a*) have shown that $\chi$, the dissipation rate of passive scalar fluctuations, displays the same multifractal characteristics as its individual terms. In these same references it was also shown that the use of Taylor's hypothesis was satisfactory. Even though it is not clear how much of this conclusion applies to $\epsilon$ (which, unlike $\chi$, has cross-terms in it), we are constrained by the present experimental technology to represent the real dissipation rate $\epsilon$ by its surrogate $\epsilon'$, where

$$\epsilon' \sim (\partial u_1/\partial t)^2. \tag{3.1}$$

| Flow: | Laboratory boundary layer | Wake of a cylinder | Atmospheric surface layer |
|---|---|---|---|
| Position of hot wire | $y/\delta = 0.2$ boundary-layer thickness: $\delta \sim 4$ cm | $x/d = 90$ Cylinder dia. $d = 1.9$ cm centreline | Height = 2 m above the roof of a 4-storey building ($h \sim 18$ m above ground level) |
| Free-stream velocity $U_\infty$ (cm/s) | 1200 | 800 | Mean velocity at hot-wire location = 600 cm/s |
| Convection velocity at hot-wire location $U_1$ (cm/s) | 900 | 720 | 600 |
| r.m.s. velocity fluctuations $u'$ (cm/s) | 50 | 26 | 42 ($\pm 30$%) |
| Taylor microscale $\lambda$ (cm) $\lambda = u'U_c/\langle(\mathrm{d}u/\mathrm{d}t)^2\rangle^{\frac{1}{2}}$ | 0.32 | 0.28 | 5.3 |
| Reynolds number $R_{\mathscr{L}} = U_\infty L/\nu$ | 32000 $L = \delta$ | 10000 $L = d$ | $7 \times 10^6$ $L = h$ |
| $R_\lambda = u'\lambda/\nu$ | 110 | 50 | 1500 ($\pm 30$%) |
| Kolmogorov microscale $\eta_k$ (cm) $\eta_k = \{\nu^2 U_c^2/15\langle u'^2\rangle\}^{\frac{1}{4}}$ | 0.016 | 0.026 | 0.07 ($\pm 7$%) |
| Longitudinal integral lengthscale $L$ (cm) from autocorrelation | 2.9 | 4.2 | >17000 |
| Data-acquisition frequency $f_s$ (Hz) | 25000 | 25000 | 6000 |
| Low-pass filter setting $f_p$ (Hz) | 12500 | 10000 | 2000 |
| Number of points | $10^7$ | $5 \times 10^6$ | $3.6 \times 10^5$ |

TABLE 1. Summary of experimental conditions

Here $u_1$ is the velocity fluctuation in the 'streamwise' direction.

### 3.1. *Experimental conditions*

Velocity measurements were made with a 5 μm diameter 0.7 mm long hot wires operated on a DANTEC 55M01 constant-temperature anemometer at an overheat ratio of 1.7. The temporal response was adjusted to be flat up to about 20 kHz. The signal was low-pass filtered (roll-off rate of 18 dB/octave) with a DANTEC 55D26 signal conditioner at a frequency $f_p$. The signal was digitized with 12-bit resolution on a MASSCOMP 5500 computer using a sampling frequency $f_s$. Details of experimental conditions are summarized in table 1. The hot wire is operated in the linear regime, so that calibration is not necessary. A voltage fluctuation $V(t_i)$, which is proportional to the velocity fluctuation $u_1$, is measured. The dissipation is then calculated using simple finite differences on the voltage $V(t_i)$. In Appendix D, we show that the results are robust with respect to different methods of evaluating the derivative. Since we normalize $\epsilon'$ by its mean, we omit multiplicative factors from the analysis and write

$$\frac{\epsilon'}{\langle\epsilon'\rangle} = \frac{[V(t_{i+1})-V(t_i)]^2}{\langle[V(t_{i+1})-V(t_i)]^2\rangle}. \tag{3.2}$$

The Kolmogorov microscale $\eta$ is calculated from the signals according to

$$\eta = \left\{\frac{\nu^2 U_1^2}{15\langle(\partial u_1/\partial t)^2\rangle}\right\}^{\frac{1}{4}} = \left[\frac{\nu^2 U_1^2 \langle V(t_i)^2\rangle}{15 f_s^2 u'^2 \langle[V(t_{i+1})-V(t_i)]^2\rangle}\right\}^{\frac{1}{4}} \tag{3.3}$$

where $U_1$ is the mean speed at the measuring station, $\nu$ is the kinematic viscosity of air and $u'$ is the root-mean-square velocity fluctuation. The resulting values of $\eta$ for different flows are shown in table 1. The Taylor microscale $\lambda$ calculated according to

$$\lambda = \frac{u' U_1}{\langle(\partial u_1/\partial t)^2\rangle^{\frac{1}{2}}} = \frac{U_1 \langle V(t_i)^2\rangle^{\frac{1}{2}}}{f_s \langle[V(t_{i+1})-V(t_i)]^2\rangle^{\frac{1}{2}}} \tag{3.4}$$

is also displayed in table 1. The integral lengthscales $L$ listed in the table were obtained from the autocorrelation of the velocity (using Taylor's hypothesis). For the atmospheric surface layer the integral scale was taken to be of the order of the height of the measuring station above ground level. The Reynolds numbers based on $u_1'$ and the integral scales $L$ and Taylor microscale $\lambda$ are also listed in table 1. Very long records of data were available for the laboratory flows ($10^7$ points for the boundary layer and $5\times10^6$ points for the wake). For the atmospheric flow, the number of points was $3.6\times10^5$.

Figures 1 (*a*) and 1 (*b*) show typical segments of $\epsilon'$ for the laboratory boundary layer and the atmospheric surface layer respectively. It is apparent that (*b*) displays more intense peaks than does (*a*). Qualitatively, since in (*b*) the scale separation between $L$ and $\eta$ is much larger than that in (*a*), one is tempted to compare them to figures 4 (*c*) and 4 (*d*) where the same multiplicative process is shown at different levels. If the process is the same, then the $f(\alpha)$ and $D_q$ curves of the measures of figures 1 (*a*) and 1 (*b*) should be the same. One of the goals of this section is to ascertain using experimental data whether this is indeed the case.

Returning to the velocity signals, figure 10 (*a*) shows the autocorrelation function of the velocity signals in the laboratory flows. The correlation remains quite substantial over distances larger than $L$. The dashed vertical lines enclose a range of scales within $r/\eta = 30$ and 300. As will be seen below when analysing the multifractal characteristics of the dissipation field, the scaling ranges for the laboratory flows are located within such a range. The autocorrelation function of the atmospheric flow decays much more slowly, and the appropriate scaling range is much larger (see below). The power spectrum of the velocity signals is shown in figure 10 (*b*) for the two laboratory flows as well as for the atmospheric one. Note that we have used $k = f/U_1$, where $f$ is the running frequency. (Using the definition $k = 2\pi f/U_1$ only shifts the curves to the right by $\log_{10}[2\pi] \approx 0.8$.) Again, the dashed lines enclose the scaling range to be used later for the laboratory flows. Also shown as a solid line is the $-\frac{5}{3}$ slope. It is clear that the spectra of the laboratory flows are slightly curved, and that no unambiguous inertial range is visible for these low-Reynolds-number flows. It will be shown in §3.2 that the scaling is somewhat better for moments of dissipation, much better when averages of the dissipation were obtained over segments of data of the order of a few integral scales only. Nevertheless, we point out that the range $r/\eta = 30$ to 300 that will be used in §3.2 is roughly consistent with a $-\frac{5}{3}$ power spectrum. In passing, we remark that the large-scale behaviour in the boundary-layer flow is consistent with a $-1$ spectrum (Perry & Abell 1975) indicated by another solid line. We further want to draw attention to the fact that for the spectrum of the wake, the best scaling range seems to be between $\log_{10}[k\eta] = -3$ and $-2$, corresponding to scales $r/\eta$ between 100 and 1000. The slope there is appreciably flatter than $-\frac{5}{3}$, but the upper bound on the scaling range is larger than where the inertial-range behaviour normally terminates. (This observation was also made in Prasad & Sreenivasan 1990*b*.) The scaling range for the atmospheric flow is sizeable and unambiguous, extending at least down to wavenumbers $\log_{10}[k\eta] \sim -4$.

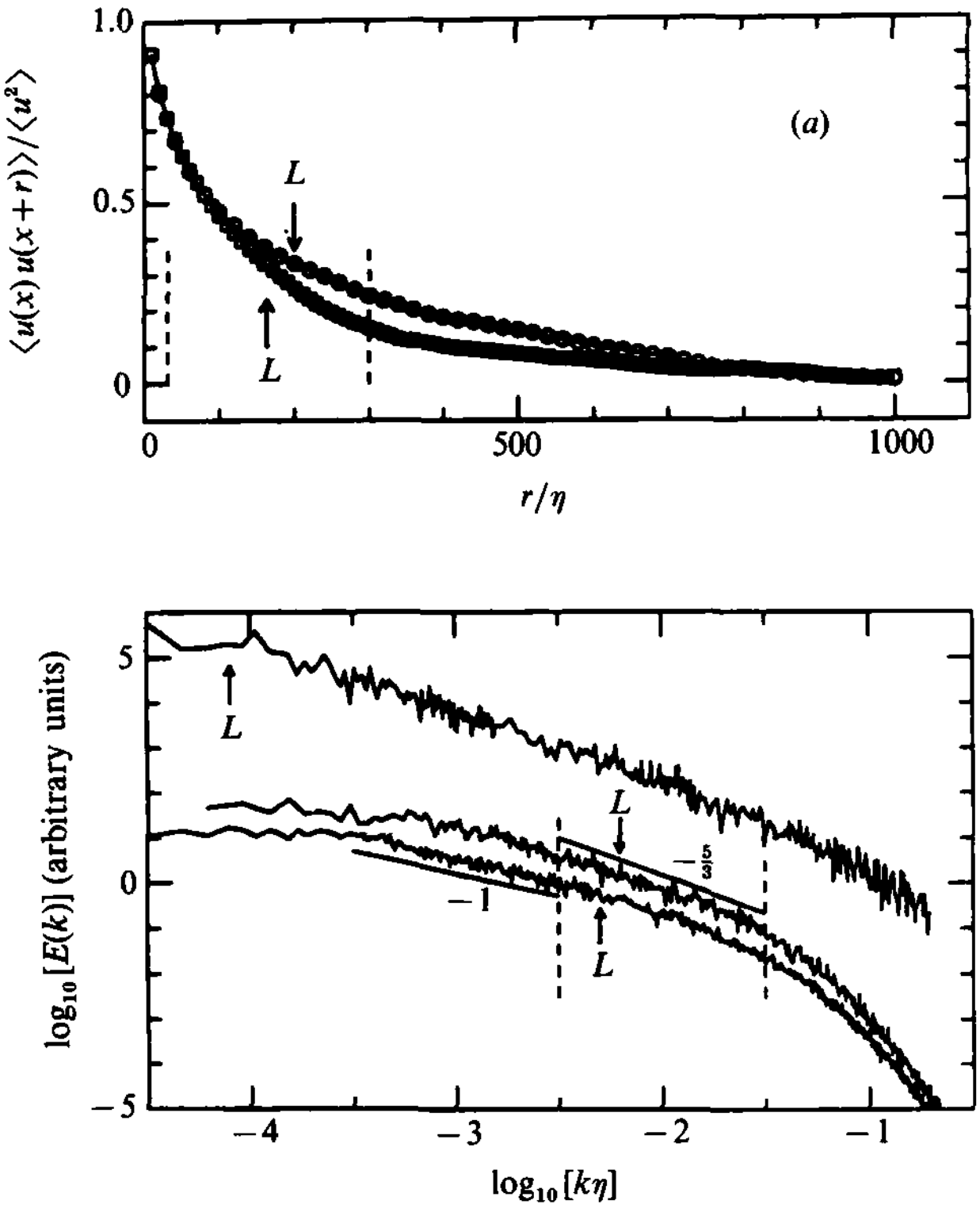

FIGURE 10. (*a*) Autocorrelation function of the velocity signal obtained in the laboratory boundary layer (circles) and in the wake of a cylinder (squares). Arrows mark the corresponding integral scales, and the dashed lines enclose the region used for finding the power-law exponents for these flows. (*b*) Power spectrum of the velocity signals obtained in the laboratory boundary layer (lower curve), in the wake of a cylinder (middle curve) and in the atmospheric surface layer (upper curve). For laboratory flows, arrows mark wavenumbers corresponding to the integral lengthscales, and dashed lines enclose the region used for finding the power-law exponents (see text). For high wavenumbers, some intermediate points are omitted to avoid cluttering.

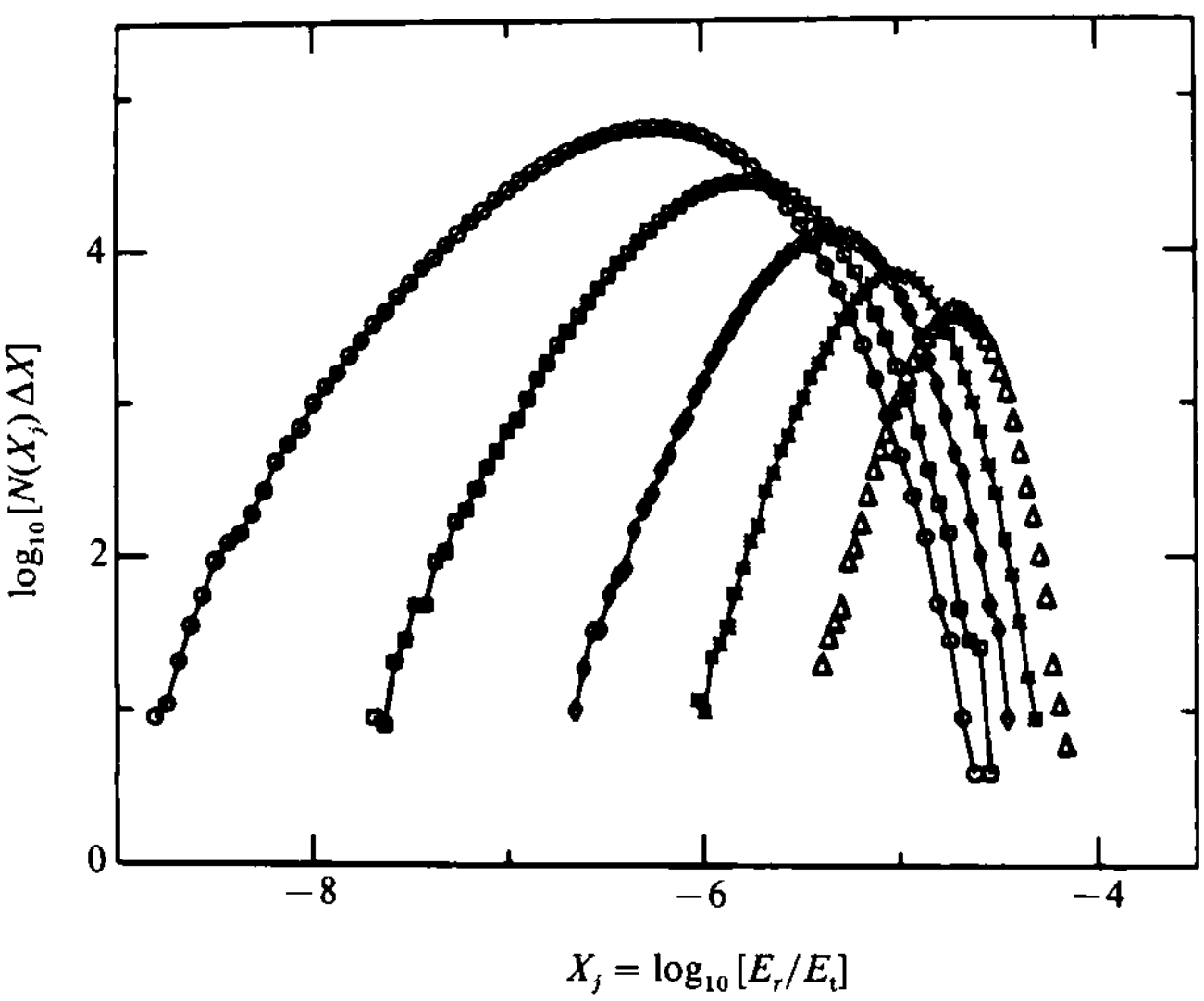

FIGURE 11. Curves based on histograms of the integrated dissipation for different box sizes obtained in the laboratory boundary layer. $N(X_j)\,\Delta X$ is the number of boxes (of size $r$), where the variable $x = \log_{10}[E_r/E_t]$ adopts values in a range $X_j \pm \frac{1}{2}\Delta X$. ○, box size $r = 16\eta$; □, $r = 40\eta$; ◇, $r = 100\eta$; *, $r = 200\eta$; △, $r = 400\eta$.

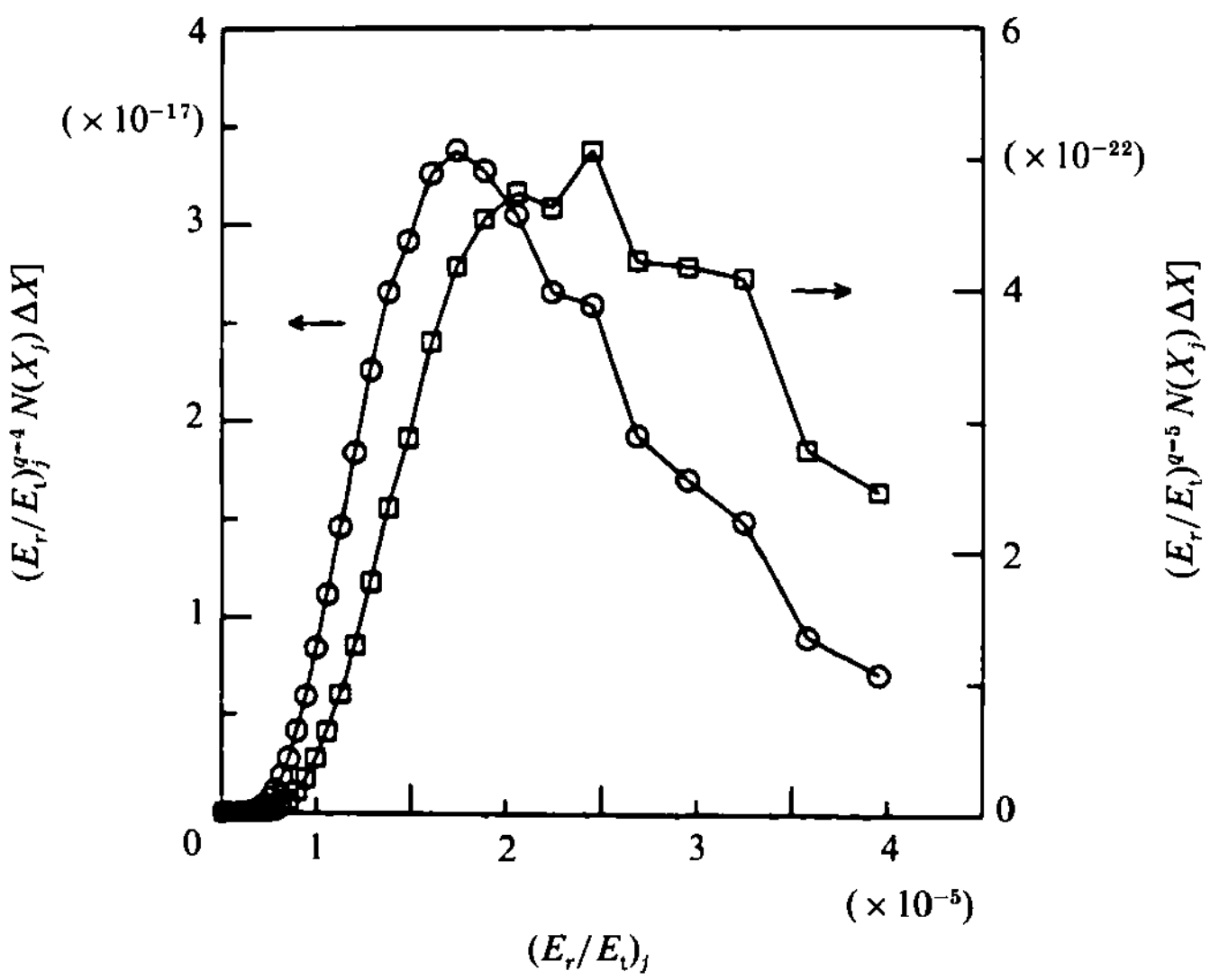

FIGURE 12. Values of $(E_r/E_t)_j$ raised to the fourth and fifth powers, weighted by their number of occurrences $N(X_j)\,\Delta X$. ○, $q = 4$; □, $q = 5$, both for a typical box size $r = 64\eta$.

### 3.2. *Measuring the $D_q$ exponents of the dissipation*

As discussed in §2.7 we consider $E_r(x_i)$, the dissipation integrated over disjoint segments of length $r$ centred around location $x_i$. For simplicity, we normalize by $E_t$, the total dissipation occurring in the entire data set. That is, we use

$$E_r(x_i) = \int_{|x-x_i|<r/2} \epsilon'(x)\,\mathrm{d}x, \quad E_t = \int_{\text{all points}} \epsilon'(x)\,\mathrm{d}x.$$

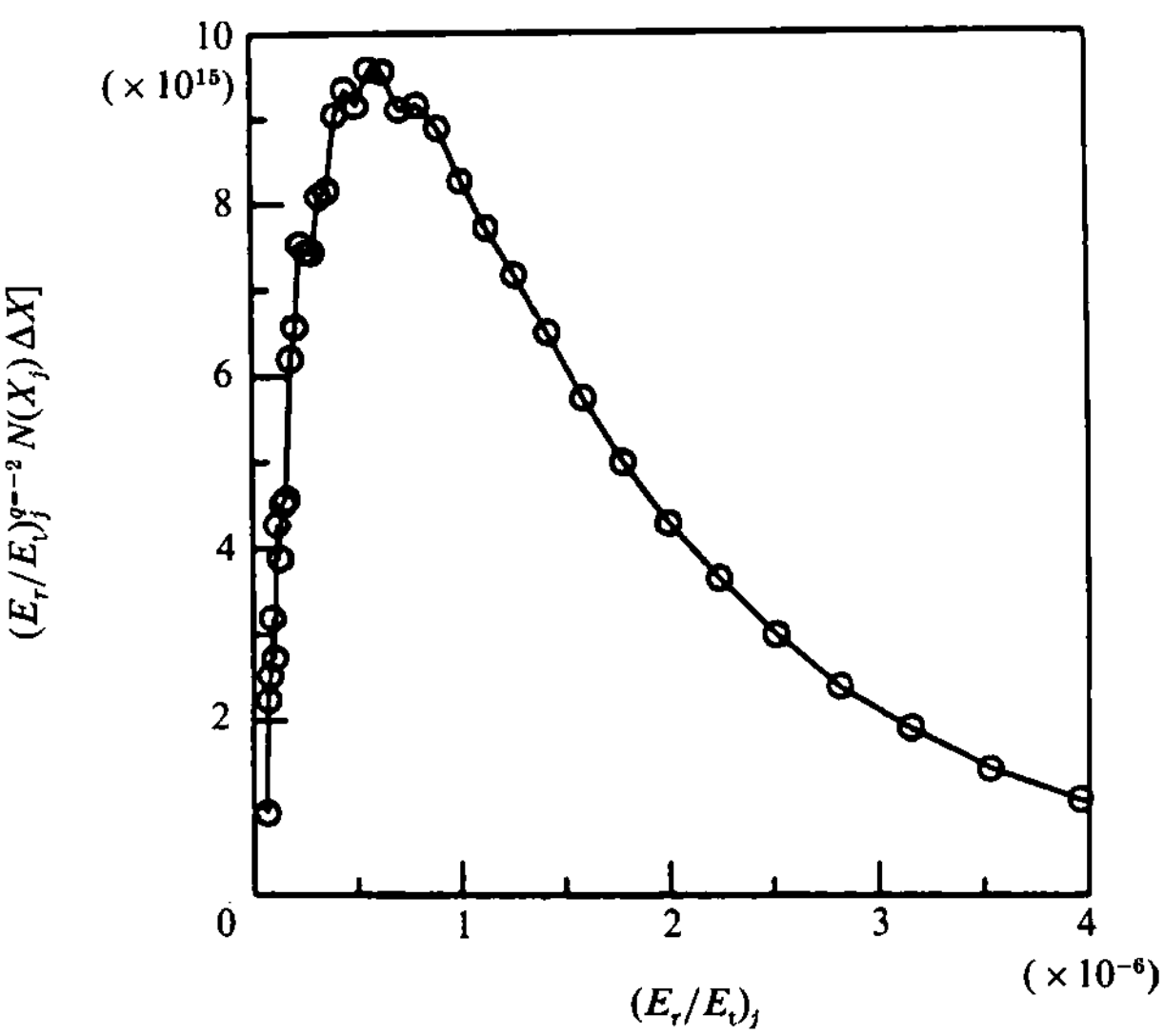

FIGURE 13. Same as in figure 12, but for the negative power $q = -2$ emphasizing low values of $(E_r/E_t)_j$.

Next one needs to computer sums $\sum_i [E_r(x_i)/E_t]^q$. For long data sets, these computations are time consuming because they have to be repeated for all values of $q$. For this and for other reasons explained below, it is convenient first to construct histograms of the random variable $X = \log_{10}(E_r/E_t)$. This was repeated for 17 values of the box size $r$, ranging from $r = 16\eta$ up to $640\eta$. Figure 11 shows curves based on histograms in logarithmic units obtained in the laboratory boundary layer for a few typical box sizes. Here $N(X_j)\,\Delta X$ is the actual number of boxes of size $r$, where $X$ takes on values in a band $X_j \pm \frac{1}{2}\Delta X$. The sums of $(E_r/E_t)^q$ are computed using the histograms as

$$\sum_i [E_r(x_i)/E_t]^q = \sum (10^{X_j})^q N_j(X_j)\,\Delta X \sim r^{(q-1)D_q}. \tag{3.5}$$

To show that data records are sufficiently long to ensure statistical convergence for values of $q$ up to 5 and down to $-2$, we show in figures 12 and 13 plots of the summands $(10^{X_j})^q N(X_j)\,\Delta X$ for $q = 5$, 4 and $-2$. The summands of (3.5) close well enough. Figures 12 and 13 are for an intermediate box size of $r = 64\eta$, but the behaviour is similar for other box sizes considered. This criterion of convergence is in fact very conservative. As discussed in §2.7, the quantities of relevance in measuring the $D_q$ exponents are the logarithm of the moments divided by $(q-1)$. To show convergence of these latter quantities, we shall present moments of $\epsilon_r = E_r/r$, normalized by $\langle\epsilon\rangle = E_{\mathscr{L}}/\mathscr{L}$, as a function of the record length $\mathscr{L}$. We consider moments of $\epsilon_r/\langle\epsilon\rangle$ rather than of $E_r/E_{\mathscr{L}}$, because the statistically stable behaviour of the latter tends to a uniformly decreasing function of $\mathscr{L}$ (because $E_{\mathscr{L}}$ increases indefinitely with $\mathscr{L}$), while moments of $\epsilon_r/\langle\epsilon\rangle$ tend to a constant value. They are trivially related by $\langle\epsilon_r^q\rangle/\langle\epsilon\rangle^q = \langle(E_r/E_{\mathscr{L}})^q\rangle(\mathscr{L}/r)^q$. Figure 14 shows wake data for $q = 4$ as a function of $\mathscr{L}$ for three typical values of the box size $r$. It is clear that there are no appreciable fluctuations over two orders of magnitude of the record length $\mathscr{L}$.

We now apply (3.5) for both laboratory flows for 15 $q$-values between $-2$ and $+5$. Figure 15($a$–$f$) shows the appropriate log–log plots for six representative values of $q$. The results for the wake are shifted from those for the boundary layer, because the records are of different lengths in the two cases. The $D_q$ exponents are obtained from the linear regions of such plots. The scaling range is not entirely unambiguous because of the low Reynolds number of the laboratory flows. However, the range between $r/\eta = 30$ and 300 appears reasonably linear. The lower limit of $r/\eta = 30$ is close to the lower limit of the inertial range used in Anselmet *et al.* (1984), but the upper limit of $r/\eta = 300$ (about $1.5L$) is considerably higher than their upper limit.

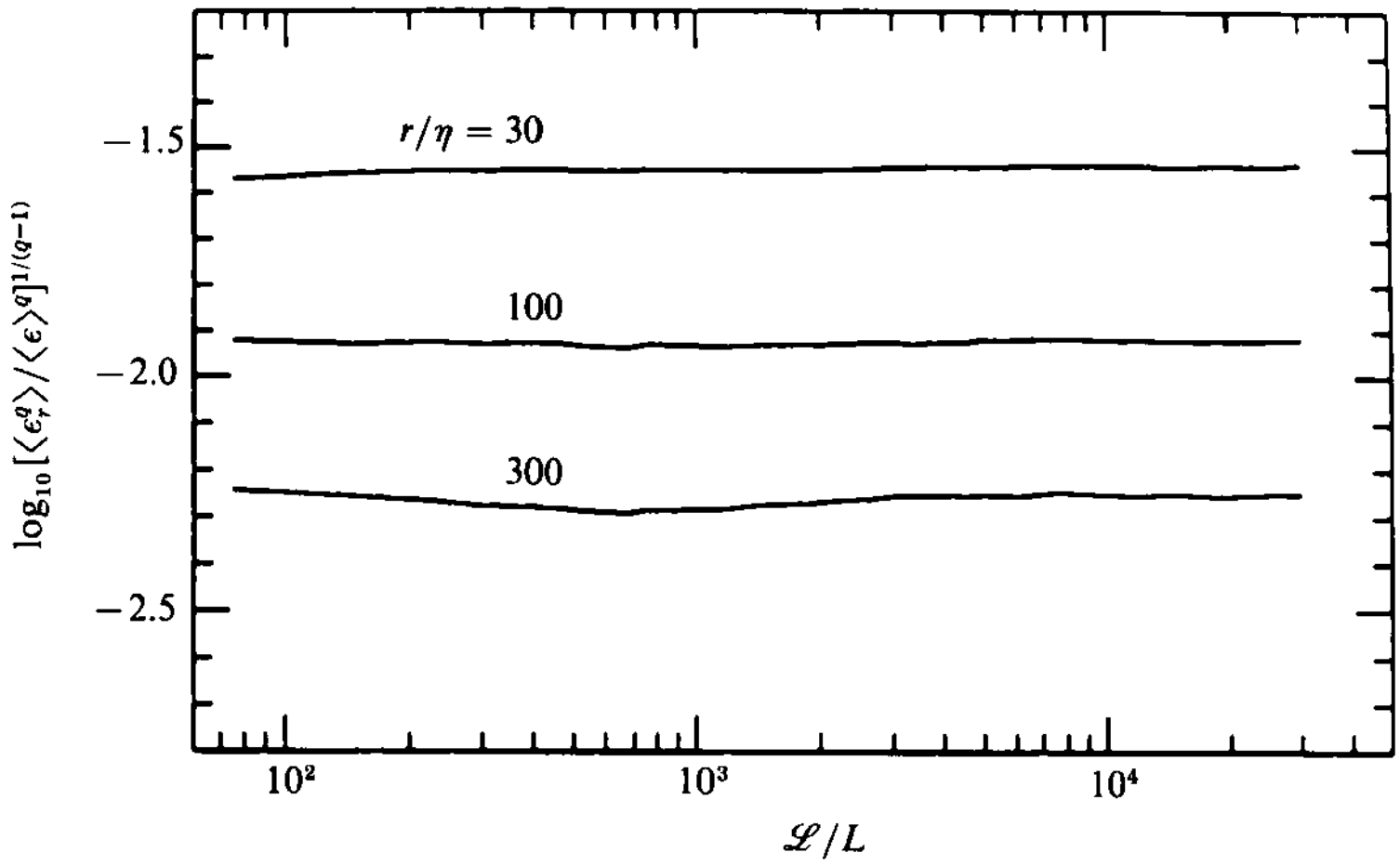

FIGURE 14. Moments of the locally averaged dissipation rate $\epsilon_r$ in the wake plotted as a function of the record length $\mathscr{L}$ for various $r/\eta$; $q = 4$.

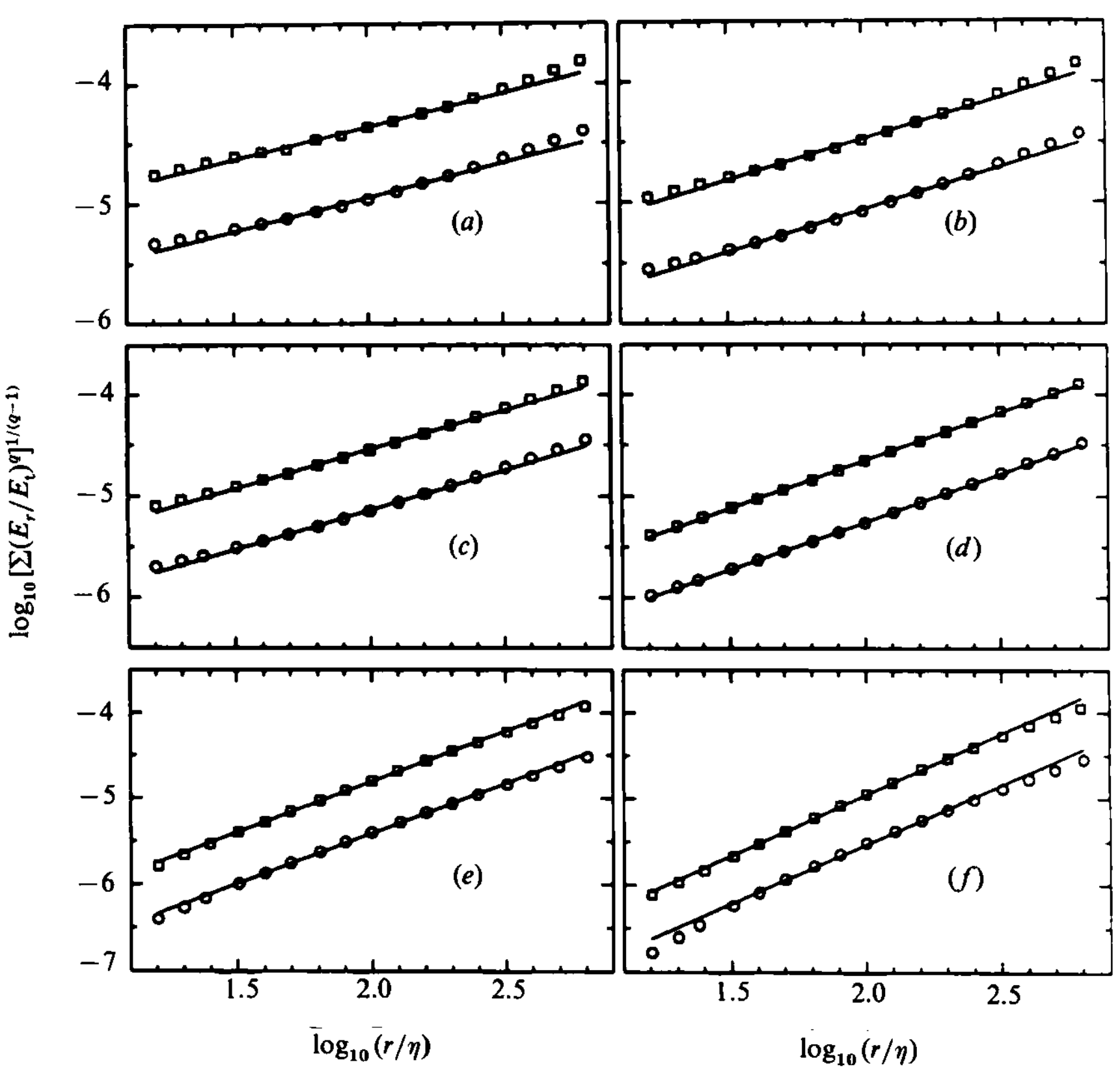

FIGURE 15. Log–log plots of $[\sum (E_r/E_t)^q]^{1/(q-1)}$ as a function of $r/\eta$ for several representative $q$-values between $-2$ and $+5$. Circles are the results for the laboratory boundary layer and squares for the wake. The solid lines are linear least-square fits between $r/\eta = 30$ to 300. The slopes of such lines give the $D_q$ exponents. (a) $q = 5$, (b) $q = 3$, (c) $q = 2$, (d) $q = 0.5$, (e) $q = -1$, (f) $q = -2$.

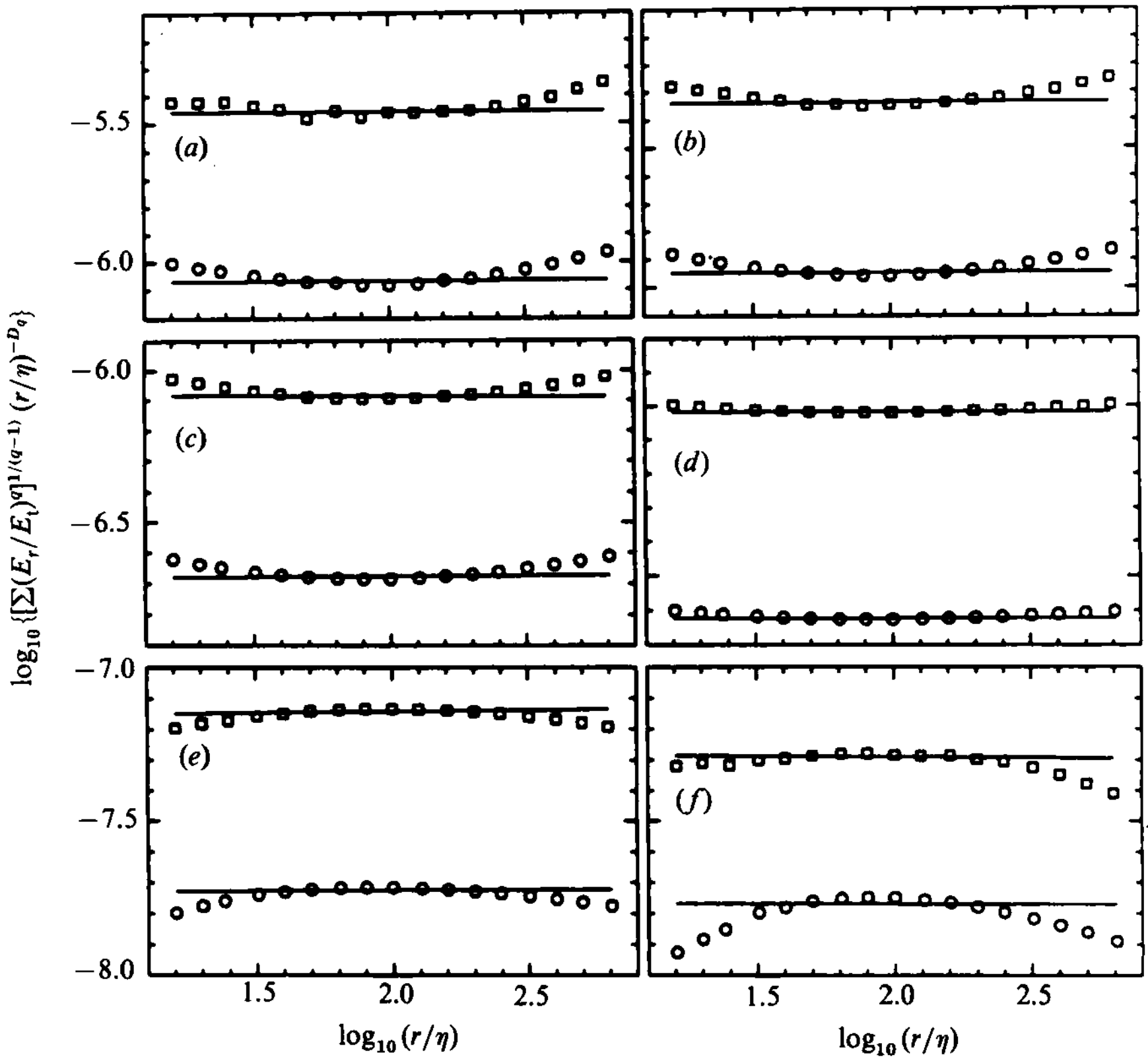

FIGURE 16. Same as figure 15, but now $[\sum(E_r/E_t)^q]^{1/(q-1)}$ is weighted by a factor $(r/\eta)^{-D_q}$, where $D_q$ are the measured slopes in figure 15. Exact power-law behaviour with exponents $D_q$ should yield a flat region. Although perfect power-law behaviour does not exist at these moderate Reynolds numbers, reasonably horizontal portions are discernible in the range $r/\eta = 30$ to 300. (*a*) $q = 5$, (*b*) $q = 3$, (*c*) $q = 2$, (*d*) $q = 0.5$, (*e*) $q = -1$, (*f*) $q = -2$.

Straight lines are drawn by least-square fitting through data points in the range $r/\eta = 30$ to 300. These are shown as solid lines, whose slope corresponds to $D_q$. In figure 16(*a–f*) the values of $[\sum(E_r/E_t)^q]^{1/(q-1)}$ weighted by $(r/\eta)^{-D_q}$ are plotted for both flows. The existence of reasonably horizontal plateaux in the range $r/\eta = 30$ to 300 points to the reasonableness of the estimated $D_q$. To test the sensitivity of the results with respect to the precise choice of the scaling range, we have obtained fits in ranges $r/\eta = 20$ to 200 and $r/\eta = 40$ to 400. Figure 17 shows the results. The sensitivity of the results to the scaling range is indicated by dashed lines, which correspond to obtaining $D_q$ in the range $r/\eta = 20$ to 200 (lower line for $q > 0$ and upper line for $q < 0$) and in the range $r/\eta = 40$ to 400 (upper line for $q > 0$ and lower line for $q < 0$). The agreement between the various results for both flows is quite good.

Given our conservative criterion for convergence, we now want to explore the effect of computing sums or moments over shorter records of data. We recall our earlier observation (Meneveau & Sreenivasan 1987*a*) that the scaling appears to be better when one computes sums over shorter records of data. (A similar observation was made for fractal interfaces in Sreenivasan & Meneveau 1986.) As discussed in Appendix A, a single realization of a cascade should contain a statistically representative sample of the $\alpha$-values in the manifest part of $f(\alpha)$. This of course is valid only asymptotically for $\eta/L \rightarrow 0$, or at very high Reynolds number, where the number of multipliers is large. For laboratory flows, averaging over a few tens of such samples improves the statistics considerably.

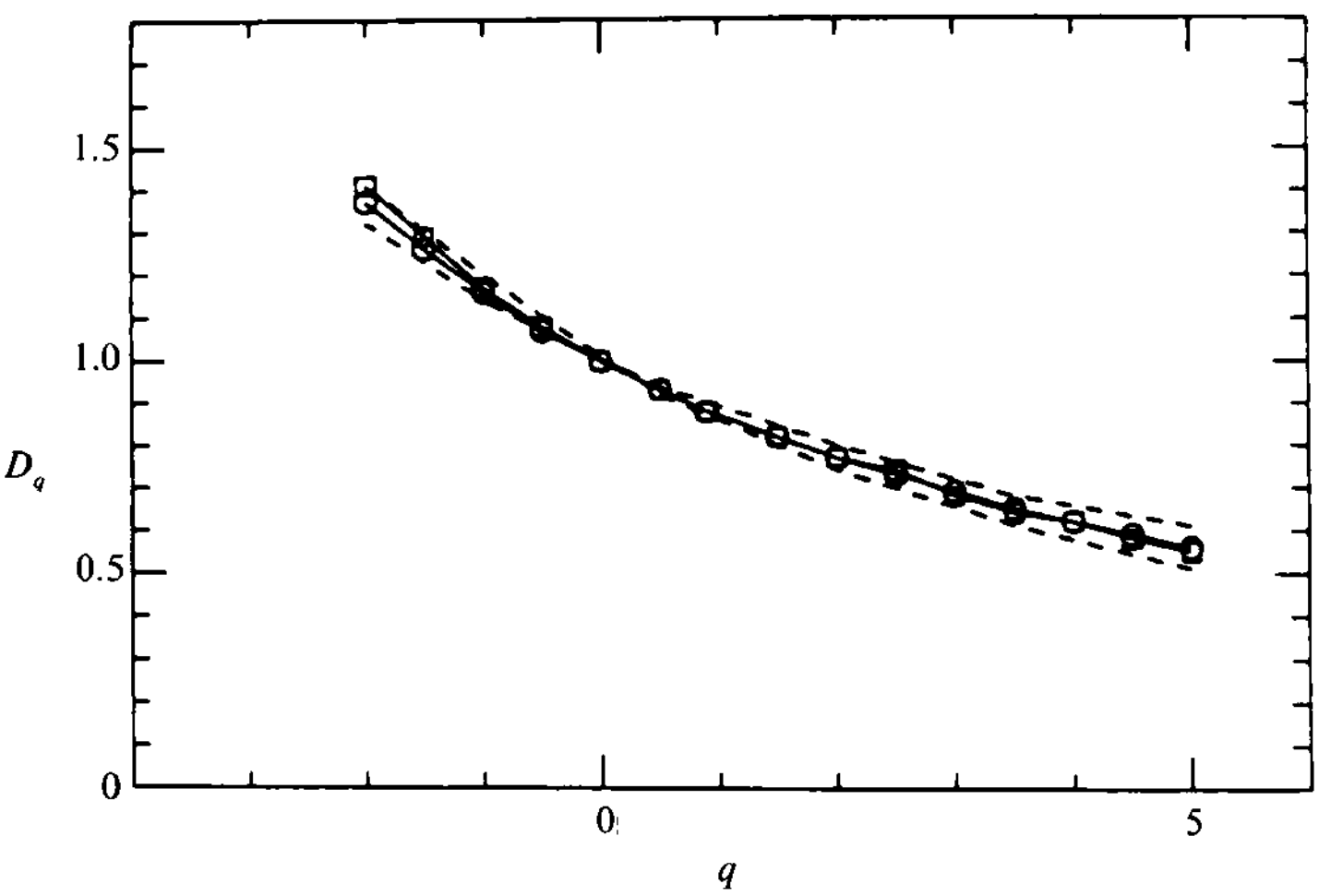

FIGURE 17. Moment exponents $D_q$ as a function of $q$ for both laboratory flows obtained from fully converged moments. The circles are for the boundary layer and squares for the wake. Dashed lines correspond to $D_q$ values obtained from using different scaling ranges. The upper curve in the region $q > 0$ (and lower curve for $q < 0$) was obtained by linear fits between $r/\eta = 40$ and 400. The lower curve in the range $q > 0$ (and upper curve for $q < 0$) was obtained by similar fits in the range between $r/\eta = 20$ and 200.

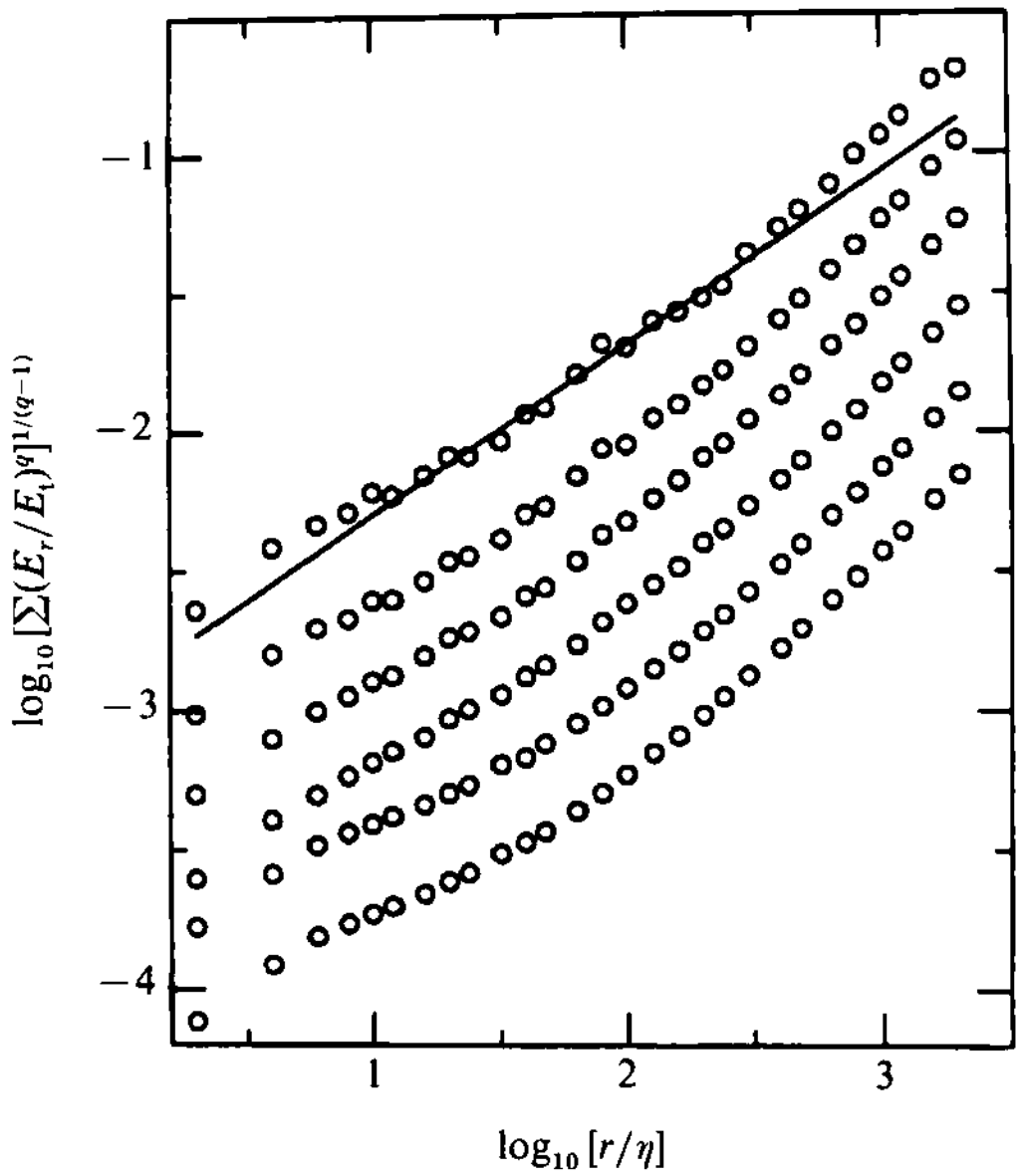

FIGURE 18. Log–log plots of $[\sum (E_r/E_t)^q]^{1/(q-1)}$ for $q = 4$ as a function of $r/\eta$, for increasing values of the record length $\mathscr{L}$ used to compute $\sum (E_r/E_t)^q$. From top to bottom, $\mathscr{L} = 50L$, $100L$, $200L$, $400L$, $800L$ and $1600L$. $L$ is the integral scale. The flow is the laboratory boundary layer. The slope of the solid line fit through the points in the range from $r/\eta = 10$ to 300 is $D_{q=4} = 0.62$.

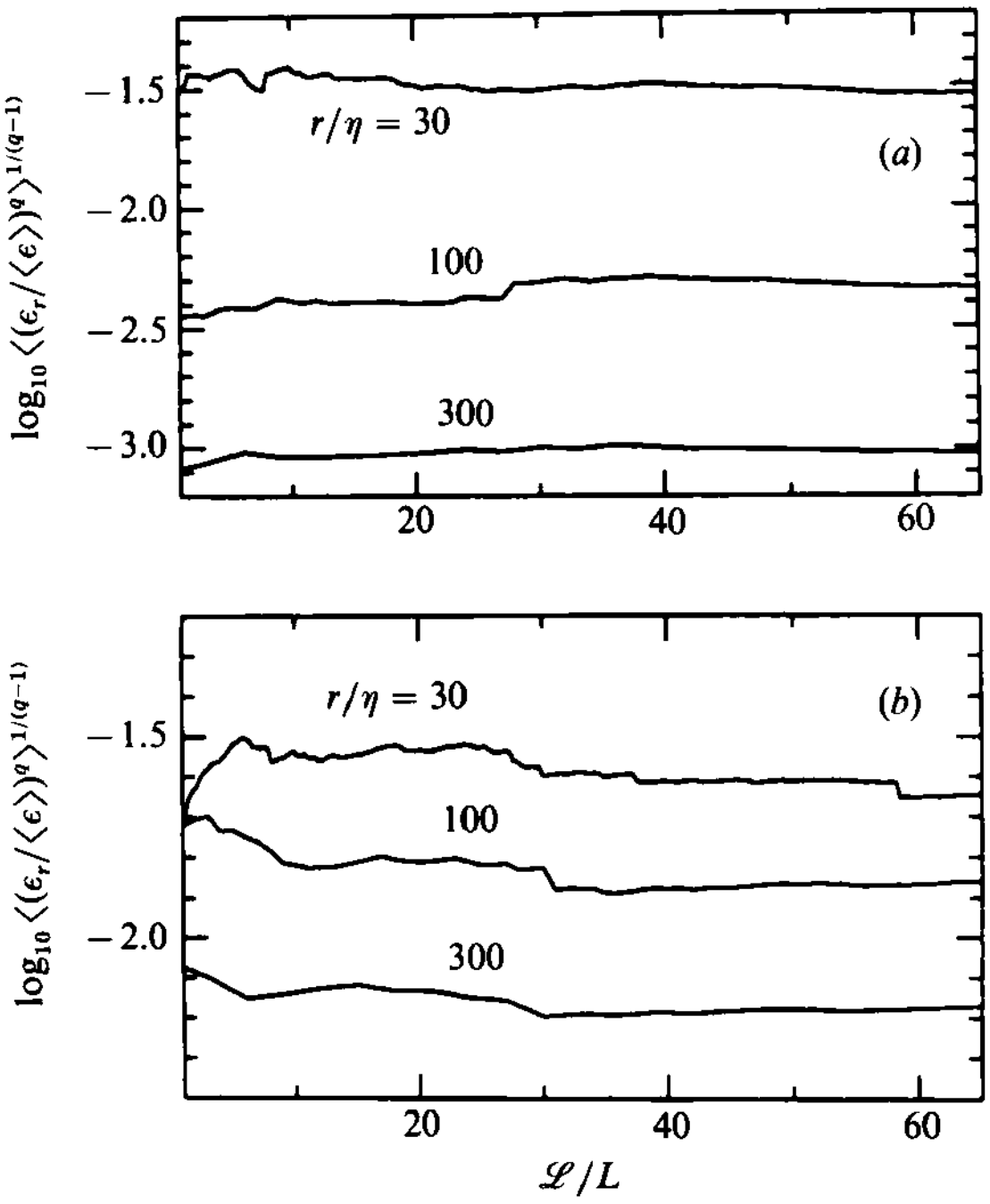

FIGURE 19. Moments of the locally averaged dissipation rate $\epsilon_r$, as a function of the record length $\mathscr{L}$ used for the averaging for various $r/\eta$, again for the boundary layer. (*a*) $q = 5$ and (*b*) $q = -5$.

Of immediate interest is the power-law behaviour under such circumstances. Figure 18 shows log–log plots of $[\sum (E_r/E_t)^q]^{1/(q-1)}$ *vs.* $(r/\eta)$ for $q = 4$ from the laboratory boundary layer (see table 1), where the sum is evaluated over increasingly long segments of data (from top to bottom: $\mathscr{L} = 50L$, $100L$, $200L$, $400L$, $800L$ and $1600L$). The solid line is a fit through the results for $50L$ in a range between $r/\eta =$ 10 to 300, and the slope is $D_{q=4} = 0.62$. The scatter disappears as $\mathscr{L}$ is increased, but the curving of the points makes it more difficult to identify power-law behaviour. It appears, therefore, that a better scaling can be observed by considering data sets of the order of a few tens of integral scales. It must be stressed that, even though moments or sums are statistically not completely converged for segments of order $10L$, the logarithms of the moments divided by $(q-1)$ do converge to reasonably stable values. This feature also permits us to compute $D_q$ for $q$ more negative than $-2$. Figure 19(*a*) shows the moments as a function of $\mathscr{L}$ for a high moment $q = 5$ for three typical values of the box size $r$. Similar results for $q = -5$ are shown in figure 19(*b*). The distance between the curves corresponding to different box sizes does not vary appreciably, meaning that the slope of their log–log plots (see below) will give good estimates of the exponents sought.

We treat these observations as empirical facts (which are not fully understood) and proceed to compute $D_q$ from short segments of data. Figure 20 shows representative log–log plots for six different segments of the boundary-layer flow for different values of $q$ (4, 0.5 and $-4$). The solid lines are fits in the best scaling ranges selected on a case-by-case basis. For $q > 0$, usually the range between $r/\eta = 12$ and 400 was used. For $q < 0$, the results at small scales $r/\eta < 40$ tend to fall-off faster than a power law. As discussed in appendix C of Meneveau & Sreenivasan (1987*a*), this is due to the influence of noise (digitizer and otherwise). The appropriate scaling range for $q < 0$ was usually between $r/\eta = 40$ to 400. Now, however, the measured values of $D_q$ fluctuate slightly from one segment of data to another, exhibiting typical standard deviations of 0.05 for $q = 4$, 0.02 for $q = 2$ and 0.05 for $q = -4$.

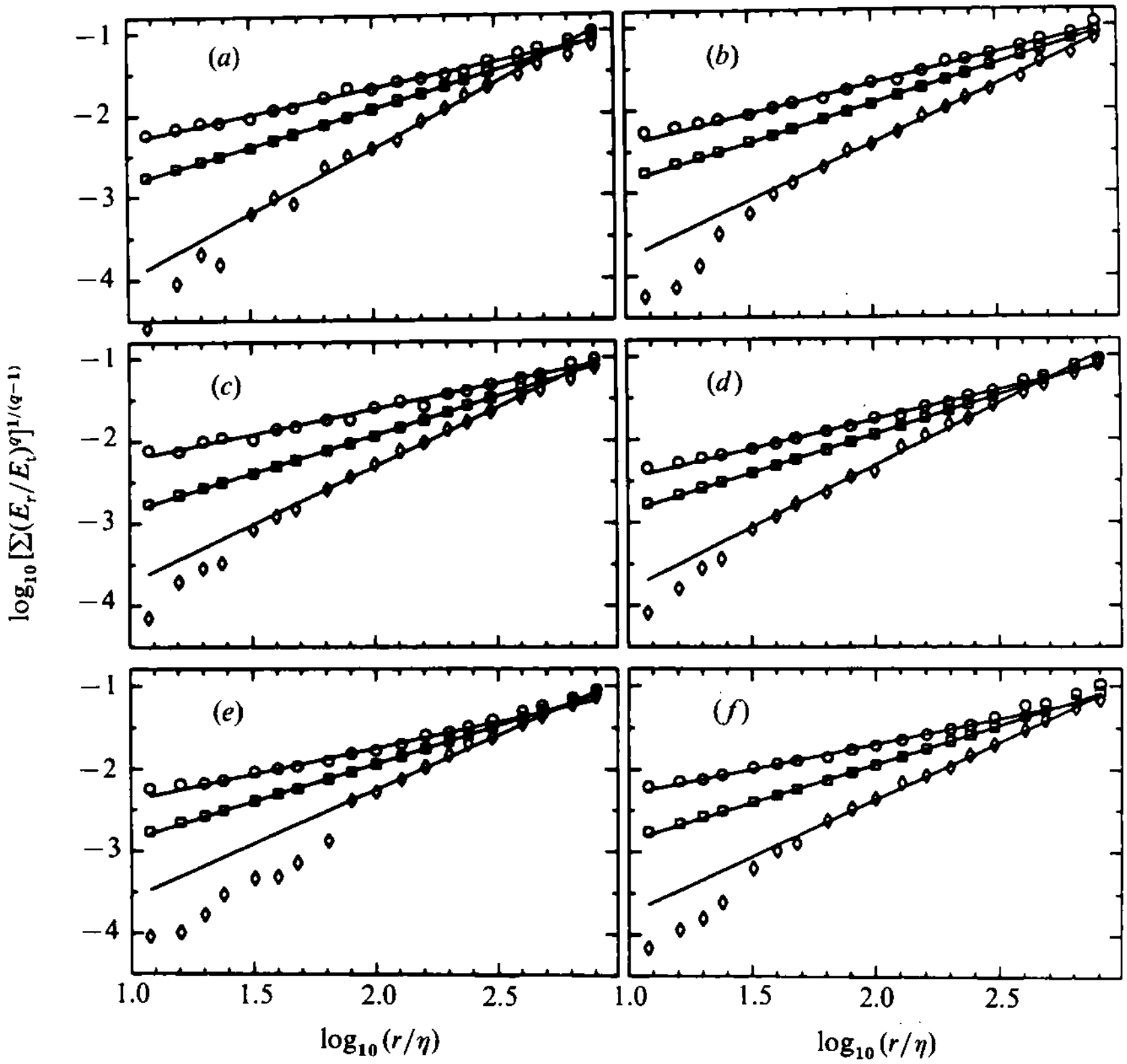

FIGURE 20. Log–log plots of $[\sum (E_r/E_t)^q]^{1/(q-1)}$ as a function of $r/\eta$ for several typical segments of length equal to $\mathscr{L} = 50L$ obtained in the laboratory boundary layer. ○, $q = 4$; □, $q = 0.5$; ◇, $q = -4$. Solid lines are linear least-square fits in a range selected on a case-by-case basis. (*a–f*) correspond to different segments of the data.

This procedure was repeated for other flows including a laboratory boundary layer at $y/\delta = 0.4$ (with $R_\lambda \sim 200$), the wake of a cylinder at a free-stream speed of 1500 cm/s, and the flow behind a grid.

For the atmospheric surface layer (see table 1), we evaluate the sum over all $3.6 \times 10^5$ data points available. This is still a relatively short segment of data because of the large integral scale of this flow. To illustrate the convergence of moments, we show in figure 21 moments evaluated as a function of the length of the data record. Stronger fluctuations of the moments can now be seen, this being so because of the much higher Reynolds number of this flow; yet, differences from one $r$ value to another remain, for the most part, essentially independent of data record length. Figure 22 shows the relevant log–log plots for six different values of $q$. Here the scaling range clearly extends over almost four decades and allows unambiguous determination of the scaling exponents.

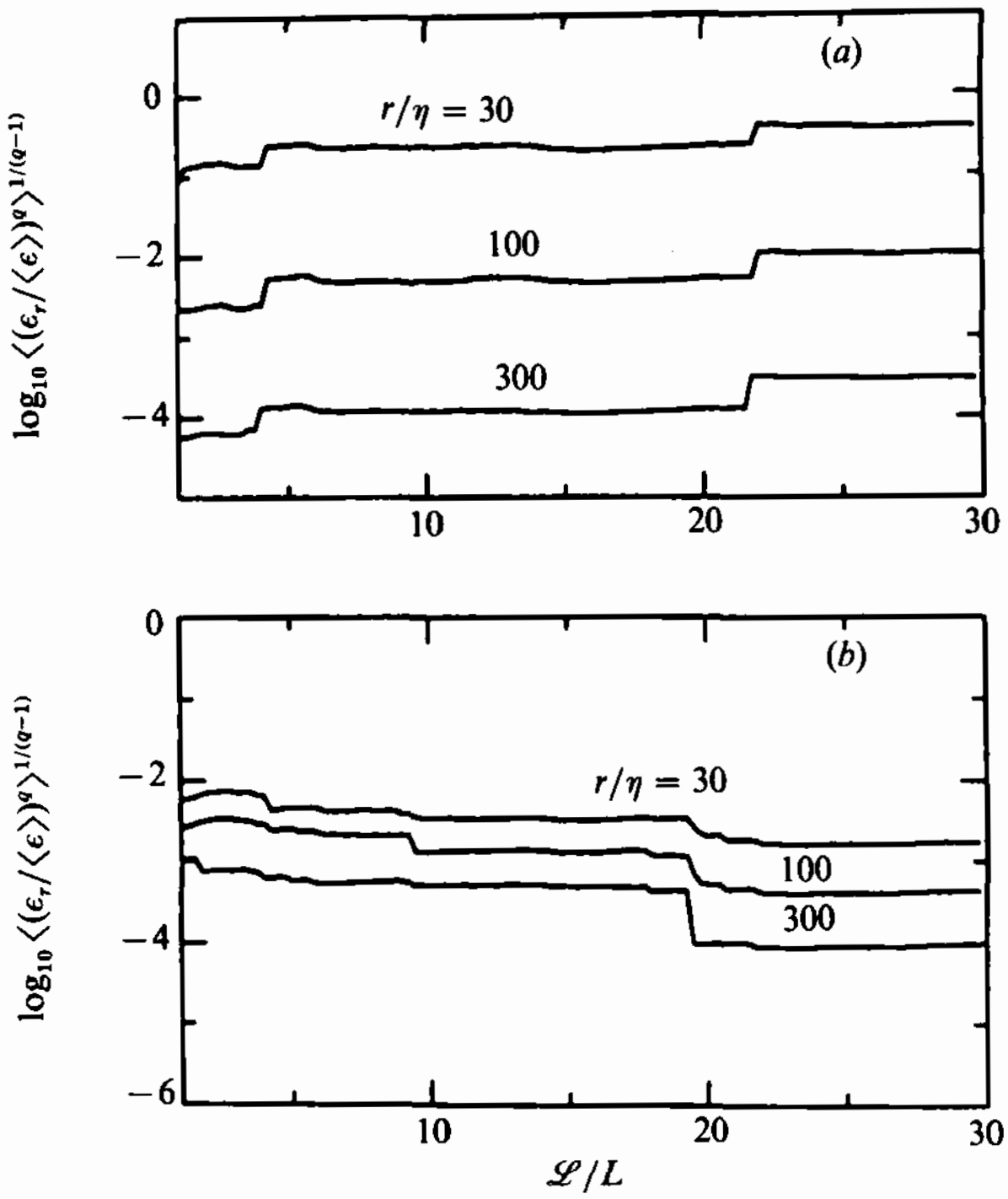

FIGURE 21. Moments of the locally averaged dissipation rate $\epsilon_r$, as a function of the record length $\mathscr{L}$ used for the averaging in the atmospheric surface layer for various $r/\eta$. (a) $q = 5$ and (b) $q = -5$.

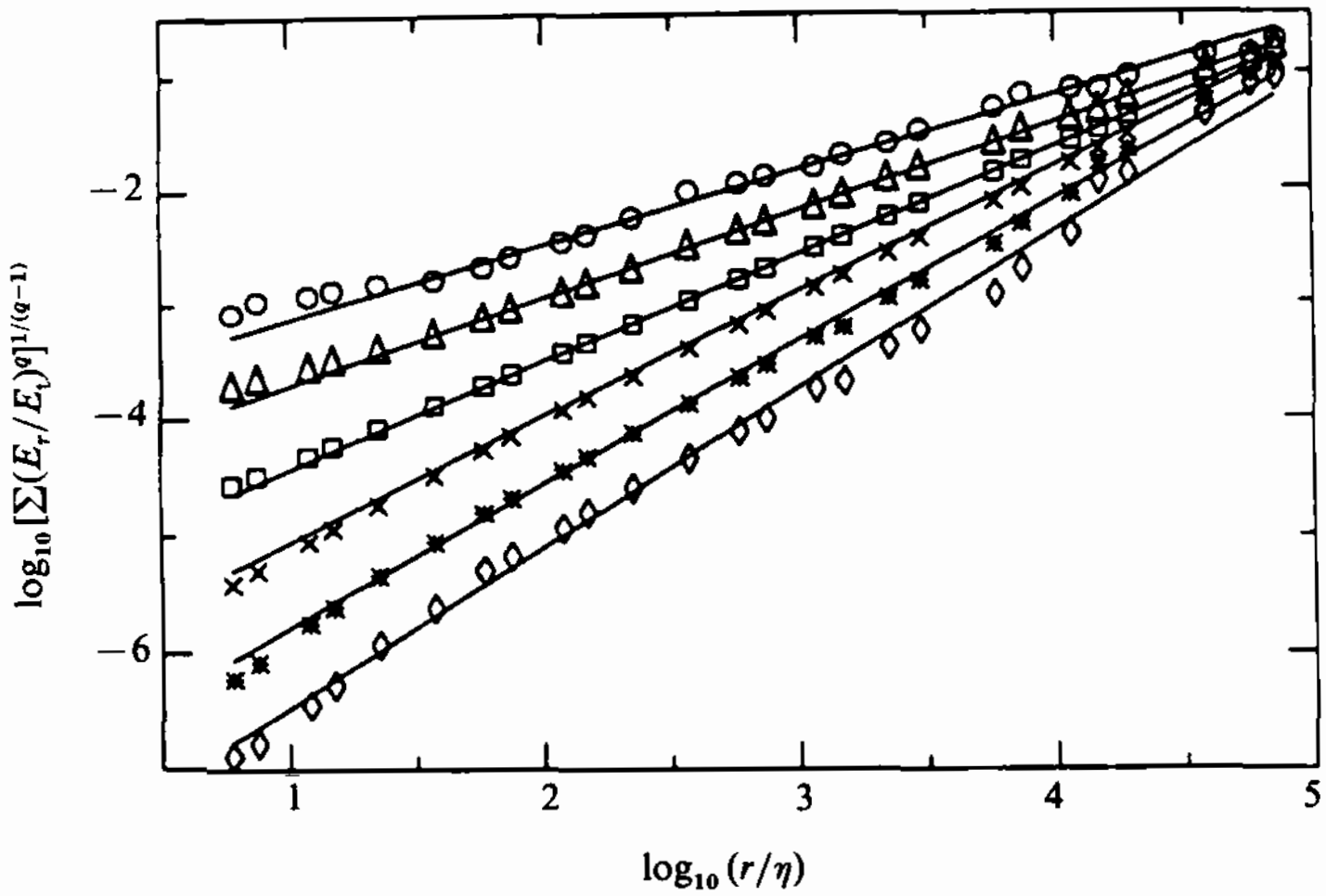

FIGURE 22. Log–log plots of $[\sum (E_r/E_t)^q]^{1/(q-1)}$ as a function of $r/\eta$ for the flow in the atmospheric surface layer. ○, $q = 4$; △, $q = 2$; □, $q = 0.6$; ×, $q = -0.6$; *, $q = -2$; ◇, $q = -4$. Solid lines are linear least-square fits in the range $r/\eta = 10$ to 30000.

The average of all the results of many segments of data in all the laboratory flows, as well as the atmosphere, give a representative $D_q$ curve. The mean curves (and their standard deviations) for each type of flow were indistinguishable from each other within experimental accuracy. The mean curve for all the flows is depicted with diamonds in figure 23, where the error bars denote standard deviations resulting from fluctuations between one segment and another. The results are indistinguishable from those obtained from the long-term averaging in the range $q > -1$. For larger negative $q$ values, we are inclined to believe that the results from the short-term averaging are the more accurate ones because of the substantially better scaling observed.

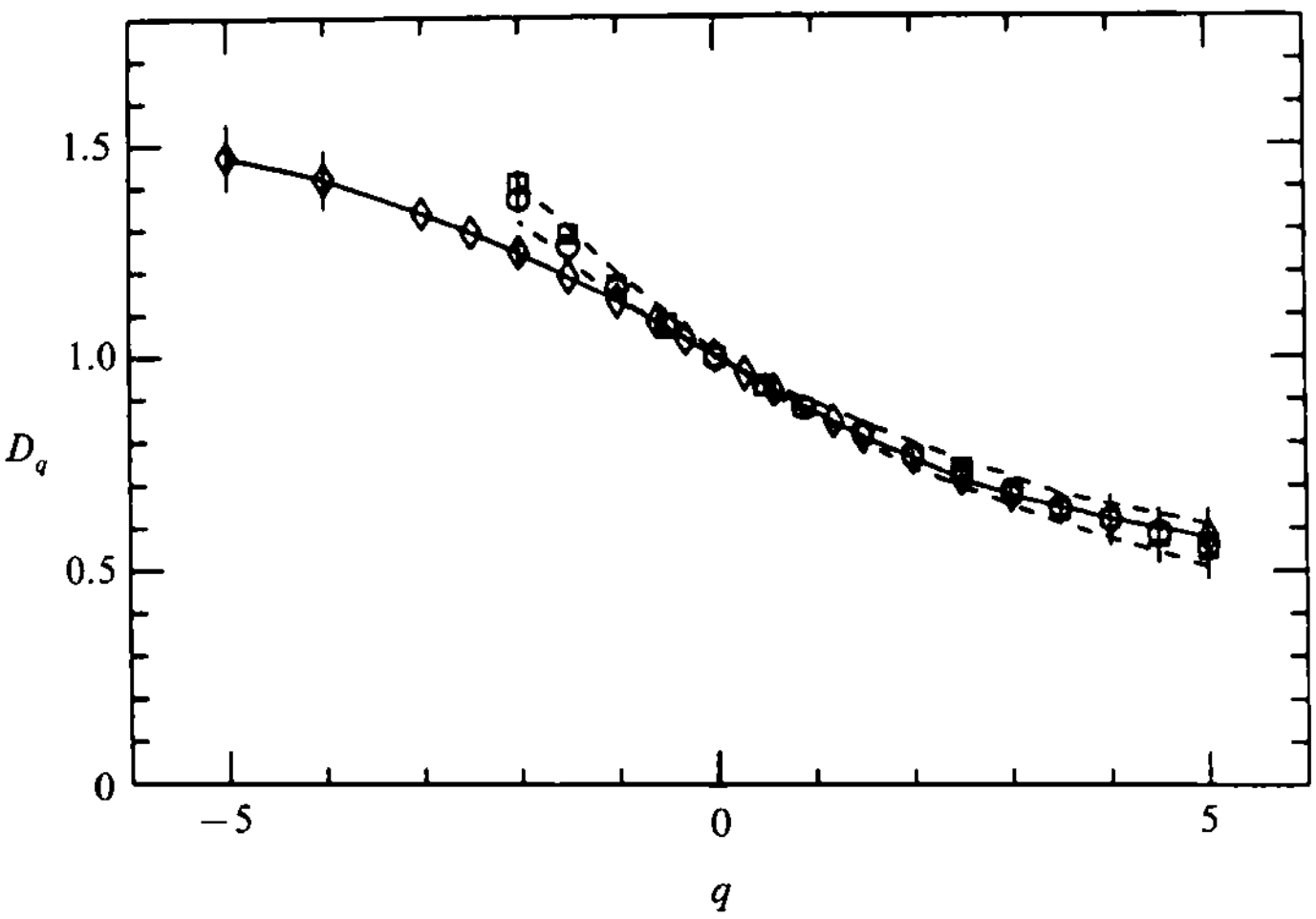

FIGURE 23. Moment exponents $D_q$ as a function of $q$. Circles, squares and dashed lines are the results from figure 17 obtained from fully converged moments. Diamonds (joined by the solid line) are the results of averaging over data segments of the order of a few tens of integral scales (see text). They represent the mean value of many such segments in a variety of flows. The error bars correspond to the standard deviation observed from different segments.

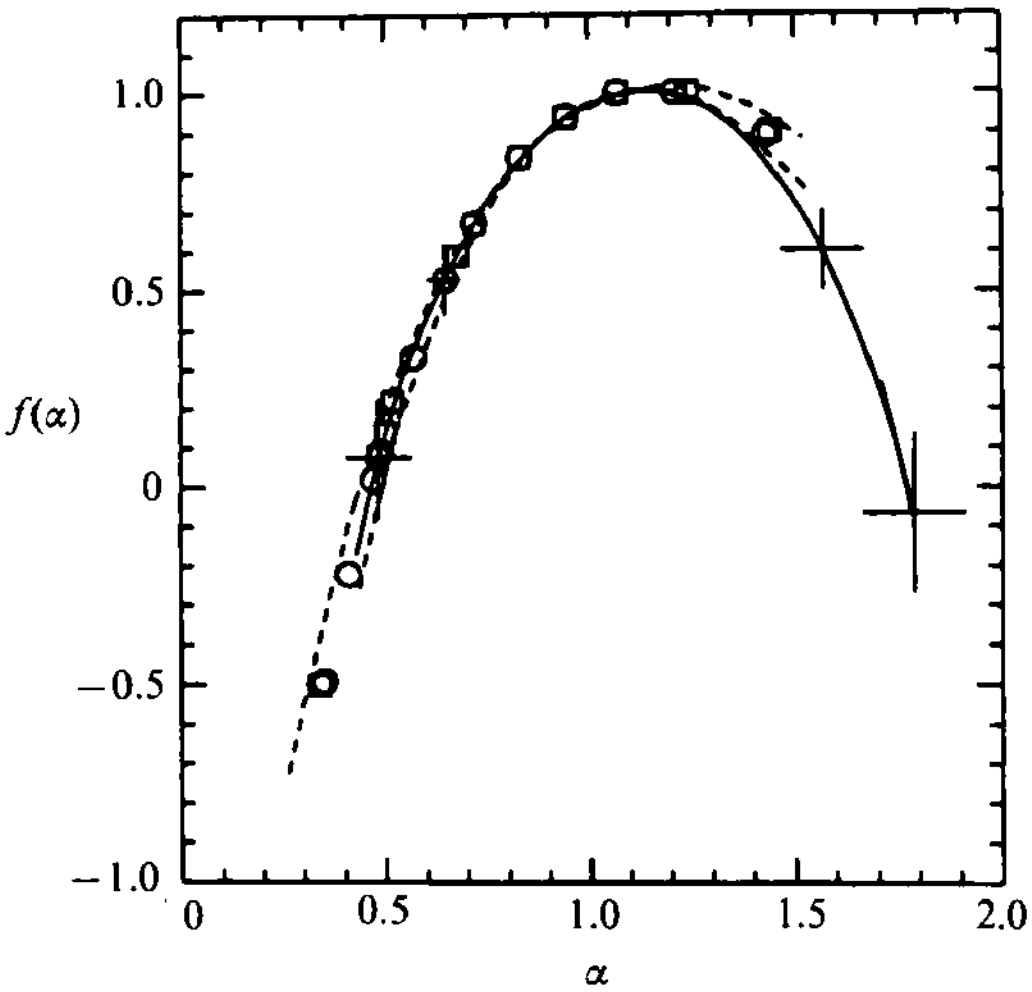

FIGURE 24. Multifractal spectrum $f(\alpha)$ obtained from Legendre transforming the results of figure 23. Symbols have the same meaning as in figure 23.

Next, the Legendre transform of $(q-1)D_q$ is computed to obtain the multifractal spectrum $f(\alpha)$; $\alpha$ is obtained by differentiating $(q-1)D_q$ using centred differences on the data of figure 23. The results, shown in figure 24, will be discussed in §4.

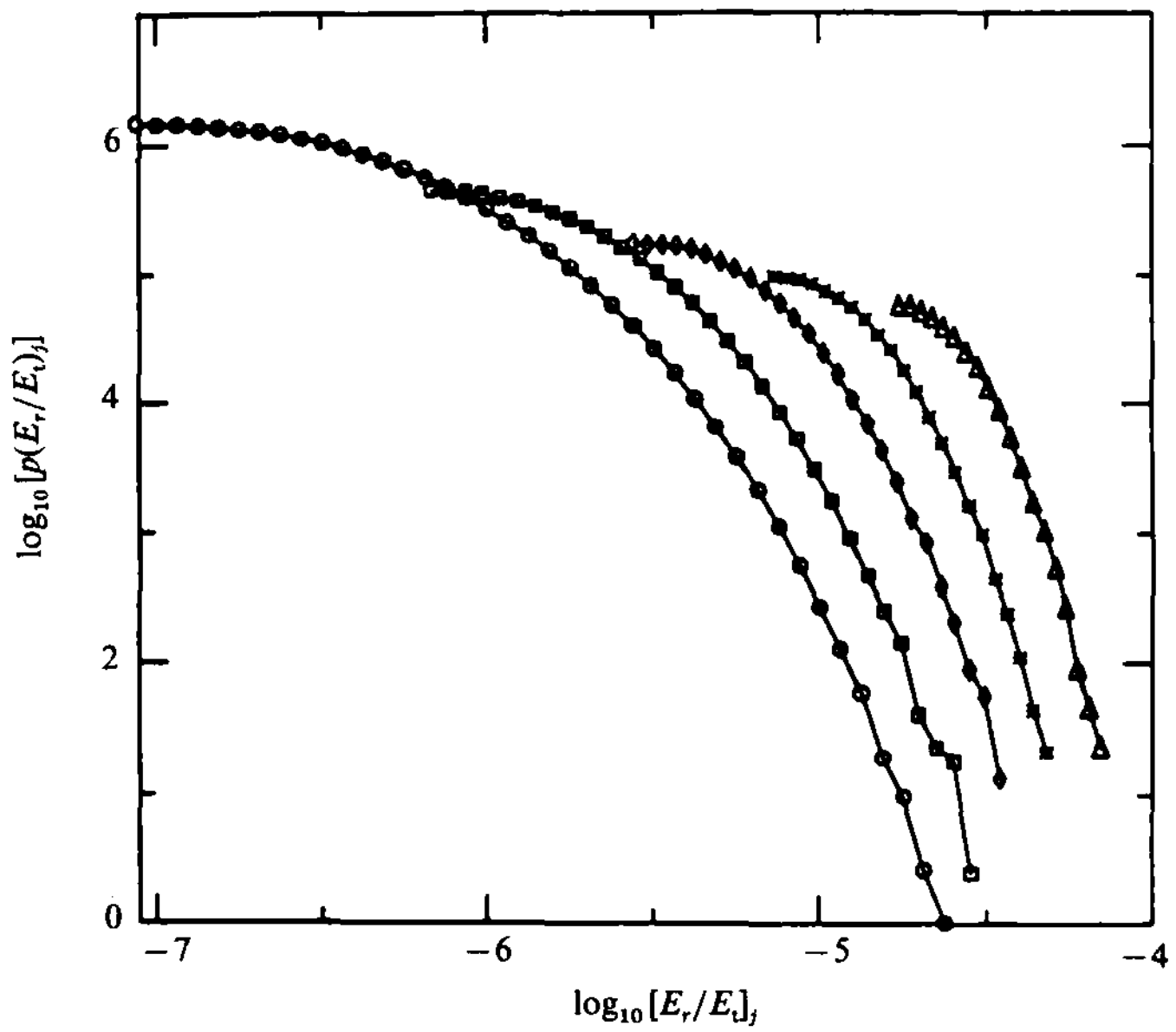

FIGURE 25. High-intensity tails of the probability density of $E_r/E_t$, plotted in log–log units. A linear behaviour at the tails would indicate a power-law (hyperbolic) distribution and divergence of high moments. Different symbols correspond to different box sizes as in figure 11.

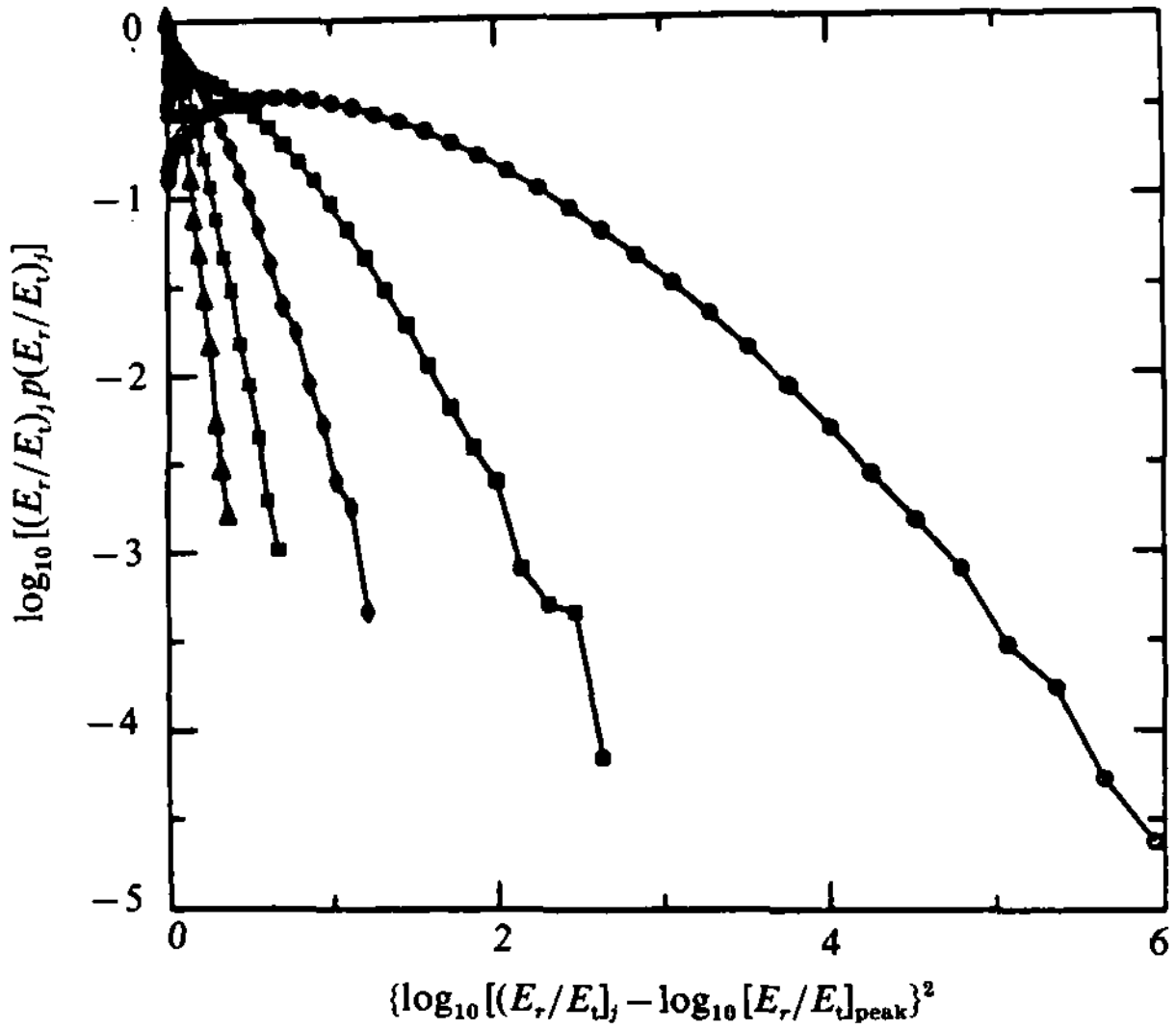

FIGURE 26. High-intensity tails of the probability density of $E_r/E_t$, plotted such that a linear behaviour at the tails indicates a lognormal distribution. Different symbols correspond to different box sizes as in figure 11.

### 3.3. *Analysis of the tails of the distribution*

In this section the possibility of extending the $D_q$ curve to $q > 5$ is considered. To do this, one needs even longer data records for proper convergence. As will be seen below, this is an inherently impossible task. Another possibility is to study in detail the tails of the distribution of $(E_r/E_t)$ in order to extend it on a rational basis to much higher values of $(E_r/E_t)$. Focusing on the probability density $p(E_r/E_t)$ of the dissipation in boxes of size $r$, we note that it is related to the earlier histograms by

$$p(E_r/E_t)_j = \frac{N(X_j)\,\Delta X}{\Delta(E_r/E_t)_j} N_r^{-1}. \tag{3.6}$$

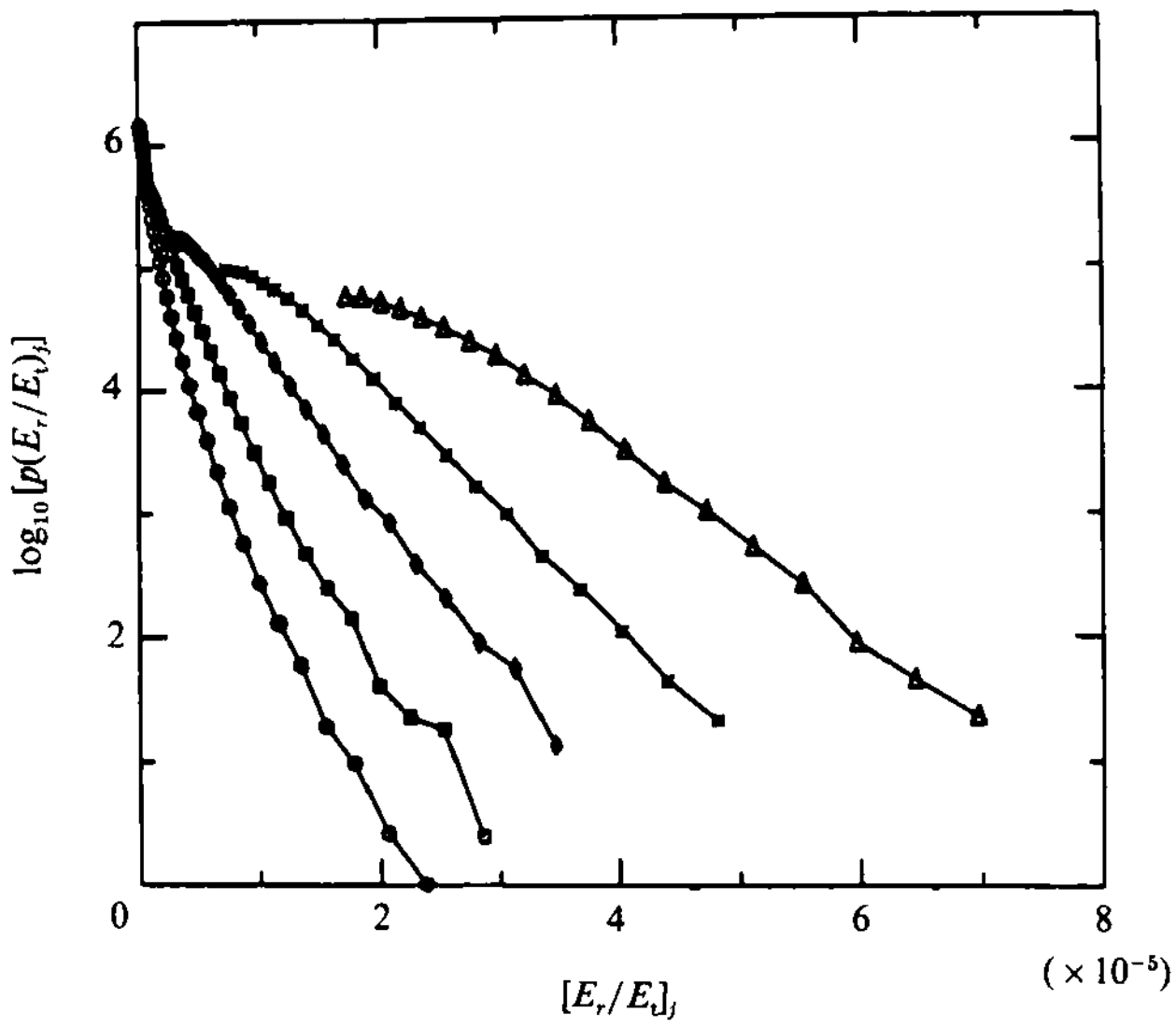

FIGURE 27. High-intensity tails of the probability density of $E_r/E_t$, plotted in log-linear units. Linear behaviour at the tails would indicate an exponential distribution. Different symbols correspond to different box sizes as in figure 11.

Here $N_r$, the number of boxes of size $r$, is equal to $\mathscr{L}/r$, and $\Delta(E_r/E_t)_j = (E_r/E_t)_{j+1} - (E_r/E_t)_j$. We now wish to distinguish among several possibilities of the high-intensity tails of $p(E_r/E_t)_j$ – namely hyperbolic, lognormal, exponential and square-root exponential.

If the tails are hyperbolic (Mandelbrot 1974, 1989; Schertzer & Lovejoy 1985), the distribution would obey

$$p(E_r/E_t) \sim (E_r/E_t)^{-\omega} \tag{3.7}$$

and yield straight lines of slope $-\omega$ on log–log plots. For such distributions moments of order higher than $\omega - 1$ do not exist. The log–log plots of tails of $p(E_r/E_t)_j$, shown in figure 25 for five different box sizes, suggest that the tails decay faster than linearly on such plots (especially for the smaller $r$). This behaviour is in agreement with the results of Anselmet *et al.* (1984) and Gagne (1987).

Next, the lognormal possibility deserves analysis, even though it is asymptotically inconsistent with multiplicative processes. For lognormal distributions (Kolmogorov 1962; Obukhov 1962), $p(E_r/E_t)$ follows

$$p(E_r/E_t) \sim (E_r/E_t)^{-1} \exp\{-\omega(r)[\log(E_r/E_t) - \log(E_r/E_t)_{\mathrm{peak}}]^2\}, \tag{3.8}$$

where $(E_r/E_t)_{\mathrm{peak}}$ is the value of $(E_r/E_t)$ at which the distribution peaks. This would imply that plots of $\log\{(E_r/E_t)\,p(E_r/E_t)\}$ *vs.* $[\log(E_r/E_t) - \log(E_r/E_t)_{\mathrm{peak}}]^2$ should yield straight lines. This is examined in figure 26, from which it appears that the measured distributions decay faster than lognormal tails, especially for the smaller boxes.

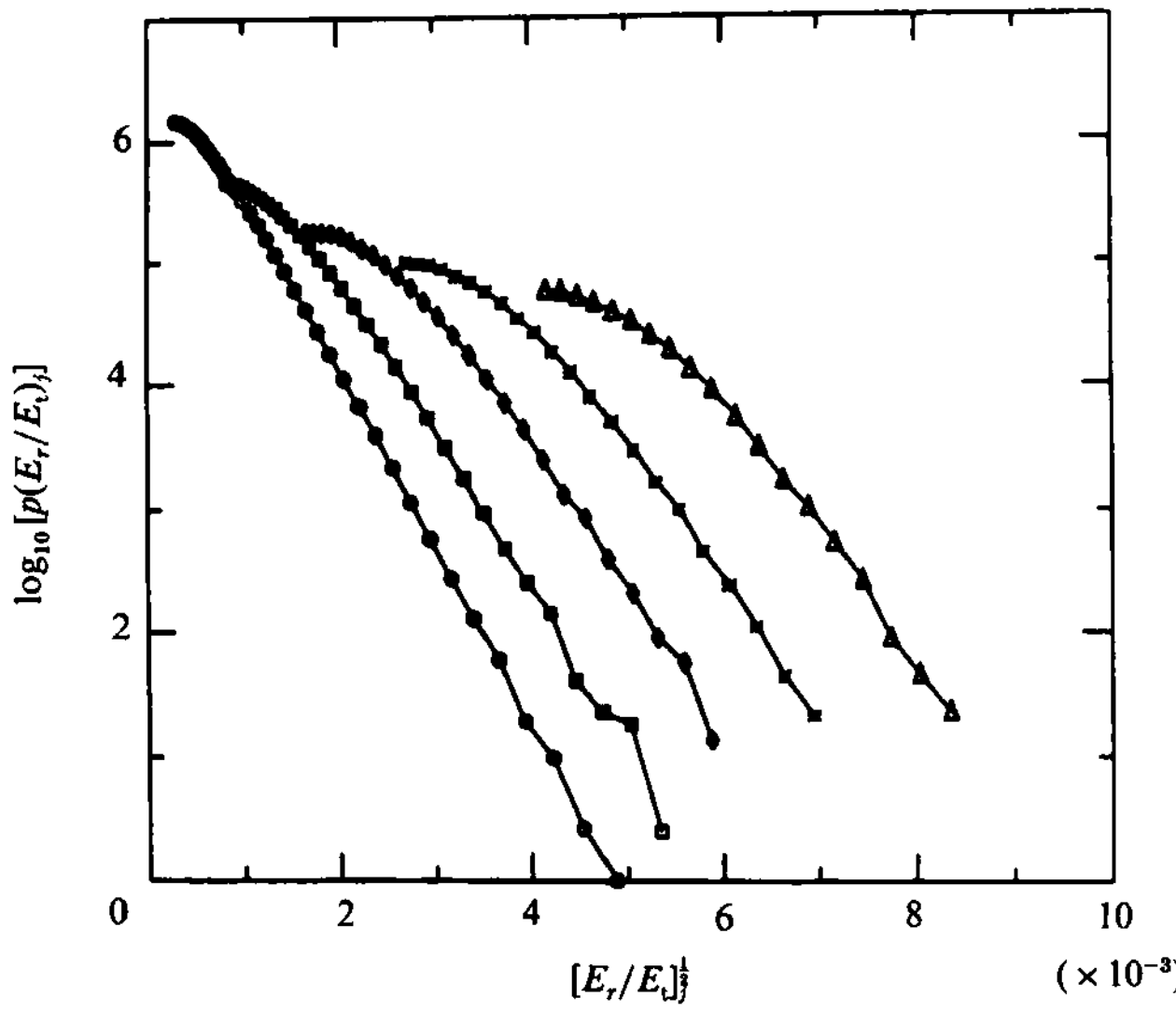

FIGURE 28. High-intensity tails of the probability density of $E_r/E_r$, plotted in logarithmic units as a function of $(E_r/E_t)^{\frac{1}{2}}$. The linear behaviour at the tails clearly shows that the tails of the distribution are of the square-root exponential type. Different symbols correspond to different box sizes as in figure 11.

A third alternative corresponds to exponential tails,

$$p(E_r/E_t) \sim \exp\{-\omega(r)(E_r/E_t)\}, \tag{3.9}$$

for which semi-logarithmic plots of $\log\{p(E_r/E_t)\}$ *vs.* $(E_r/E_t)$ should show linear behaviour. This is tested in figure 27. Here, unlike the two previous cases, it is apparent that for small box sizes the tails decay slower than the proposed distribution. This type of behaviour was also noticed by Gagne (1987) for velocity differences.

Finally we examine the possibility that the tails of the probability density are square-root exponential. This possibility has been suggested by Gagne (1987) for velocity differences (also, see §2.7). For such tails, one has

$$p(E_r/E_t) \sim \exp\{-a(r)(E_r/E_t)^{\frac{1}{2}}+b(r)\}. \tag{3.10}$$

By plotting $\log[p(E_r/E_t)]$ *vs.* $(E_r/E_t)^{\frac{1}{2}}$, one should observe straight lines of slope $-a(r)$ and intercept $b(r)$. It is apparent from figure 28 that such linear behaviour indeed exists for all box sizes. The magnitude of the slopes $a(r)$ is a slowly decreasing function of $r$, and the intercept $b(r)$ increases with $r$. The same behaviour is observed for the tails in the wake flow. We conclude that square-root exponential tails are the best candidate for extrapolation.

The actual extrapolation of the distribution $p(E_r/E_t)$ is performed as follows. First, $a(r)$ and $b(r)$ are estimated by linear least-square fitting through the seven right-most points of $p(E_r/E_t)$. Then 30 more points along that straight line are added to the distribution. (The extent of $p(E_r/E_t)$ is thereby increased by about 10 orders of magnitude, which is why we commented earlier that the required measurements are inherently impossible.) Then the moments are computed using

$$\sum_j [E_r(y_i)/E_t]^q = \sum_j (E_r/E_t)_j^q N_r p(E_r/E_t)_j \Delta(E_r/E_t)_j, \tag{3.11}$$

where the sum on the right-hand side includes all points added to the distribution by extrapolation. This is repeated for $q$-values ranging from 4 to 10 for both the boundary layer and the wake. Figure 29 shows the resulting log–log plots used to obtain $D_q$ with the fits (again in a range $r/\eta = 30$ to 300) indicated by solid lines. Figure 30 shows the $D_q$ curve, with points now extending up to $q = 10$.

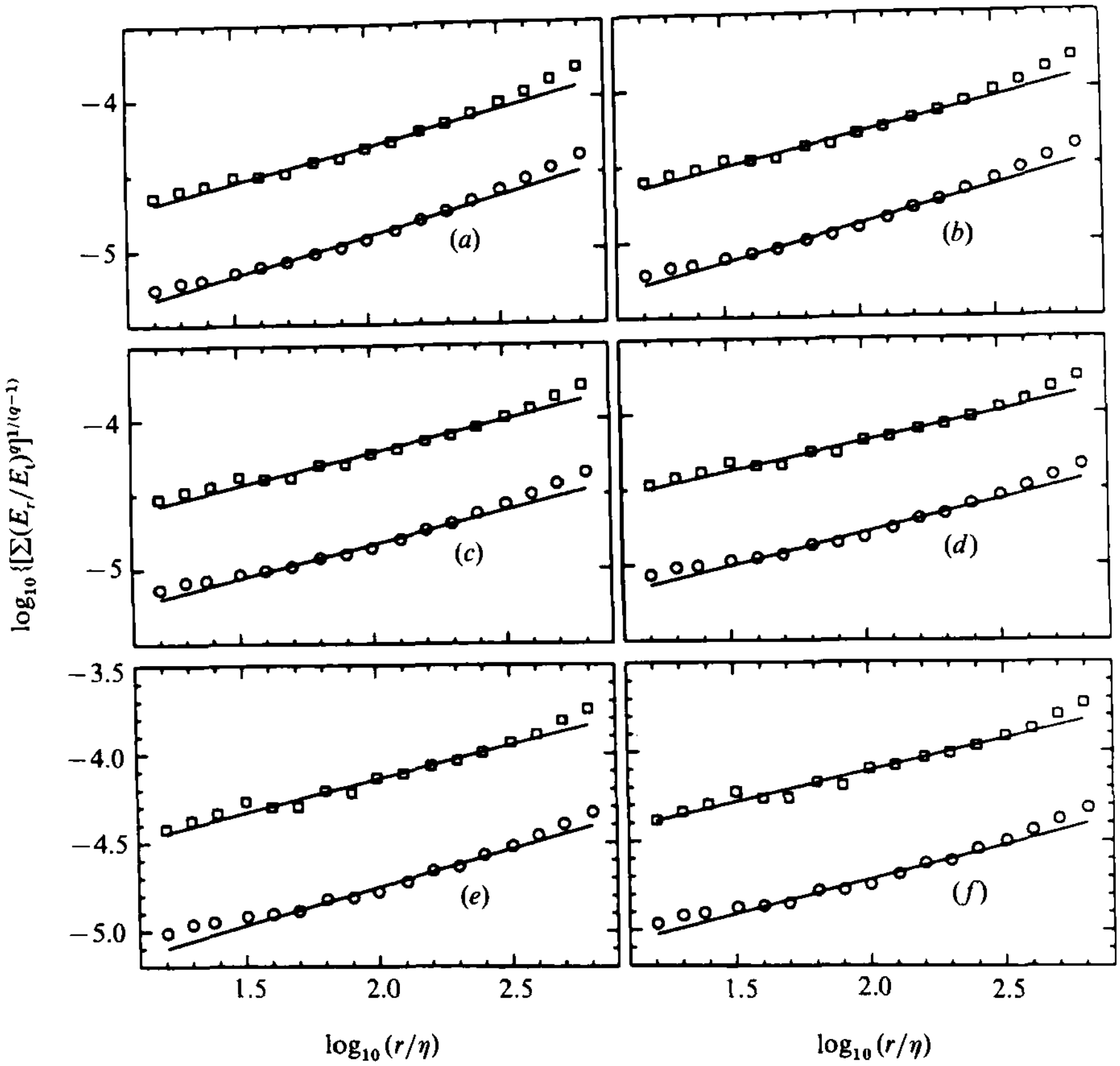

FIGURE 29. Log–log plots of $[\sum (E_r/E_t)^q]^{1/(q-1)}$ as a function of $r/\eta$ for several representative $q$-values between 5.5 and 10. The sums are computed using the distribution of $(E_r/E_t)$ extrapolated according to square-root exponential tails. Circles are for the laboratory boundary layer and squares for the wake. Solid lines are linear least-square fits in the range $r/\eta = 30$ to 300. Their slopes correspond to $D_q$ exponents obtained by this extrapolation procedure. (a) $q = 5.5$, (b) $q = 6$, (c) $q = 7$, (d) $q = 8$, (e) $q = 9$, (f) $q = 10$.

In order to find the asymptotic value of $D_q$ for $q \to \infty$, it is convenient to compute $\sum_i (E_r/E_t)^q$ as the product of the total number of boxes and the mean value of $(E_r/E_t)^q$ according to

$$\sum_i [E_r(x_i)/E_t]^q = N_r \langle (E_r/E_t)^q \rangle = N_r \int_0^\infty (E_r/E_t)^q p(E_r/E_t)\, \mathrm{d}(E_r/E_t). \tag{3.12}$$

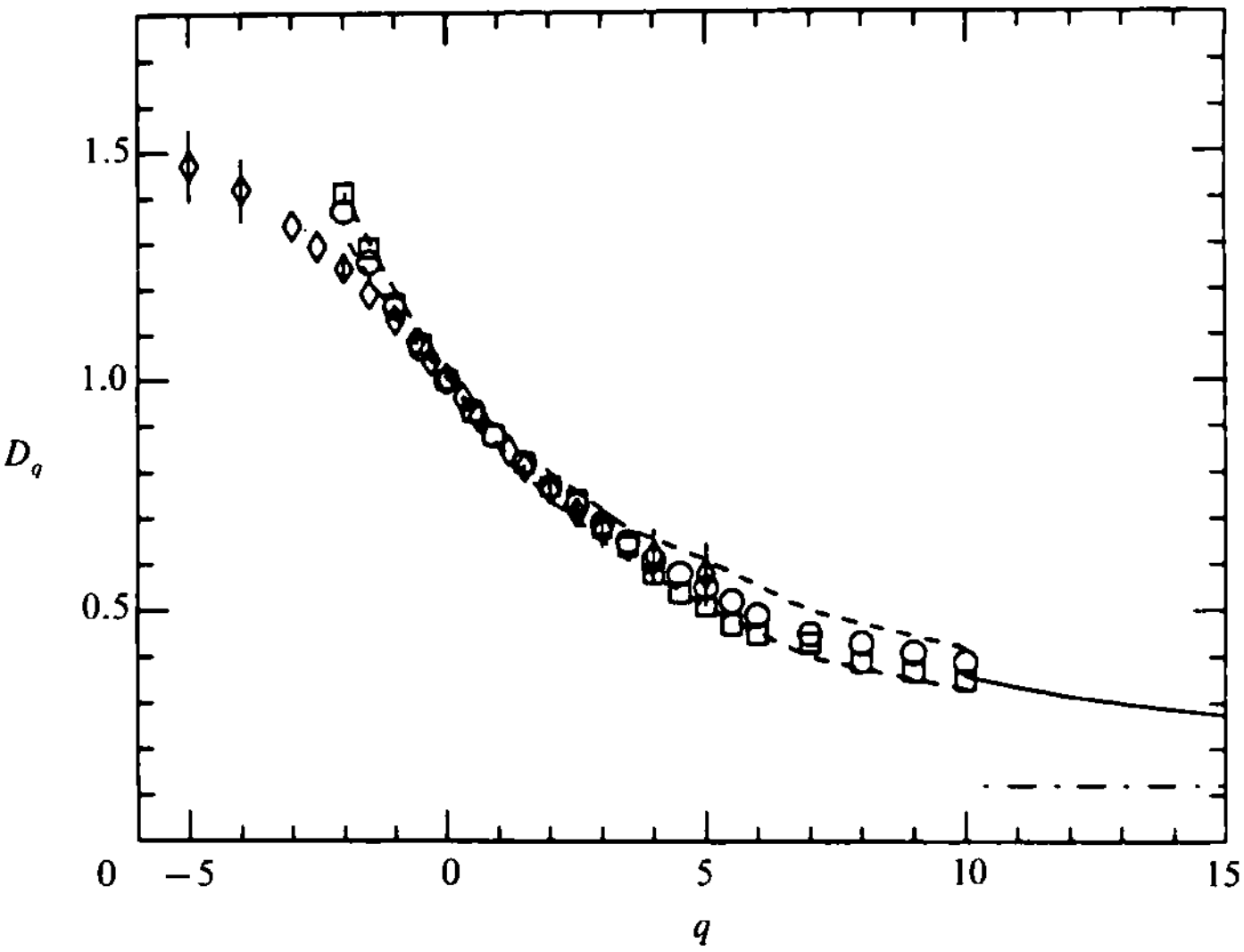

FIGURE 30. Moment exponents $D_q$ as a function of $q$ for both laboratory flows, where the results for $q \geqslant 5$ are obtained from extrapolated moments. Circles correspond to the boundary layer and the squares to the wake. The dashed lines represent $D_q$ values obtained from different scaling ranges (same as in figure 17). The solid line is the $D_q$ curve obtained from purely square-root exponential tails (see text), which asymptotes at a slow rate to a $D_\infty$ value of 0.12 indicated by the dot-dashed line.

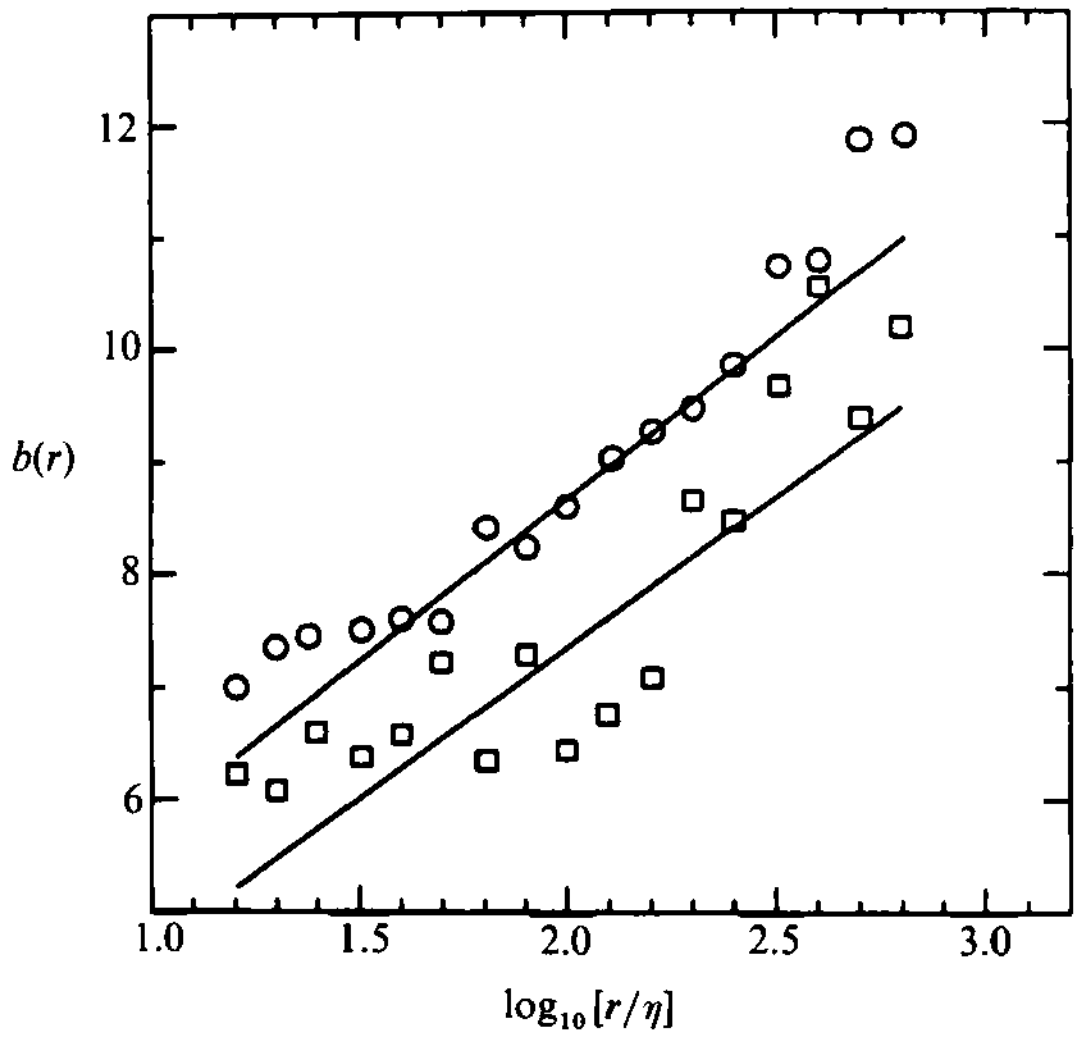

FIGURE 31. Values of the intercept $b(r)$ from the extrapolated distribution functions as a function of the box size $r$. Circles correspond to the boundary layer and squares to the wake. Solid lines are fits in the range $r/\eta = 30$ and 300. The slopes $\theta$ (roughly the same for the two cases) are estimated to be about $2.9 \pm 0.6$.

Replacing $p(E_r/E_t)$ by (3.10) and using $N_r = \mathscr{L}/r$ one obtains

$$\sum (E_r/E_t)^q = 2\mathscr{L}\Gamma(2q+2)\,e^{b(r)}a(r)^{-2(q+1)}r-1. \tag{3.13}$$

In order for this to obey a power-law with $r$, $b(r)$ and $a(r)$ have to be of the form

$$b(r) = \theta \log(r) + c, \quad a(r) \sim r^{-\phi}. \tag{3.14}$$

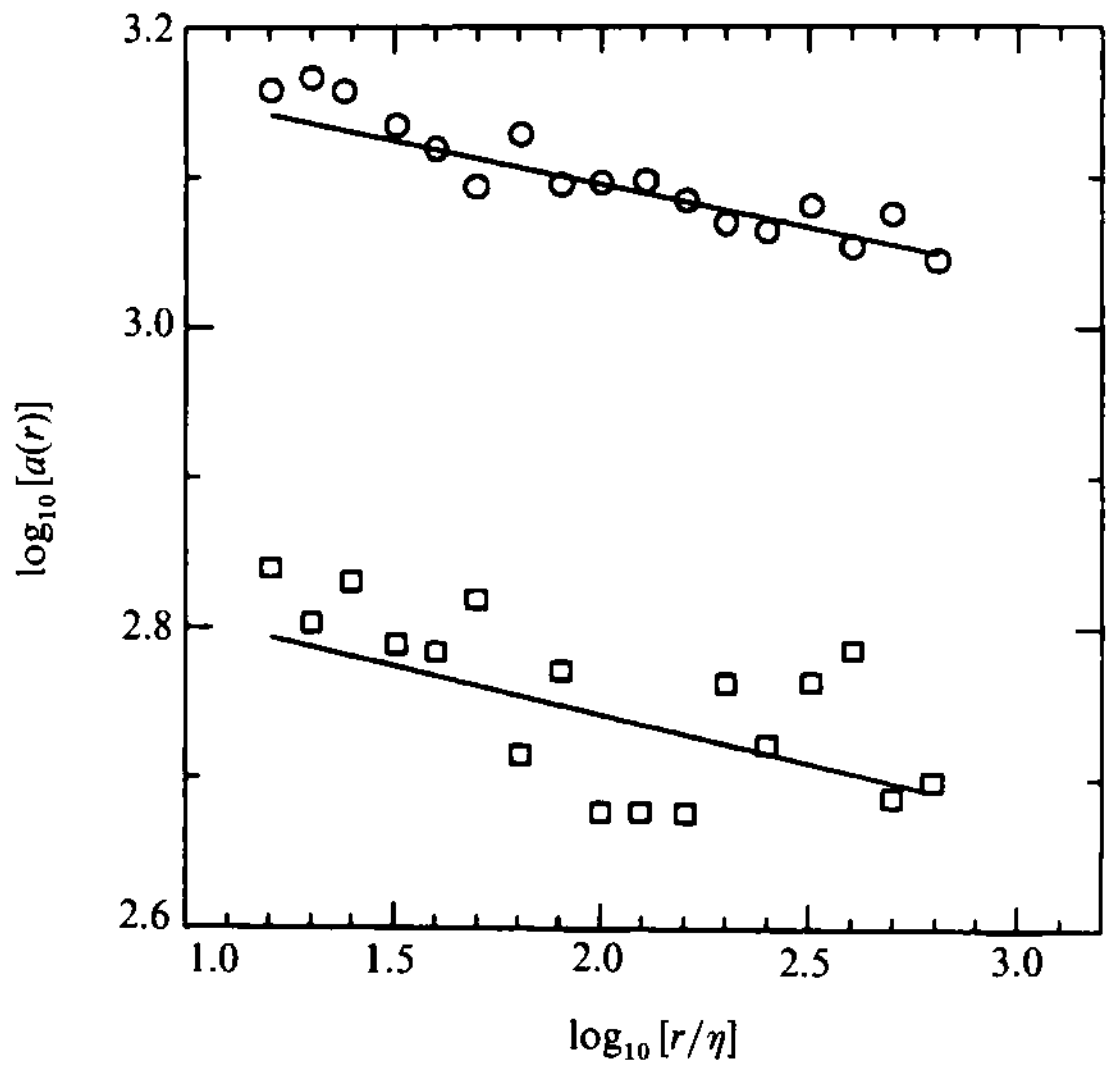

FIGURE 32. Values of the slopes $a(r)$ from the extrapolated distribution functions as a function of the box size $r$, plotted in log–log units. Circles correspond to the boundary layer and squares to the wake. Solid lines are fits in the range $r/\eta = 30$ and 300. The (negative) slopes $\phi$ are estimated to be about $0.06 \pm 0.04$.

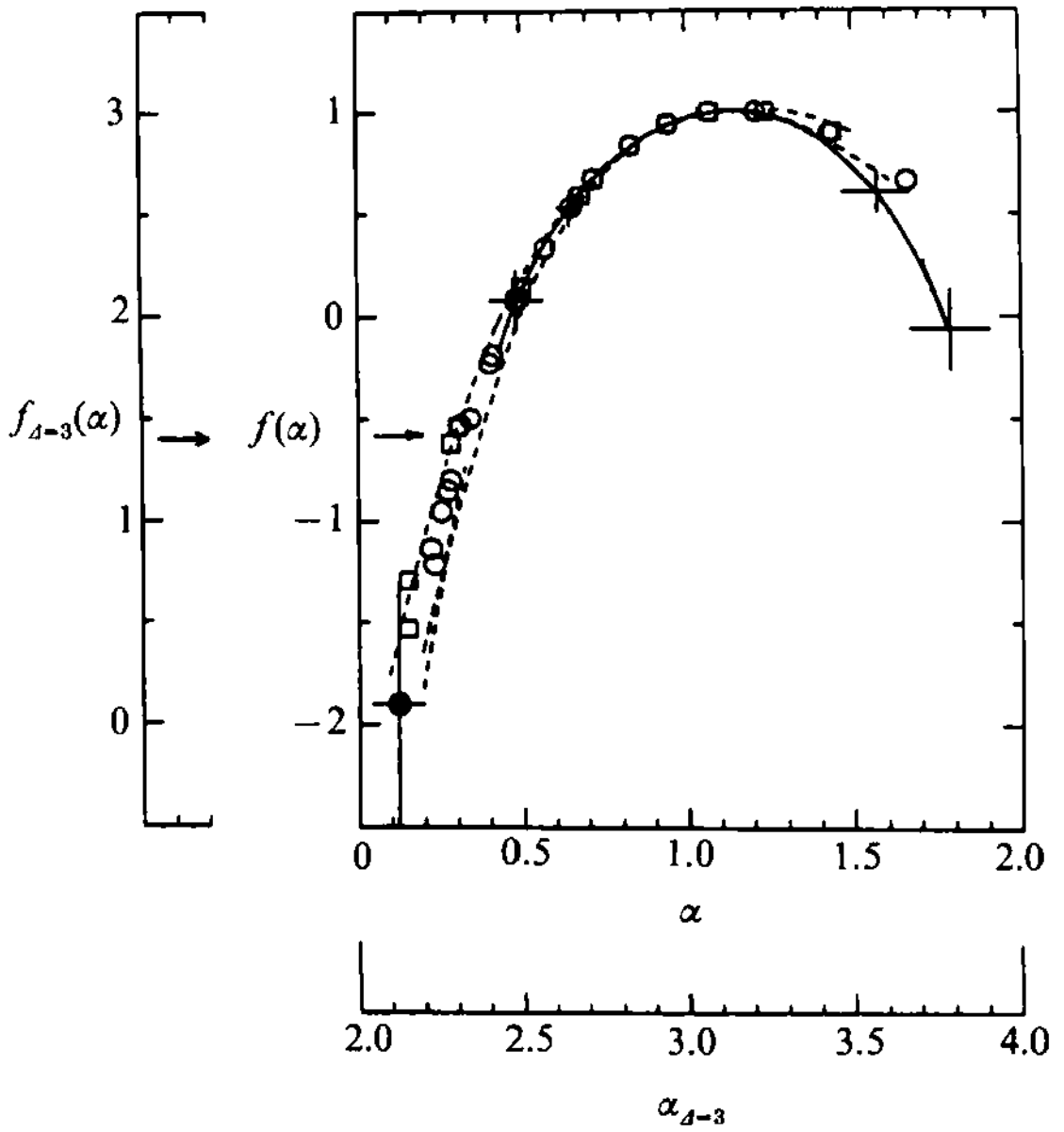

FIGURE 33. Multifractal spectrum $f(\alpha)$ obtained from Legendre transforming the results of figure 30. The solid circle at the lower left corner of the $f(\alpha)$ curve corresponds to the square-root exponential behaviour with the measured values of $\theta$ and $\phi$ (see text). The secondary scales are for the multifractal spectrum $f_3(\alpha)$ corresponding to the three-dimensional situation, obtained by adding 2 to $\alpha$ and $f(\alpha)$ from the experimental results. The solid line and the error bars are the mean and standard deviation of the various results from short-term averaging. Circles, squares and dashed lines correspond to the analysis of very long records of data in moderate-Reynolds-number flows (boundary layer and wake). The arrow marks the location down to where $f(\alpha)$ was computed without extrapolation. Lower values entail the extrapolation procedure. A detailed analysis of this curve is relegated to §4.

Substituting this into (3.13), and using $\sum (E_r/E_t)^q \sim r^{(q-1)D_q}$, we obtain

$$D_q = [2\phi(q+1)+\theta-1]/(q-1), \tag{3.15}$$

and in the limit, $D_\infty = 2\phi$. Figures 31 and 32 show $b(r)$ and $\log_{10}[a(r)]$ as functions of $\log_{10}[r/\eta]$ for both the boundary layer and the wake. The plots are consistent with a linear behaviour, substantiating relations (3.14). $\theta$ and $\phi$ are obtained from these plots by fitting straight lines through the data in the range $r/\eta = 30$ to 300. The scatter, relatively large especially for the wake data, should be kept in mind when interpreting the results. The mean values are

$$\theta \approx 2.9 \pm 0.6, \quad \phi \approx 0.06 \pm 0.04. \tag{3.16}$$

This implies that $D_\infty \approx 0.12 \pm 0.08$. Relation (3.15) is shown as the solid line in figure 30, and the dot-dashed line indicates the asymptotic value $D_\infty$.

Finally, $f(\alpha)$ is computed from the $D_q$ exponent obtained from the extrapolation procedure. The results are shown in figure 33. The $f(\alpha)$ curve was computed without any extrapolation of the distributions (figure 24) down to the arrow. Lower values are the results of extrapolation.

Asymptotically for $q \rightarrow \infty$, it is clear from (3.15) that

$$\alpha(\infty) = D_\infty = 2\phi \tag{3.17}$$

and

$$f[\alpha(\infty)] = -[2\phi+\theta-1]. \tag{3.18}$$

In the last step, (2.20) has been used. This asymptotic state is shown as the filled circle in figure 33 for the estimated values of $\phi$ and $\theta$. The termination of the $f(\alpha)$ curve at that point arises because of the rapid fall-off of the square-root exponential tail. A detailed interpretation of the $f(\alpha)$ curve is given in the next section.

## 4. Discussion of results and comparison with models

### 4.1. *Results*

The curve $f_3(\alpha_3)$ corresponding to the three-dimensional situation is obtained according to (2.34) by adding 2 to the values of $\alpha$ and $f(\alpha)$ obtained from one-dimensional cuts (figure 33). For $f_3(\alpha) > 2$, the curve seems fairly symmetric, with a maximum occurring at $\alpha = \langle\alpha\rangle = \alpha(q=0) \approx 3.13$ and $f_3(\alpha)_{\max} = 3.0$. The curve has unit slope (or $\alpha = f_3(\alpha)$) at the point $\alpha(q=1) = f_3(\alpha) = D_{\Delta=3, q=1} \approx 2.87$, this being the dimension of the set where all of the dissipation is concentrated asymptotically (Sreenivasan & Meneveau 1988). As remarked by Chhabra & Jensen (1989), $D_1$ is the dimension of the measure-theoretic support of the measure. On the other hand, $f_3(\alpha_3 = \Delta = 3) \approx 2.96$. This is the dimension of the set where all the singularities ($\alpha_3 < 3.0$) of the dissipation are located (Sreenivasan & Meneveau 1988). The fact that $f_3(\alpha_3)$ is larger than $f_3[\alpha_3(q=1)]$ (which is true quite outside of experimental uncertainty) means that the mean dissipation is dominated by some set where the dissipation is singular (but not extremely so!). This conclusion may have some bearing on closure models.

By computing the second derivative of $\tau(q) = (q-1)D_q$ at $q = 0$ (using centre differences on the data obtained from the short-term averaging) one obtains that $d^2\tau/dq^2 \approx -0.26 \pm 0.03$. From (2.23) we obtain the intermittency exponent $\mu \approx 0.26 \pm 0.03$. Also, remembering from the previous section that $D_2 \approx 0.76 \pm 0.02$ (for $d = 1$), we obtain that $\mu' = d - D_2 \approx 0.24 \pm 0.02$, comparable with $\mu$ within experimental accuracy. The small difference between the two results arises because of the multifractal nature of the dissipation.

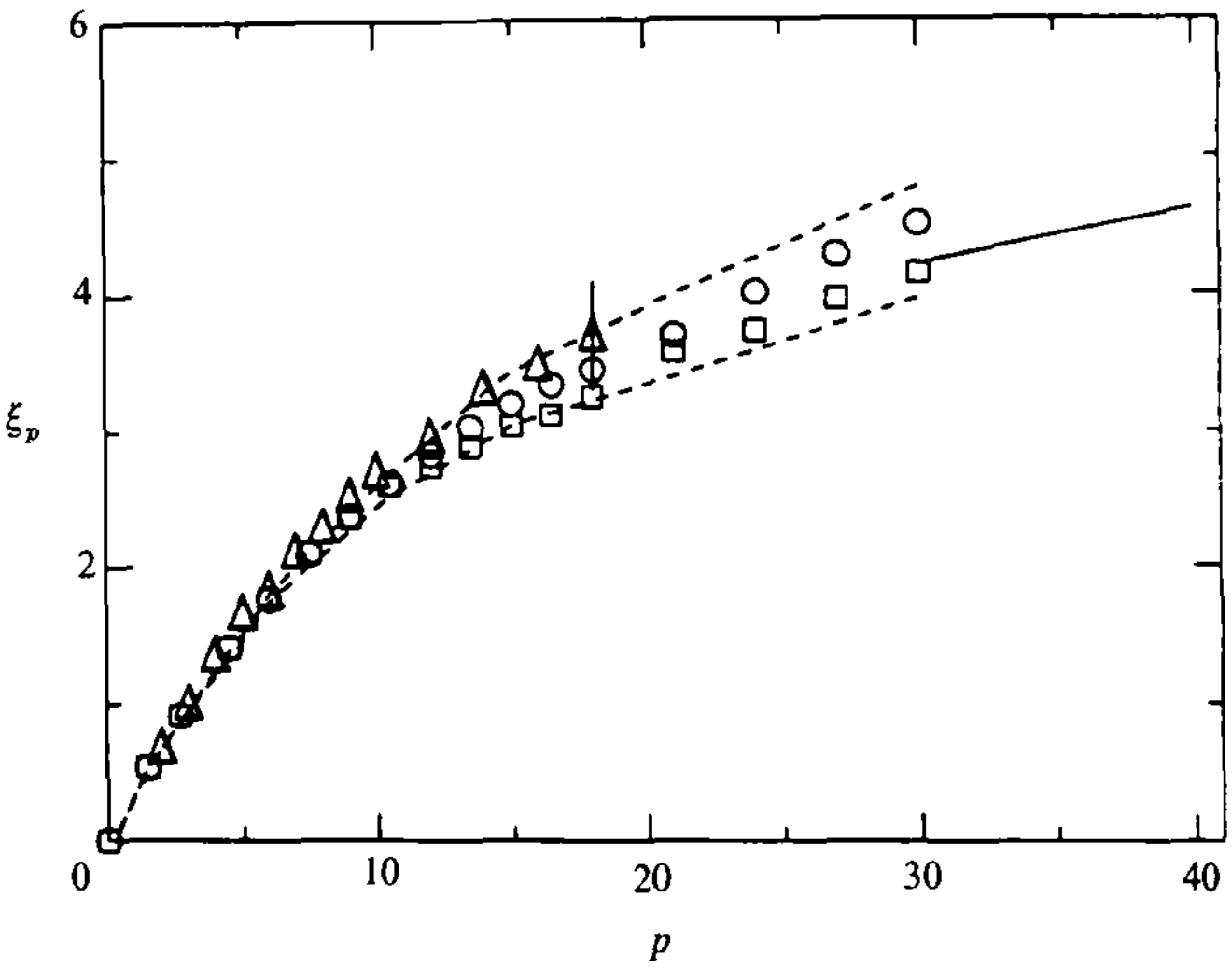

FIGURE 34. Scaling exponents $\xi_p$ of velocity structure functions. Triangles (and error bars) are from experimental results by Anselmet *et al.* (1984). The other symbols, dashed lines and the solid line are the results of figure 30, related to $\xi_p$ via (C 11). Both sets of results agree within experimental uncertainty.

It is now also possible to compare these results to those of Anselmet *et al.* (1984) on the velocity structure-function exponents. To do this, we compute $\xi_p$ from our $D_q$ curve using relation (C 11) of Appendix C. This relation assumes that $(r\epsilon_r)$ and the cube of $\Delta u_r = |u(x) - u(x+r)|$ have the same scaling laws. There is no direct evidence for this, the only rigorous result from the Kármán–Howarth equation being the equality of their mean values. The results of the comparison are shown in figure 34. The present results fall a little lower for high moments but the agreement is quite good considering the overall experimental uncertainty. Furthermore, using the asymptotic results corresponding to the square-root exponential tails, one obtains for high $p$ the result that

$$\xi_p = \tfrac{2}{3}\phi p + 2\phi + \theta, \tag{4.1}$$

which depends linearly on $p$ with a slope of $\frac{2}{3}\phi \approx 0.04$. This is depicted as solid line in figure 34.

### 4.2. *Comparison with models*

In this section, these experimental results are compared to models summarized in §2. Figure 35 ($a$, $b$) shows the present experimental results as small circles. The non-intermittent theory of Kolmogorov (1941), the $\beta$-model with $D = 2.87$, and Nakano & Nelkin's (1985) temporal wavepacket model (with $z = 0.84$, Nakano 1988*b*) are depicted using large symbols in ($a$) for the $f(\alpha)$ curve, and different lines in ($b$) for the $D_q$ curve. The $D_q$ curve in ($b$) corresponds to the three-dimensional case by using (2.34). The lognormal model with $\mu = 0.26$ is shown by the dashed line in both ($a$) and ($b$). As expected from the analysis of the tails of the individual distributions in §3, the experimental $f(\alpha)$ falls off faster than for lognormality.

The solid lines in figure 36 ($a$, $b$ correspond to the random $\beta$-model of Benzi *et al.* (1984) with their proposed binomial distribution of the random variable $\beta$ (see Appendix C):

$$p(\beta) = P\delta(\beta - 0.5) + (1-P)(\beta - 1). \tag{4.2}$$

From (C 9) it can be shown that $D_1 = 3 - P$, so that a $D_1$ of 2.87 selects $P = 0.13$. ($P$ was called $x$ in Benzi *et al.* 1984.) This model assumes that sheet-like structures are created with probability 0.13, while space-filling eddies are generated with probability 0.87. The model always yields $f(\alpha)_{\max} < 3.0$ (in this case 2.9), stemming from the assumption that some eddies receive no dissipation. As seen in figure 36 (*a*), the model works reasonably well for the left-most part of the $f(\alpha)$ curve (high-intensity dissipation) or at the higher moments. There is some disagreement around the peak and left part of the distribution, which is highlighted for the moment exponents $q < 0$ in figure 36 (*b*).

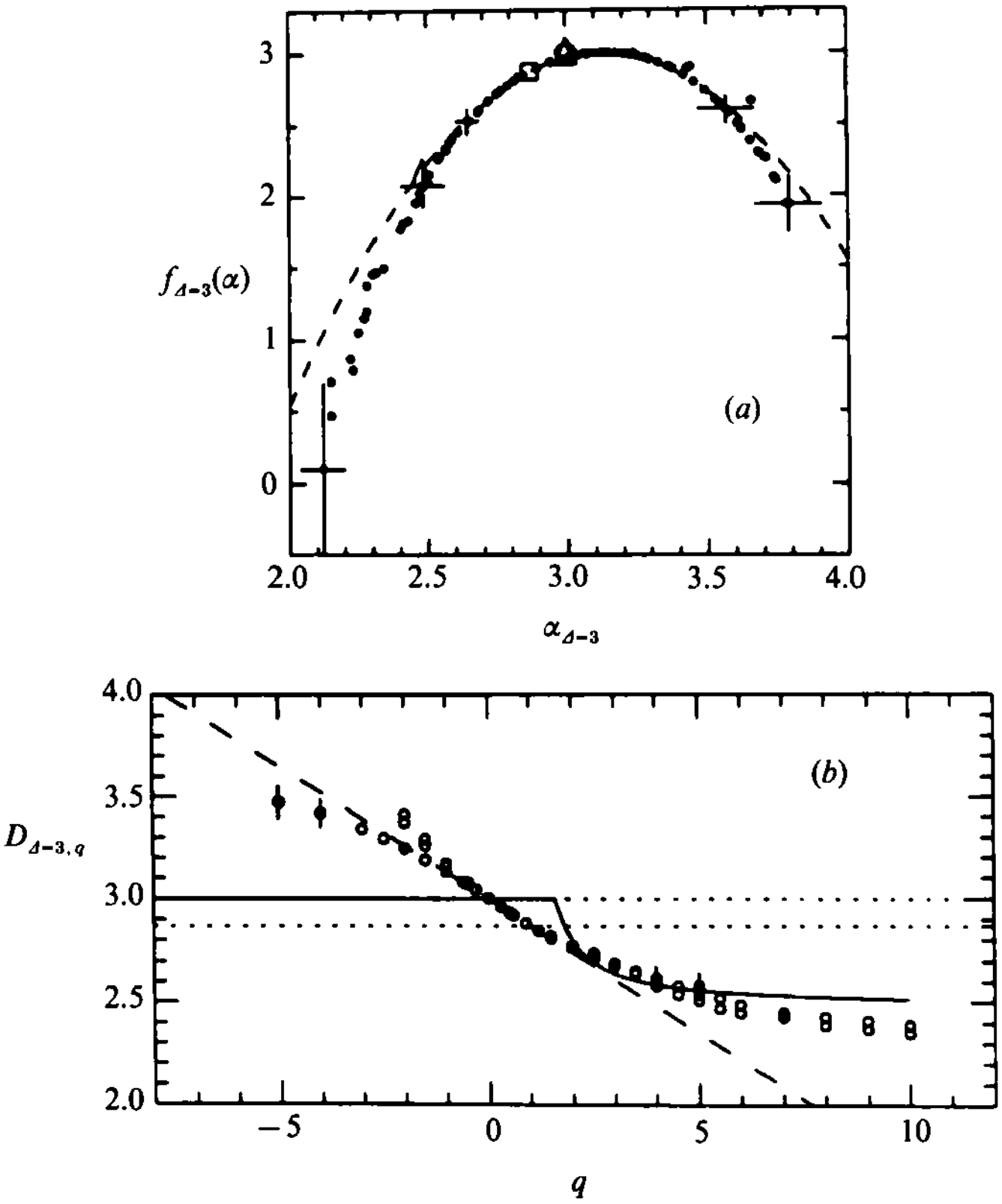

FIGURE 35. (*a*) Comparison of the $f_3(\alpha)$ curve from the experimental results (small circles and error bars) with several cascade models. The circle at $f = \alpha = 3$ is the original non-intermittent Kolmogorov (1941) theory. The square at $f = \alpha = 2.87$ corresponds to the fractally homogeneous $\beta$-model of Frisch *et al.* (1978). The two triangles correspond to the temporal wavepacket model (Nakano 1988*b*). The dashed line is a parabola corresponding to the lognormal distribution with $\mu = 0.26$. (*b*) Comparison of the $D_{3,q}$ curve from experimental results (small circles and error bars) with several cascade models. The dotted lines at $D_{3,q} = 3$ and 2.87 correspond to the Kolmogorov (1941) theory and the $\beta$-model respectively. The solid line corresponds to the temporal wavepacket model (with $z = 0.84$, Nakano 1988*b*). The dashed line is tangent to the measured $D_{3,q}$ curve at $q = 0$ corresponding to the lognormal distribution with $\mu = 0.26$.

In figure 37 (*a*, *b*) we illustrate some results of the $\alpha$-model of Schertzer & Lovejoy (1985). Here the random multipliers $M$ are assumed to have a distribution

$$p(M) = P\delta(M-M_0)+(1-P)\,\delta(M-M_1). \tag{4.3}$$

Forcing the curve to pass through the measured values of $\alpha_\infty = 2.12$ and $f(\alpha_\infty) \approx 0$, one obtains that $M_0 = 2^{-\alpha_\infty} \approx 0.23$ and $P \approx 8^{-1}$. $M_1$ is obtained from the normalization condition. The resulting $f(\alpha)$ curve is shown as solid line in figure 37 (*a*). The corresponding $D_q$ curve is shown in figure 37(b). Since this model involves two free parameters, there are other possibilities as well. As mentioned in §2.7 and discussed in detail in Appendix A, this model can also produce divergence of moments on linear cuts. From (A 8), we require $M_0 > \frac{1}{4}$ for this to occur. For instance, the choice $M_0 = 0.26$ with $P = 0.4671$ produces divergence of moments for $q \geqslant \frac{5}{3}$ on the linear cuts, which is the critical value $q_{cr}$ proposed in Schertzer & Lovejoy (1985). The $f(\alpha)$ and $D_q$ curves corresponding to this choice of parameters are shown by the dashed lines in figure 37(a) and (b) respectively. Notice that $D_{3,q} = 2$ or $D_{1,q} = 0$ when $q = \frac{5}{3}$. Other combinations of $M_0$ and $P$ giving divergence of moments of order $\frac{5}{3}$ can be readily found. However, since this always implies that the curve crosses the axis $\alpha = 2$ with a slope smaller than $\frac{5}{3}$ (see Appendix A), this is not compatible with present experimental results. In other words, since the $D_{3,q}$ curve must go through both $D_0 = 3$ and $D_{\frac{5}{3}} = 2$, it will fall far from observations (see figure 37b). Nevertheless, the model can be made more general by relaxing the condition of divergence of a specific moment on the linear cut. In fact, we shall see below that by assuming that $P = 0.5$, one can obtain good fits to the manifest part of $f(\alpha)$.

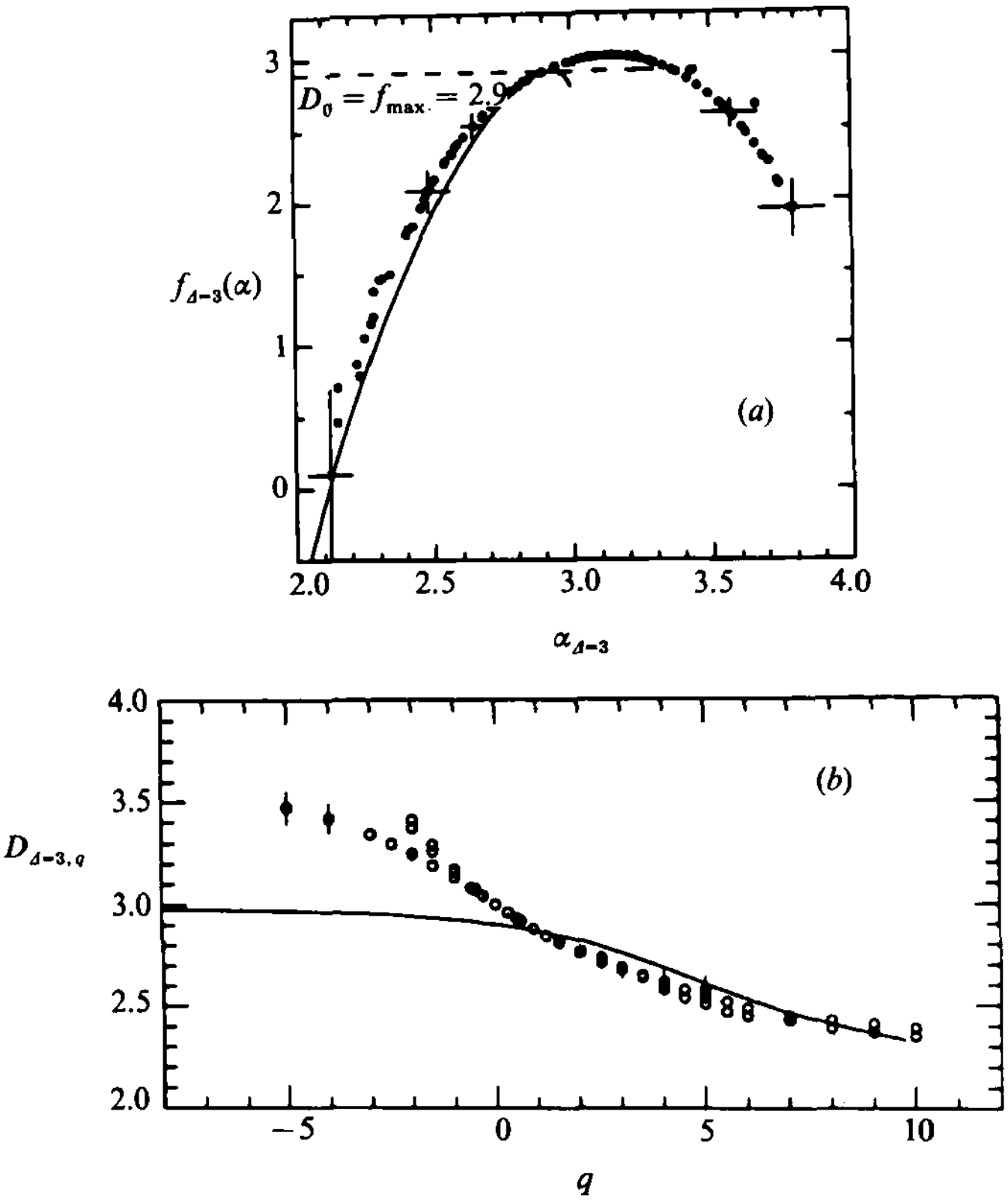

FIGURE 36. (a) Comparison of the measured $f_3(\alpha)$ curve (small circles) with the prediction of the random $\beta$-model of Benzi *et al.* (1984), using a binomial distribution for $\beta$ and fitting it at the point where $f = \alpha = 2.87$. (b) Comparison of the measured $D_{3,q}$ curve (small circles) with the prediction of the random $\beta$-model (solid line) with the same distribution as in (a).

The solid line of figure 38(a) in the range $f_3(\alpha) \geqslant 2$ corresponds to the binomial model described in §2.8, with $p_1 = 0.7$ $(= 4M_0)$. The agreement between the data and the model is quite good in the range $f_3(\alpha) \geqslant 2$. But owing to the assumption that at every stage the newly generated eddies receive exactly the same amount $M_0 = \frac{1}{4}p_1$ or $M_1 = \frac{1}{4}(1-p_1)$, both with probability $\frac{1}{2}$, this model does not produce singularities distributed on sets of dimension less than 2: singularity sheets are the sparsest sets that can be produced by this type of model. This can be seen in figure 38(b), where the $D_q$ curve agrees with experiments for $q$-values between $-3$ and 4. Higher moments emphasize singularities with $f_3(\alpha) < 2$, and give lower values for $D_{3,q}$ than the binomial model.

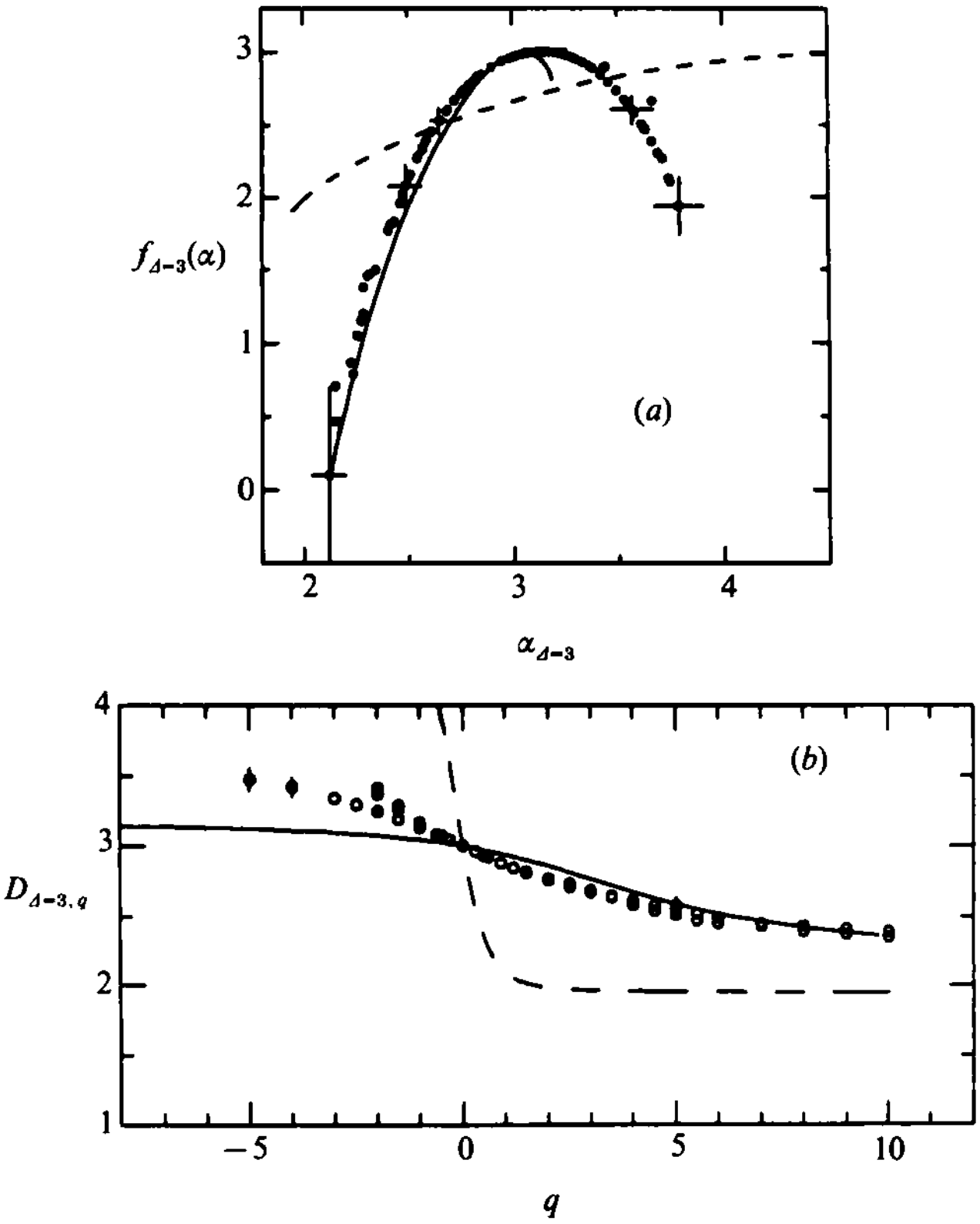

FIGURE 37. (a) Comparison of the measured $f_3(\alpha)$ curve (small circles) with two out of several possibilities arising in the $\alpha$-model of Schertzer & Lovejoy (1985). The model assumes that the multipliers $M$ can adopt two values with different probabilities, and has two free parameters. They can be selected for instance by forcing the curve to go through the lower-left most point of the curve (solid line), or to produce divergence of the $\frac{5}{3}$-order moments on the linear cuts (dashed line). (b) Comparison of the measured $D_{3,q}$ (small circles) with the outcomes of the $\alpha$-model. The legend is the same as in (a).

To model the entire range of $f_3(\alpha) \geqslant 0$, one can generalize the binomial model to a 'multinomial' one in which the number of free parameters can be made arbitrarily large. This restricts the usefulness of such a procedure. For completeness, we observe that (e.g.) a probability distribution where the multipliers can take on three distinct values with different probabilities according to

$$p(M) = P_0\,\delta(M-M_0)+P_1\,\delta(M-M_1)+P_2\,\delta(M-M_2), \tag{4.4}$$

produces, with $M_0 = 0.235$, $M_1 = 0.119$, $M_2 = 0.052$, $P_0 = \frac{1}{8}$, $P_1 = \frac{3}{4}$ and $P_2 = \frac{1}{8}$, the dashed lines of figure 38(a, b) – in good agreement with experiments.

Finally, it should be noted that the probabilistic model of Chhabra & Sreenivasan

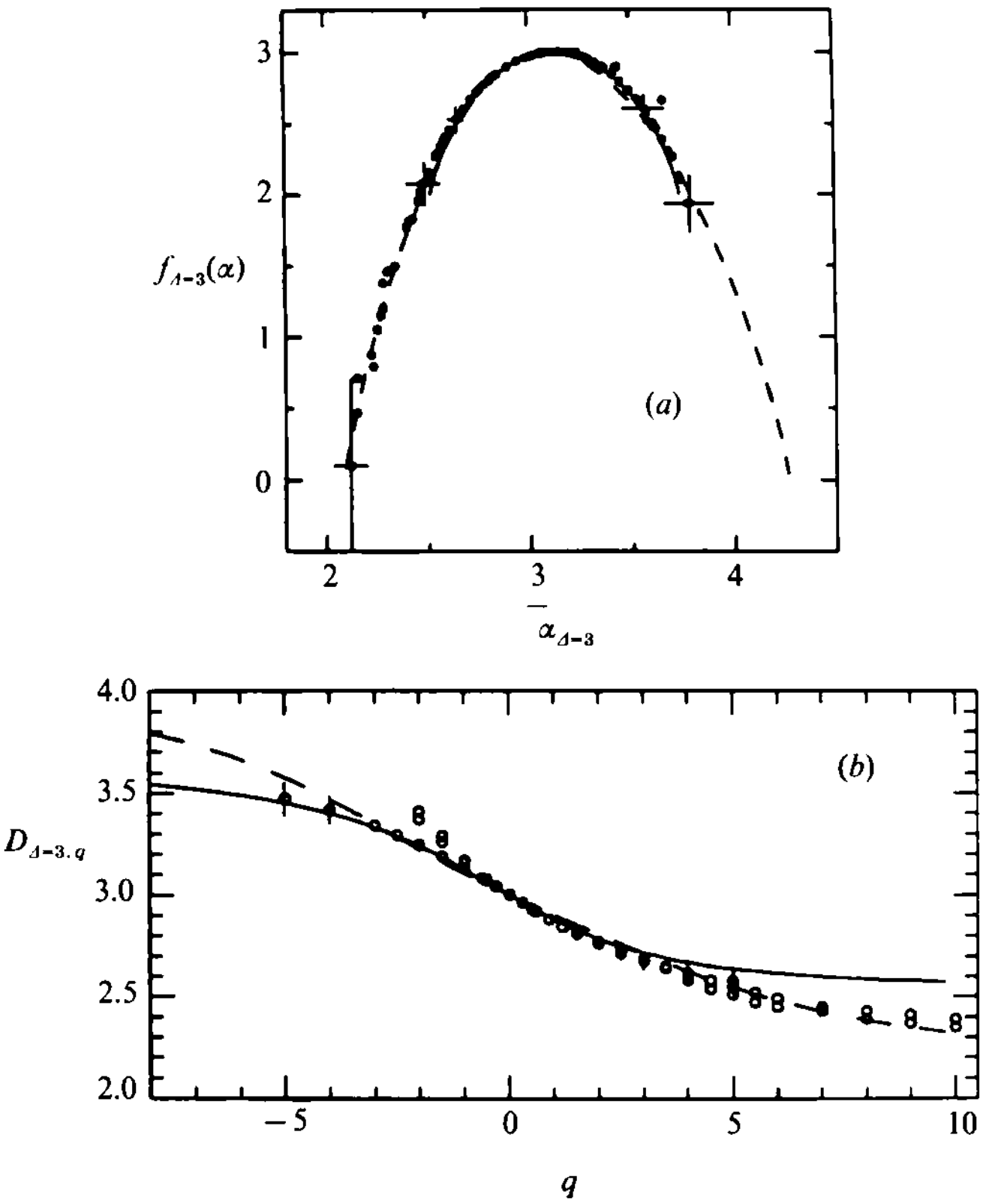

FIGURE 38. ($a$) Comparison of the measured $f_3(\alpha)$ curve (small circles) with a binomial distribution (solid line) of the multipliers (with $4M_0 = p_1 = 0.7$ and $p_2 = 0.3$), but each value occurs with the same probability, $\frac{1}{2}$. This model (designated the $p$-model in Meneveau & Sreenivasan 1987$b$) agrees well with the data in the range $f_3(\alpha) \geqslant 2$. If the multipliers are allowed to assume three values, each with different probabilities, it is easy to fit the entire curve (e.g. the dashed line shows such a fit). ($b$) Comparison of the measured $D_{3,q}$ curve (small circles) with the $p$-model (solid line) and with its multinomial extension (dashed line).

(1990) – to which we have already referred – agrees well with the measured $f(\alpha)$ curve over the entire range.

## 5. Conclusions

The main conclusions are now summarized. The observation that the dissipation field $\epsilon'$ has a multifractal distribution supports the notion of a self-similar multiplicative fragmentation process occurring in turbulent flows. Using concepts from the theory of random curdling, it was shown that one could in principle use linear cuts to obtain information on $f_3(\alpha)$ of the three-dimensional distribution. We point out that recent analysis in three dimensions of direct numerical simulations of homogeneous shear flows (Deane & Keefe 1988) and isotropic turbulence (Hosokawa & Yamamoto 1990) give $f(\alpha)$ curves that are in good overall agreement within experimental accuracy. The only difference is that they show a slightly smaller degree of intermittency than does our mean $f(\alpha)$. Quantitative results on high-order moments are not very accurate because the scaling range is modest at moderate Reynolds numbers, but are of sufficient quality to highlight much of the high-density tail of $f_3(\alpha)$, with an error margin of about $\pm 15\%$. It was found that the probability distributions of the dissipation rates exhibit square-root exponential tails. By extrapolating this behaviour, we were able to infer the asymptotic values of moment exponents and the $f(\alpha)$ distribution. The asymptotic value of $D_q$ for $q \to \infty$ appears to be somewhat larger than zero on the linear cuts which, according to Appendix A, implies that there is no divergence of moments. This result is based on laboratory flows only. For the atmospheric flow, the number of points needed to explore this issue satisfactorily can be shown to be prohibitively large. If we wanted to 'capture' singularities with $f_3(\alpha) \sim 0$ using a flow where $L/\eta \sim 10^4$ (as in the atmospheric flow), we would need $(10^4)^3$ points – several years of data acquisition! Perhaps the only way of obtaining useful results there is via the multiplier method used by Chhabra & Sreenivasan (1990). This method takes explicit advantage of scale similarity at various levels and averages information over them. The method also gives $D_\infty > 0$.

The present results are related to inertial-range exponents such as structure-function exponents, and are essentially the same as previous results of Anselmet *et al.* (1984). We emphasize that this means that the inertial-range scaling can be deduced (at least to a good approximation) by examining the scaling of the dissipation rate $\epsilon$ when averaged over inertial-range boxes.

Comparing measurements with several models of intermittency, it was concluded that scaling models with single exponents, lognormal and $\beta$-models are not satisfactory in general. In this sense, $f(\alpha)$ is a useful characterization of intermittency, since it permits one to establish the validity of cascade models. On the other hand, it was shown that simple versions of random curdling (binomial or multinomial models) could account for observations in the manifest part of the $f(\alpha)$ curve. However, owing to the degeneracy of the multifractal formalism (Feigenbaum, Jensen & Procaccia 1986; Chhabra *et al.* 1989) one cannot claim that the turbulent fragmentation process actually proceeds according to these simple models, but it is worth noting that spatial fluctuations of $\epsilon$ can be well quantified by the multipliers 0.7 and 0.3. These numbers have to be understood in the following sense. Dynamically, we lack a convincing model for the spatial characteristics of the flux of kinetic energy to small scales. If such a process were to occur, it must exhibit fluctuations – this being the origin of intermittency. The multipliers 0.7 and 0.3 correspond to the simplest possible fluctuations that will reproduce most of the observations. Indeed, all the practically important moments are sufficiently low that they can be obtained by knowing the positive part of $f(\alpha)$ on the linear cuts only. The merit of the simple binomial model is that, unlike lognormality, its high-order moments are consistent with a multiplicative process, even though it reproduces the observations only over single 'typical' cascades on the linear cuts. To reproduce the more infrequent events occurring on sets of dimension smaller than two (corresponding to the latent part of $f(\alpha)$), one needs to invoke more general processes such as the multinomial process of §4 or the probabilistic model of the type discussed by Chhabra & Sreenivasan (1990).

It is important to stress that the multifractal nature of the dissipation implies a non-trivial spatial structure, which can be seen for instance in the behaviour of two-point correlation functions of multifractals. It was shown (Meneveau & Chhabra 1990) that there are interesting spatial correlations in the local exponents $\alpha$, stemming from the fact that the measure at two nearby points will share more common 'history' of the multiplicative process than those that are far apart. This reasoning can be made precise (Cates & Deutsch 1987; Meneveau & Chhabra 1990), and might lead to improved statistical treatment of the fine-structure of turbulence.

It is of interest to highlight other questions concerning the multifractal description of turbulence. For instance, the degree of correlation existing among joint distributions of intermittent quantities in turbulence, such as the dissipation of kinetic energy and the dissipation of passive scalar fluctuations, or the squared vorticity, can be well described by extending the multifractal formalism to more than one variable (Meneveau *et al.* 1989). Another interesting problem addressed in Ramshankar (1988) and Tong & Goldburg (1988) concerns the behaviour of multifractal scaling exponents during the transition to fully developed turbulence. In addition, the multifractal nature of the dissipation has implications for the number of degrees of freedom (Meneveau & Nelkin 1989) as well as for the fractal dimension of interfaces (Meneveau & Sreenivasan 1990) in turbulent flows. Another interesting problem is the extension of the multifractal formalism to non-isotropic fields (Schertzer & Lovejoy 1985).

Finally, we note that all these models involved the binary base (i.e. $b = 2$). Other bases can be shown to make no difference to the scaling properties embodied in the $f(\alpha)$ curve. However, these models assume that all offspring are of the same size. It turns out that fluctuations in the size of the new pieces created during the cascade also typically lead to multifractal distributions (this is what typically leads to multifractal measures of attractors in phase space). The statistics of such fluctuations can, under certain conditions, be related to expansion and contraction rates of fluid elements. (For a discussion of this approach in the context of passively convected vector and scalar fields, see Finn & Ott 1988; Ott & Antonson 1989.) In turbulence, we suspect that a mixture of fluctuating length and measure multipliers is the most likely possibility. As mentioned before, this is impossible to discern among the plethora of possibilities using the $f(\alpha)$ curve alone (Chhabra *et al.* 1989; Chhabra 1989. It is interesting to recall the demonstration of Chhabra *et al.* that it is in general not necessary to consider variation in both length and measure multipliers.) Other data processing techniques such as wavelet transforms (Grossmann & Morlet 1984; Everson, Sirovich & Sreenivasan 1990; Meneveau 1990), detailed flow visualization, analysis of full numerical simulations, etc., may go some way to clarifying dynamical details leading to small-scale intermittency.

We wish to thank A. B. Chhabra, R. V. Jensen, B. B. Mandelbrot and M. Nelkin for many stimulating discussions. We are especially thankful to Benoit Mandelbrot for drawing attention to the notion of negative dimensions and their relation to his early work. This work was supported by DARPA (URI) and AFOSR.

## Appendix A. Sections through fields generated by random curdling

In this appendix we examine the relation between the multifractal features of $d$-dimensional intersections through intermittent fields generated by random curdling in $\Delta$-dimensions. We start by noting that *densities* of the measure, or *averages* of the dissipation rate, are the same in a given box, whether one obtains it in the $\Delta$-dimensional domain, or on a $d$-dimensional cut. This also holds for the ratios of densities. Therefore, the ratios of the total measure or energy flux $M$ on the $\Delta$-dimensional domain (denoted henceforth by $M_{(\Delta)}$) can be related to the ratios of total measure $M_{(d)}$ on the $d$-dimensional cut by equating the corresponding densities

$$M_{(d)}\, b^{d} = M_{(\Delta)}\, b^{\Delta}. \qquad \text{(A 1)}$$

Therefore, whenever the multiplier in a $\Delta$-dimensional domain is $M_{(\Delta)}$, the multiplier

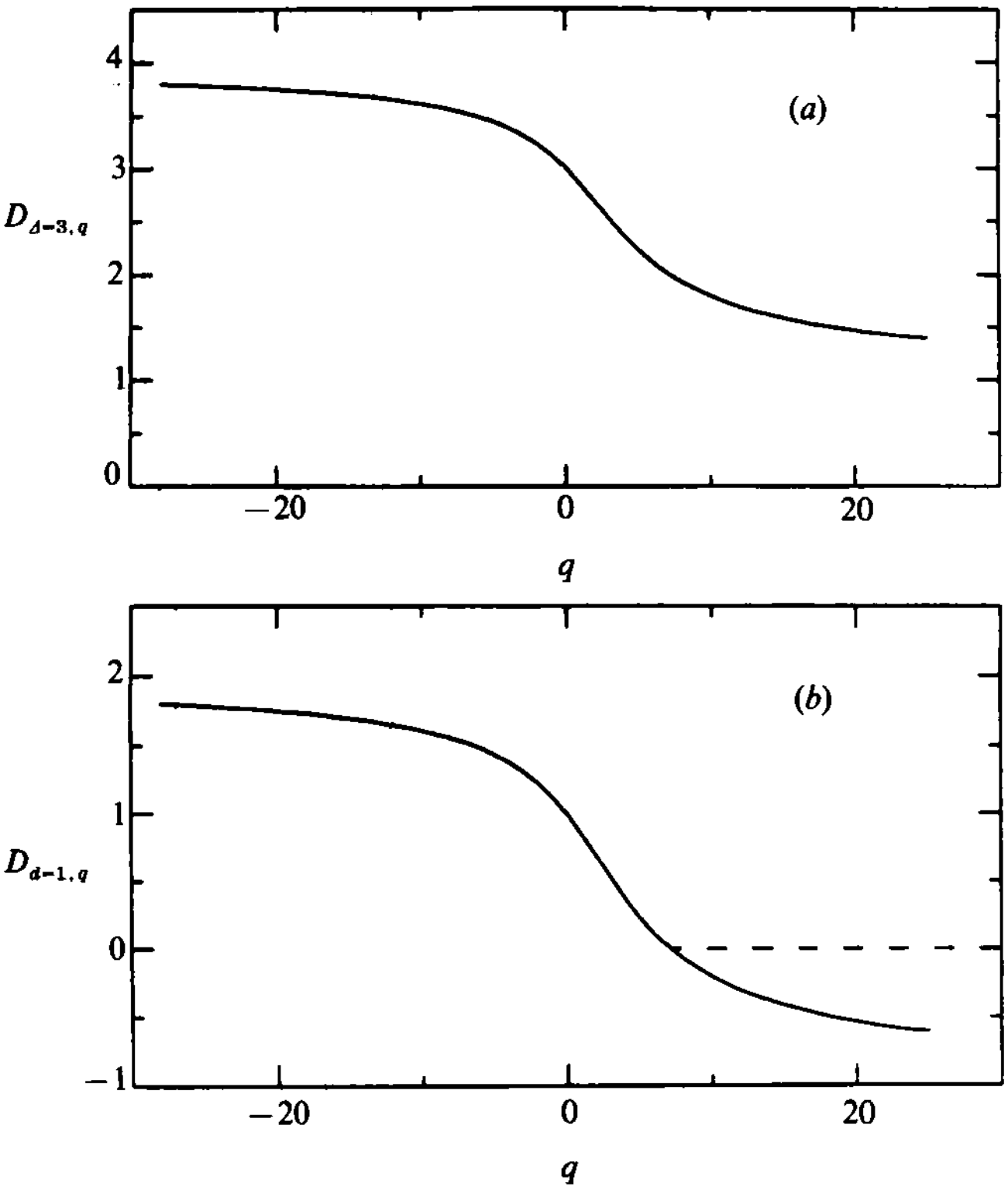

FIGURE 39. (a) $D_{3,q}$ *vs.* $q$ for the multiplicative process in three dimensions using the exponential distribution of multipliers. The mean of $M_{(\Delta)}$ is $2^{-3}$ and the minimum and maximum values it can acquire are $\frac{1}{16}$ and $\frac{1}{2}$ respectively. The distribution is $p(M_{(\Delta)}) - 42.958 \exp[-15.89361 M_{(\Delta)}]$ for $\frac{1}{16} < M_{(\Delta)} < \frac{1}{2}$ and zero otherwise. $D_{3,q}$ is computed according to $D_{3,q} = \log_2[8\langle M^q \rangle]/(1-q)$. (b) $D_{1,q}$ for a $d = 1$ dimensional cut through the three-dimensional process, obtained by subtracting 2 from (a) (see text). The dashed line shows the result that one would measure from an 'experimental' one-dimensional cut, where only the last stage of the cascade is known (see end of this section).

$M_{(d)}$ on the $d$-dimensional cut is $M_{(\Delta)}\, b^{\Delta-d}$. One now considers a multiplicative process on a $d$-dimensional domain with multipliers given by $M_{(d)}$, where $M_{(d)}$ has the same statistics as $M_{(\Delta)}\, b^{\Delta-d}$. In particular, the condition of normalization implies that

$$\langle M_{(d)} \rangle = b^{-d}, \tag{A 2}$$

but the local condition of conservation is relaxed on the cut. Therefore, one now concentrates on a $d$-dimensional, non-conservative, multiplicative process with base $b$ and multipliers $M_{(d)}$ obeying the properties (A 1) and (A 2).

We focus again on $E_{d,r}$, the total dissipation contained in a $d$-dimensional box of size $r$. As before, we have

$$r/L = b^{-k}, \quad \eta/L = b^{-n}, \tag{A 3}$$

where it is presumed that the cascade stops once a box size $\eta/L$ is reached after $n$ stages. The total dissipation in a box of size $r/L$ after the cascade has proceeded $k$ steps only, will be given by a certain sequence of multipliers $M_{(d)}$ according to

$$E_{d,r}/E_d = \prod_{j=1}^{k} M_{(d),j} = E_{\Delta,r}/E_\Delta\, b^{\Delta-d}. \tag{A 4}$$

From this it follows that the $D_q$ exponents, as well as $\alpha$ and $f(\alpha)$ of the distribution in $d$-dimensions are simply related to those in $\Delta$-dimensions according to (2.34). Thus by knowing the exponents in $\Delta$-dimensions, one can obtain the corresponding ones in $d$-dimensional cuts, but the question of more practical interest is the inverse problem of obtaining the exponents in the $\Delta$-space from those in the $d$-dimensional cut.

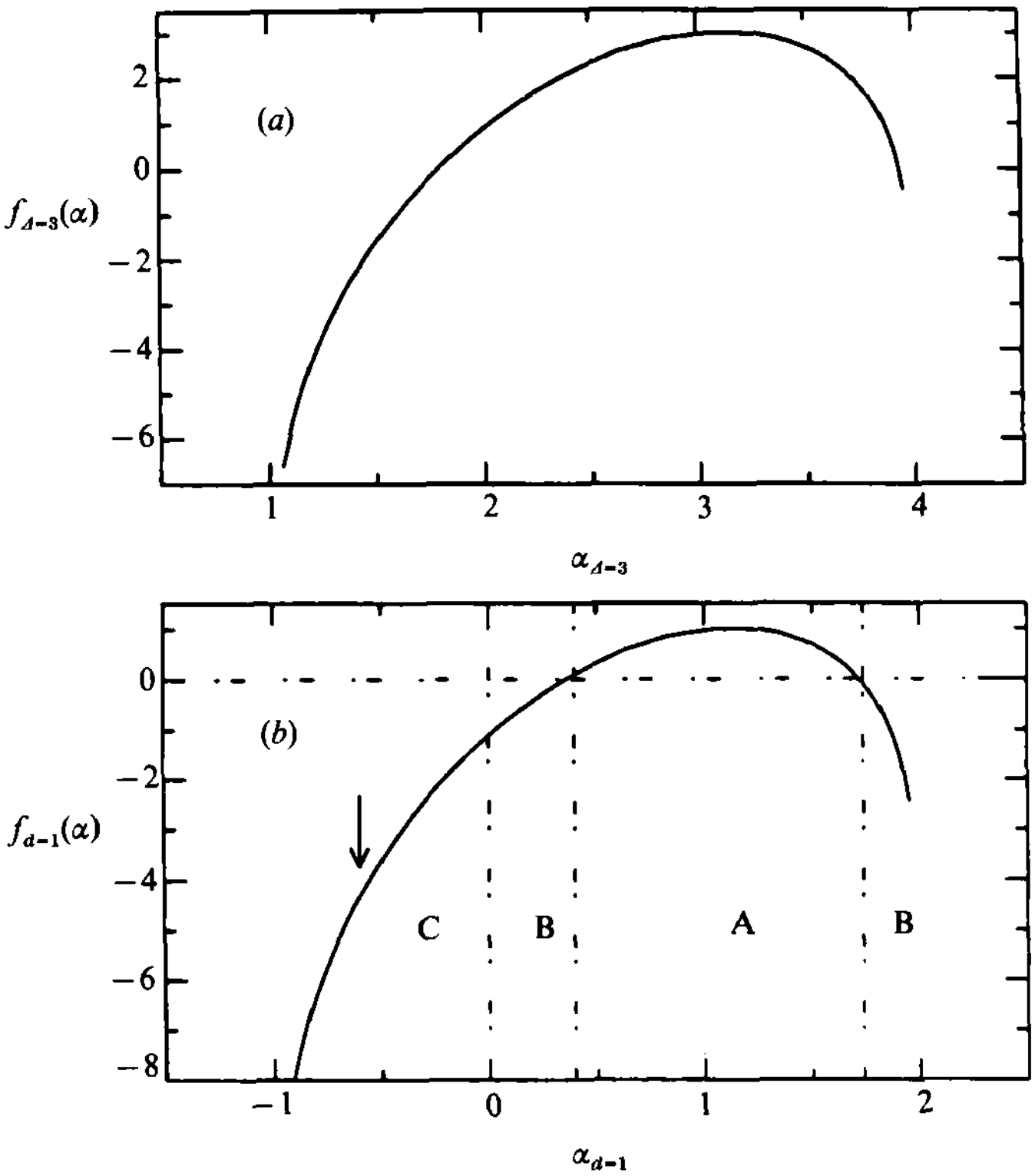

FIGURE 40. (*a*) $f_3(\alpha)$ of the measure in three dimensions obtained by applying the Legendre transforms to figure 39(*a*). (*b*) $f_1(\alpha)$ for a one-dimensional cut through the measure of (*a*), obtained by subtracting 2 to both $\alpha$ and $f(\alpha)$ of (*a*). Regions A, B and C correspond to manifest, latent and virtual singularities (see text). The arrow shows the limit of $f(\alpha)$ corresponding to the dashed line of figure 39(*b*). The point where $f_1(\alpha)$ reaches this limit has a tangent that goes through the origin.

Before considering this, a specific example might be helpful in illustrating the ideas presented so far. Let us consider a process in three dimensions ($\Delta = 3$) and with base $b = 2$, where the multipliers obey the following distribution:

$$p(M_{(\Delta)}) = \begin{cases} A\,\mathrm{e}^{-BM_{(\Delta)}} & \text{for } \frac{1}{16} < M_{(\Delta)} < \frac{1}{2} \\ 0 & \text{otherwise.} \end{cases} \tag{A 5}$$

The constants $A$ and $B$ are obtained by normalizing $p(M_{(\Delta)})$ and requiring that $\langle M_{(\Delta)} \rangle = 2^{-3}$.

Figure 39(*a*) shows the $D_{3,q}$ curve in three dimensions obtained by applying (2.31) to this process. Figure 39(*b*) shows the $D_{1,q}$ curve for the corresponding process on a one-dimensional cut ($d = 1$) obtained from (2.34). Similarly, figure 40(*a*) and 40(*b*) show respectively the $f_3(\alpha)$ curve in three-dimensions ($\Delta = 3$) (obtained from the $D_{3,q}$ curve using the Legendre transforms), and the $f_1(\alpha)$ curve on a one-dimensional cut ($d = 1$) through the three-dimensional distribution; (2.34) has been used again. Since $M_{\min} = \frac{1}{16}$, we have that $D_{3,-\infty} = \alpha_{3,\max} = \log_2[M_{\min}]^{-1} = 4$. Also, since $M_{\max} = \frac{1}{2}$, we have that $D_{3,\infty} = \alpha_{3,\min} = \log_2[M_{\max}]^{-1} = 1$. Since $M$ is never zero, $D_{3,0} = \Delta = 3.0$. Another interesting property arising from the continuous probability density $p(M)$ is that the probability of $M$ being exactly $M_{\min} = \frac{1}{16}$ or $M_{\max} = \frac{1}{2}$ is zero. Both $f(\alpha_{\min})$ and $f(\alpha_{\max})$ are related to the probability of $M$ being exactly $M_{\max}$ or $M_{\min}$ at evey step in the cascade according to $f_3(\alpha) = \log_b[P(M)] + \Delta$. This shows that the value of $f_3(\alpha)$ tends to $-\infty$ at the tails of the curve, consistent with figure 40($a$).

Following Mandelbrot (1989), it is convenient to organize a more detailed discussion of $f_1(\alpha)$ into three separate cases; whether all or some of them occur in practice depends on the precise statistics of the multipliers $M$.

*Manifest Singularities:* This corresponds to a range of $\alpha$-values ($\alpha$ now stands for the singularity strength on the $d$-dimensional cut, i.e. $\alpha_d$) such that $f_d(\alpha) \geqslant 0$, or $f_\Delta(\alpha) \geqslant \Delta - d$. This is shown as region A in figure 40($b$). In this range, $f_d(\alpha)$ can be interpreted as a dimension, and there are no problems when going from $d$ dimensions to $\Delta$. Also, a single cut, or a single realization of the cascade in $d$ dimensions will typically capture all the singularities that are densely distributed such that $f_\Delta(\alpha) > \Delta - d$. This is obvious since $f_d(\alpha) > 0$ means that there is more than one box where $\alpha$ has a certain value. This number becomes larger and larger as $r$ decreases, or as the level $k$ increases, and remains of order unity if $f_d(\alpha) = 0$.

*Latent Singularities:* This corresponds to a range of $\alpha > 0$ where $f_d(\alpha) < 0$ or $f_\Delta(\alpha) < (\Delta - d)$. This region is denoted by B in figure 40($b$). The condition $f_d(\alpha) < 0$ means that there is typically less than one box in a typical sample with those values of $\alpha$. Since the formulation is probabilistic, it is convenient to write that the probability of $\alpha$ occurring in a band $\mathrm{d}\alpha$ (dropping normalization constants) is

$$\Pi_r(\alpha)\,\mathrm{d}\alpha \sim b^{-kd} b^{kf_d(\alpha)}\,\mathrm{d}\alpha. \tag{A 6}$$

This is smaller than $b^{-kd}$ whenever $f_d(\alpha) < 0$ (when $\alpha$ is within region B). Therefore, a typical $d$-dimensional cut will miss these $\alpha$-values. However, since $\Pi_r(\alpha)\,\mathrm{d}\alpha$ is small but non-zero, if one takes many cuts or many realizations of the cut, one will inevitably encounter such rare $\alpha$-values. One consideration of interest in §3 is the number of cuts one has to take to be able to detect an $\alpha$-value whose $f_\Delta(\alpha) = 0$, or $f_d(\alpha) = d - \Delta$. According to (A 6), the probability of a box having such an $\alpha$ is $\Pi_r(\alpha)\,\mathrm{d}\alpha \sim b^{-k\Delta}\,\mathrm{d}\alpha$. Since there are $b^{kd}$ boxes on a single $d$-dimensional cut, the probability of encountering such a value on the entire cut is $\sim b^{k(d-\Delta)}$. It follows that one would need $\sim b^{k(\Delta-d)}$ such cuts to have a probability of detecting such an $\alpha$-value of order one. Therefore, *latent singularities* can be detected by increasing the number of cuts at a given resolution $r = b^{-k}$. It is important to realize that increasing the resolution $r$ or $k$ does in principle decrease the probability of encountering the rare events on a $d$-dimensional cut. Also, note that, in the example, there are latent singularities even on the $\Delta = 3$ dimensional domain, meaning that the high values of $M$ occur so rarely that even a single realization of the three-dimensional multiplicative process will not always contain the most intense singularity corresponding to $\alpha_{3,\min} = 1$.

*Virtual Singularities:* This is the region shown as C in figure 40($b$). Here $\alpha_d < 0$ or $\alpha_\Delta < d - \Delta$. Since $f(\alpha) \leqslant \alpha$ always, here $f_d(\alpha) < 0$ also. $\alpha_d < 0$ means that there are points where

$$E_{d,r1}/E_d = (b^{-k})^\alpha < E_{d,r2}/E_d = (b^{-(k+1)})^\alpha, \tag{A 7}$$

where $r_1/L = b^{-k}$ is larger than $r_2/L = b^{-(k+1)}$. This means that the dissipation in one of the offspring is larger than the total dissipation received by its predecessor. Naturally, this is possible only if the cascade is non-conservative in $d$-dimensions.

Another interesting range of singularities appears when $D_{d,q} \leqslant 0$, corresponding to $\langle \sum E^q_{d,r1} \rangle < \langle \sum E^q_{d,r2} \rangle$ whenever $r_1 > r_2$. The critical value of $q$ at which this happens is denoted by $q_{cr}$, and from (2.31), we see that the condition for $D_{d,q} = 0$ is

$$\langle M^{q_{cr}}_{(d)} \rangle = b^{-d}. \tag{A 8}$$

From figure 39(*a*) we see that $D_{d,q} = 0$ occurs near $q_{cr} \approx 7$ for the example (A 5). Following Mandelbrot (1974), we note that

$$M_{max} = \lim_{q \to \infty} \langle M^q_{(d)} \rangle_{1/q} = \lim_{q \to \infty} [b^{-d/q}] = 1. \tag{A 9}$$

This means that as soon as a multiplier $M_{(d)}$ becomes larger than 1, there will be some value of $q_{cr}$ above which $D_{d,q} < 0$ because $D_{d,q \to \infty} = 0$ implies that $M_{max} = 1$. Mandelbrot (1974) shows that the condition (A 9) is both necessary and sufficient for the existence of a $q_{cr}$.

Returning to figure 39(*a*), Mandelbrot (1984, 1989) and Schertzer & Lovejoy (1985) have remarked that one can now define the exponents $D_{\Delta,q}$ as the dimension of a set $S(D_q)$ which, when used to intersect the original measure in $\Delta$ dimensions, will produce a $q_{cr} = q$. The interpretation of $D_q$ as a dimension is thus justified.

We have so far illustrated the relationship between the exponents on different dimensions $\Delta$ and $d$. It was shown that $D_{d,q}$ can become negative, at least in principle. The question now is whether this is possible in practice. At this point it is important to realize that if one were measuring $E_{d,r}$ from an experiment where the cascade had proceeded down to the $n$th cascade level (box size $\eta/L = b^{-n}$), one would instead measure $E_{d,r}$ as the sum of all $E_{d,\eta}$ contained in the original box of size $r/L$. In that case one would obtain

$$\frac{E_{d,r}}{E_d} = \sum_{i=1}^{r/\eta} \prod_j^n M_{(d),j,i}. \tag{A 10}$$

If the cascade was conservative in $d$ dimensions it is easy to show that this would always be equivalent to (2.28), i.e.

$$\sum_i^{r/\eta} \prod_j^n M_{(d),j,i} \Big/ \prod_j^k M_{(d),j} = 1. \tag{A 11}$$

For non-conservative cascades, let us call this ratio $\Omega_{k,n}$. It is a fluctuating quantity that varies from box to box, but it is straightforward to prove that $\langle \Omega_{k,n} \rangle = 1$. Other interesting properties of $\Omega$ are that $\Omega_{n,n} = 1$ always, and that

$$\Omega_{k-1,n} = \sum_i^{b^d} \Omega_{k,n} M_{(d),k,i}, \tag{A 12}$$

locally. If one now measures the $D_q$ exponents from a $d$-dimensional cut one obtains (combining (2.28) and the definition of $\Omega_{k,n}$):

$$\langle \sum (E_{d,r}/E_d)^q \rangle = b^{kd} \left\langle \Omega^q_{k,n} \prod_{j=1}^k M^q_{(d),j} \right\rangle. \tag{A 13}$$

Using the assumption that the $M$ on different cascade levels are uncorrelated, and using the recursion relation (A 12), one obtains, after some manipulations, two asymptotic scaling regimes for $\langle \sum E_{d,r}/E_d)^q \rangle$. The precise cross-over depends on the statistics of $M$, or on the $D_q$ values. If

$$\langle M^q_{(d)} \rangle < b^{-d},$$

one obtains (for $n \gg k$)

$$\langle \sum (E_{d,r}/E_d)^q \rangle \sim [b\langle M_d^q \rangle]^k \sim (r/L)^{(q-1)D_{d,q}}, \tag{A 14}$$

so that the $D_{q,d}$ measured from the non-conservative cascade agree with the $D_{q,d}$ that one would obtain from the additive relations (2.34). On the other hand, if $\langle M_{(d)}^q \rangle > b^{-d}$, one obtains

$$\langle \sum (E_{d,r}/E_d)^q \rangle \sim b^{nd} \left\langle \prod_{j=1}^{n} M_{(d),j}^q \right\rangle \sim (\eta/L)^{(q-1)D_{q,d}}. \tag{A 15}$$

The cross-over occurs at $\langle M_{(d)}^q \rangle = b^{-d}$, which happens exactly when $D_{q,d} = 0$, or when $q = q_{cr}$. Also, for $q > q_{cr}$, $\langle \sum (E_{d,r}/E_d)^q \rangle$ no longer depends on $r/L$, but is a constant for a given $\eta/L$. (This constant diverges with $\eta/L$ since $D_{d,q} < 0$.) However, according to (A 15), if we were to measure $D_{d,q}$ from the measure at step $n$ by using boxes of varying sizes $r/L$, one would obtain the result that $D_{d,q} = 0$ (Meneveau 1989) for all values of $q > q_{cr}$. (This is valid asymptotically for $n \gg k$, or $\eta \ll r$.)

The dashed line in figure 39 ($b$) corresponds to $D_{d,q} = 0$ for $q > q_{cr}$ which would be the result of measurements on the $d$-dimensional cut performed after the cascade has proceeded to some high number of steps. The arrow in figure 40 ($b$) shows the corresponding position on the $f_d(\alpha)$ curve.

## Appendix B. Intermittency exponents

Let us consider moments of the local scaling exponent $\alpha$ itself. By considering the generating function

$$G(q) = \langle (E_r/E_L)^q \rangle, \tag{B 1}$$

and using the definition (2.8) of $\alpha$ with unity prefactor, we see that

$$dG(q)/dq|_{q=0} = \ln(r/L) \langle \alpha \rangle. \tag{B 2}$$

The spatial average $\langle \alpha \rangle$ in (B 2) is taken over all non-empty boxes. On the other hand it follows from (2.14) that

$$G(q) = (r/L)^{\tau(q)+D_0}, \tag{B 3}$$

and evaluating the derivatives of $G(q)$ at $q = 0$, we obtain

$$\langle \alpha \rangle = d\tau(q)/dq|_{q=0} = \alpha_0, \tag{B 4}$$

$$d^2G(q)/dq^2|_{q=0} = (\ln[r/L])^2 \langle \alpha^2 \rangle = (\ln[r/L])^2[(\ln[r/L])^{-1} d^2\tau/dq^2 + (d\tau/dq)^2]_{q=0}. \tag{B 5}$$

From this it follows that $\sigma_\alpha^2$, the variance of $\alpha$, is given by

$$\sigma_\alpha^2 = \langle (\alpha - \alpha_0)^2 \rangle = (\ln[r/L])^{-1} d^2\tau(q)/dq^2|_{q=0}. \tag{B 6}$$

We conclude that for a given $\tau(q) = (q-1) D_q$ curve, the variance of the variable $\alpha$ is a function of $r$, and decreases as $r$ decreases. For future convenience, we now focus on the variance $\sigma_{\ln E}^2$ of $\ln(E_r/E_L)$. Since $\ln(E_r/E_L) \sim \alpha \ln(r/L)$, it is clear that

$$\sigma_{\ln E}^2 = -d^2\tau(q)/dq^2|_{q=0} \ln(L/r). \tag{B 7}$$

Therefore, for multifractal measures, the variance of the logarithm of the measure in a box of size $r$ increases with decreasing box size. Comparing this result with (2.22), it follows that the intermittency exponent is given by

$$\mu = -d^2\tau(q)/dq^2|_{q=0}. \tag{B 8}$$

Taking higher-order derivatives of $G(q)$, it is easy to show (Meneveau 1989) that the $n$th centred moments of $\alpha$ (and of $\ln(E_r)$) are given in terms of higher-order derivatives of $\tau(q)$.

We remark that in Meneveau & Sreenivasan (1987*a*), we had defined an intermittency exponent in terms of the slope of $D_q$ at $q=0$ as $\mu = -2\,\mathrm{d}D_q/\mathrm{d}q|_{q=0}$. Around $q=0$, $\mathrm{d}^2D_q/\mathrm{d}q^2$ is usually quite small so that both definitions are numerically close, but conceptually not equivalent. We employ in this paper the definition of $\mu$ given in (B 8).

## Appendix C. Relation between the multifractal description and early cascade models

Early cascade models can be shown to correspond to special cases of multifractal distributions. The smooth non-intermittent character of Kolmogorov's (1941) theory implies that

$$D_{q,3} = 3 \tag{C 1}$$

for all $q$. This means that $\epsilon$ is space filling with no intermittency. Alternatively, we get from (2.20) and (2.21) that the $f(\alpha)$ curve degenerates to the point $\alpha_3 = 3, f_3(\alpha) = 3$.

On the other hand, recalling that $\alpha$ is proportional to $\ln(E_r/E_L)$ and that $f(\alpha)$ is proportional to the logarithm of the probability density function of $\alpha$ or $\ln(E_r)$, it is easy to realize that $f(\alpha)$ must be parabolic if the distribution of $\ln(E_r)$ is Gaussian. This corresponds to the lognormal model. Denoting by $m_{\ln E}$ and $\sigma^2_{\ln E}$ the mean and variance of $\ln(E_r/E_L)$ respectively, it is straightforward to show (Meneveau & Sreenivasan 1987*a*) that the lognormal distribution corresponds to

$$f(\alpha) = d-(\alpha-\alpha_0)^2/(2\mu), \tag{C 2}$$

where
$$\alpha_0 = \langle\alpha\rangle = m_{\ln E}[ln\,(r/L)]^{-1}, \quad \mu = \sigma^2_{\ln E}[\ln\,(L/r)]^{-1}. \tag{C 3}$$

For lognormal distributions, the conservation of the measure imposes a relation between its mean and variance, which can be expressed from (2.20) as a relation between $\alpha_0$ and $\mu$ by requiring that

$$f = \alpha \quad \text{when} \quad \partial f/\partial\alpha = q = 1. \tag{C 4}$$

The result is that
$$\alpha_0 - d = \tfrac{1}{2}\mu. \tag{C 5}$$

Applying the Legendre transforms to (C 2), we obtain the $\tau(q)$ curve for lognormal distributions to be

$$\tau(q) = (q-1)\,[d+(\mu/2)\,q], \tag{C 6}$$

giving
$$D_q = d-\tfrac{1}{2}\mu q. \tag{C 7}$$

It is clear that moments of order $q$ higher than $(2d)/\mu$ become negative. According to (2.15), if $D_q$ were negative, $\sum(E_r/E_L)^q$ would increase as the box size decreases, which is not possible in practice. Given that $f(\alpha)$ is related to the logarithm of the probability density of the dissipation normalized by $\ln(r/L)$, as one proceeds to smaller $r$-values (or more steps in the cascade) it is continually emphasizing the tails of the distribution for which the central-limit theorem does not hold. As more steps are taken into account (larger $n$), one would expect lognormality to become a better approximation over larger and larger regions of the distribution of $E_r$, yet not so for the logarithm of the distribution divided by $\ln(r/L)$. Thus the central-limit theorem does not apply for scaling exponents in the multifractal analysis, even asymptotically.

Despite these inadequacies of lognormality, it works well for low-order moments or the central part of the $f(\alpha)$ curve. This is because any reasonably smooth $f(\alpha)$ curve is well approximated around its maximum by its second-order expansion. Of course, the value of $\mu$ (related to the curvature of $f(\alpha)$ at its maximum) depends on the variance of the multipliers $M_j$, and cannot be determined from central-limit-type arguments. Thus, any general multifractal distribution has a 'universal' parabolic shape near the maximum of $f(\alpha)$ where the central-limit theorem applies, but the tails depend strongly on the details of the distribution of the multipliers.

For the $\beta$-model one obtains

$$D_q = D(\beta) = d + \log_b \beta \tag{C 8}$$

independent of $q$. It is easy to show that $f(\alpha)$ consists of a single point at $\alpha = f(\alpha) = D(\beta)$ for this model. One shortcoming of the $\beta$-model is that the dissipation has exactly the same value in all non-empty regions. The random $\beta$-model (Benzi *et al.* 1984), whose physics of eddy breakdown is basically the same as in the standard $\beta$-model ($0 < \beta \leqslant 1$), allows for fluctuations in the intensity of the (non-zero) values of the dissipation. For this model, one obtains

$$D_q = d + \log_b \langle \beta^{1-q} \rangle / (1-q), \tag{C 9}$$

leading to a non-trivial $f(\alpha)$ curve whose maximum is less than $d$.

The $\tau(q)$ or $f(\alpha)$ curves of the dissipation field can also be related to other inertial-range exponents if one estimates the local flux of kinetic energy at a particular scale $r$ by $\Delta u_r^3/r$ and assumes this to have statistics similar to $\epsilon_r$. It follows (Meneveau & Sreenivasan 1987*a*) that the $n$th-order velocity structure functions obey

$$\langle \Delta u_r^n \rangle \sim [\langle \epsilon \rangle L]^{n/3} (r/L)^{\xi_n}, \tag{C 10}$$

where

$$\xi_n = \tfrac{1}{3}n + (\tfrac{1}{3}n - 1)(D_{n/3} - d). \tag{C 11}$$

For $n = 2$, (C 11) can be shown to imply (with $d = 3$) that the energy spectrum has the form

$$\Phi(k) \sim k^{-[\frac{5}{3} + (3 - D_{\frac{2}{3}})/3]}, \tag{C 12}$$

which, for any $D_{\frac{2}{3}} < 3$, is steeper than the $-\frac{5}{3}$ spectrum predicted by Kolmogorov's 1941 theory (Mandelbrot 1976).

## Appendix D. Methods for evaluating velocity derivatives

This appendix contains a summary of the sensitivity studies with respect to different methods of evaluating the velocity derivatives used to compute $\epsilon'$. We compare typical log–log plots of $\sum (E_r/E_t)^q$ for $q = 2$ and $-2$ where $\epsilon'$ is obtained using the simple method (3.2) as well as three different alternatives. We apply these different methods to a segment of atmospheric data consisting of 80000 points. Figure 41 shows plots of $\log_{10}[\sum (E_r/E_t)^q]^{1/(q-1)}$ *vs.* $\log_{10}[r/\eta]$, where $E_r$ has been computed using the different methods of differentiation. Circles correspond to (3.2), and squares to $\epsilon'$ evaluated by taking derivatives as differences over distances larger than the sampling interval, namely over five data points as $\epsilon' = [u(t_{i+5}) - u(t_i)]$. The next method consists in evaluating the derivatives using a differencing scheme of fourth-order accuracy according to $\epsilon' = [8u(t_{i+1}) - 8u(t_{i-1}) - u(t_{i-2}) + u(t_{i-2})]$. The results are shown as triangles in figure 41. Finally, we employ a smoothing technique to the velocity signal, which consists in least-square fitting a parabola through five points around every data point of the velocity signal. Subsequent data processing was done according to (3.2). The asterisks show the results of that procedure.

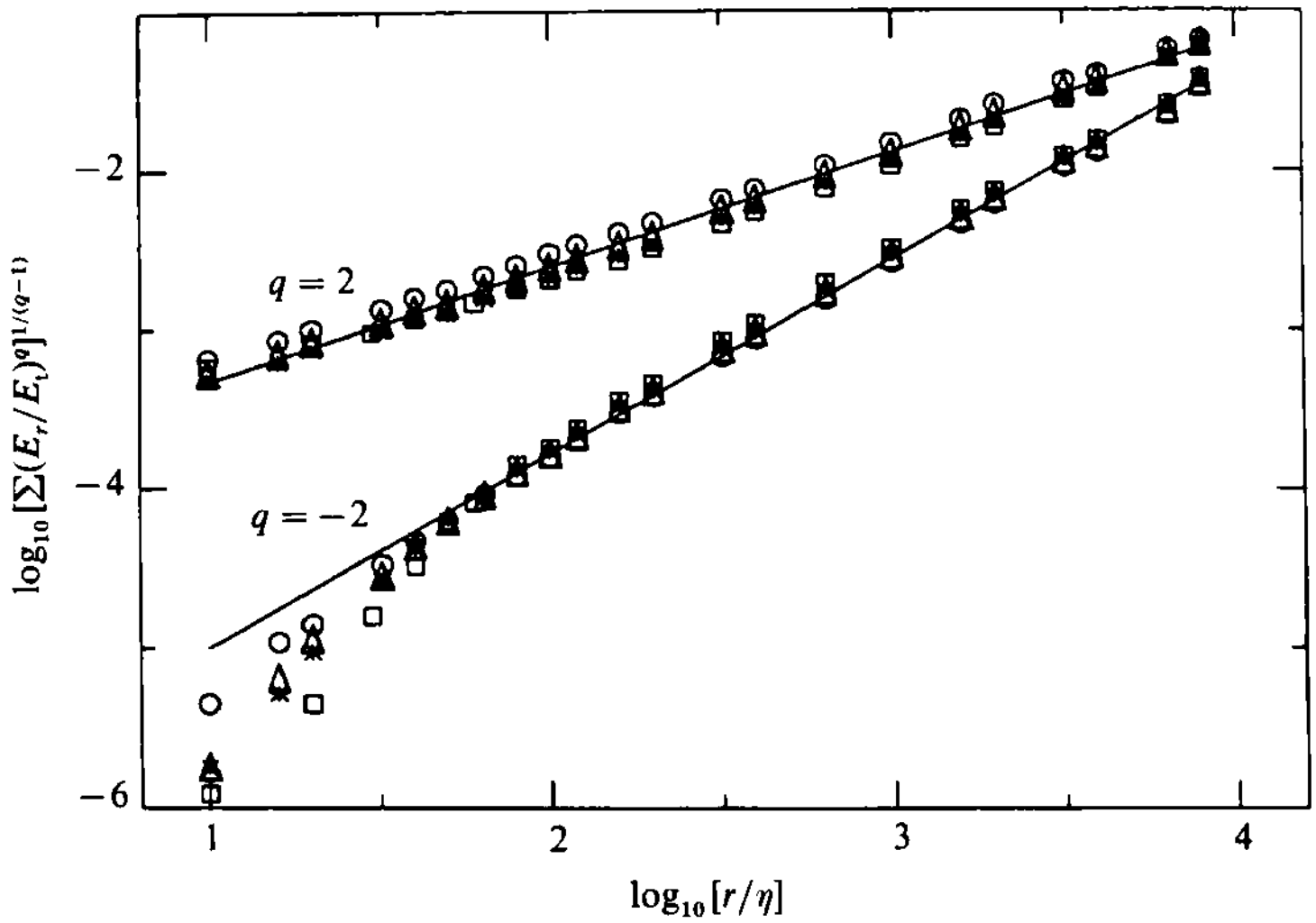

FIGURE 41. Log–log plots of $[\sum (E_r/E_t)^q]^{1/(q-1)}$ for the atmospheric surface layer as a function of $r/\eta$ for two representative $q$-values (2 and $-2$). Different symbols correspond to different methods of evaluating velocity derivatives and $\epsilon'$. Circles correspond to finite differencing between two neighbouring data points. Squares correspond to finite differences over points separated by five data points. Triangles correspond to a differencing formula of fourth-order accuracy, while the asterisks correspond to a smoothing technique consisting of least-square parabolic fit using five points around every data point. The results are quite robust with respect to the precise differencing technique. (For $q < 0$, this is true for box sizes large than $30\eta$.)

As is obvious from figure 41, the curves are at most shifted by small amounts, but the slopes are essentially unchanged. This is valid for both positive and negative values of $q$. Similar conclusions are obtained for other values of $q$, as well as for the other flows studied. We conclude that the results are robust with respect to the method of differentiation.

*Note added in proof*: A few additional remarks concerning the computation of the $f(\alpha)$ curve (figure 33) may be useful. One can take advantage of the thermodynamic analogy of multifractals and partially account for finite-size effects by employing 'Boltzmann weights' in computing $f$ and $\alpha$. This so-called canonical method (Chhabra & Jensen 1989) yields results in agreement with figure 33. One can also compute the $f(\alpha)$ curve by the multiplier distributions in the inertial range; see equations (2.31)–(2.33). In particular, this method has been shown by Chhabra & Sreenivasan (1990) to be capable of yielding negative dimensions reliably. The results from the multiplier method are also consistent with figure 33. The one unresolved issue is the relation between the present $D_q$ exponents for $q < 1$ and those of the inertial range quantity, namely the scale-to-scale energy flux.